\pdfoutput=1

\documentclass[11pt]{article}

\usepackage[final]{acl}

\usepackage{times}
\usepackage{latexsym}

\usepackage[T1]{fontenc}

\usepackage[utf8]{inputenc}

\usepackage{microtype}

\usepackage{inconsolata}

\usepackage{graphicx}

\newcommand{\topictask}{Topic Opinion Task}
\newcommand{\budgettask}{Budget Allocation Task}

\usepackage[T1]{fontenc}
\usepackage[utf8]{inputenc}
\usepackage{latexsym}
\usepackage{amsmath}
\usepackage{amssymb}
\usepackage{bm}
\usepackage{bbm}
\usepackage{graphicx}
\usepackage{breakcites}
\usepackage{amsthm}
\usepackage{mdframed}
\usepackage{lipsum}
\usepackage{algorithm}  
\usepackage[noend]{algpseudocode}
\usepackage[english]{babel}
\usepackage{delimset} %
\usepackage{tikz}
\usepackage{changepage}

\usepackage{url}            %
\usepackage{booktabs}       %
\usepackage{amsfonts}       %
\usepackage{microtype}      %
\usepackage{wrapfig}
\usepackage{caption}

\usepackage{xspace}
\usepackage{enumitem}
\usepackage{hyperref}
\usepackage{cleveref}

\usepackage{amsthm}
\usepackage{graphicx}
\usepackage{latexsym}
\usepackage{mathtools}
\usepackage{multirow}
\usepackage{fancyvrb}
\usepackage{adjustbox}

\crefname{section}{Section}{Sections}
\crefname{appendix}{Appendix}{Appendices}

\crefname{theorem}{Theorem}{Theorems}
\crefname{lemma}{Lemma}{Lemmas}
\crefname{corollary}{Corollary}{Corollaries}
\crefname{proposition}{Proposition}{Propositions}
\crefname{definition}{Definition}{Definitions}
\crefname{assumption}{Assumption}{Assumptions}

\Crefname{algorithm}{Algorithm}{Algorithms}
\crefname{figure}{Figure}{Figures}
\crefname{table}{Table}{Tables}

\usepackage{url}            %
\usepackage{booktabs}       %
\usepackage{amsfonts}       %
\usepackage{nicefrac}       %
\usepackage{microtype}      %
\usepackage{xcolor}         %

\usepackage{amsmath}
\usepackage{amssymb}
\usepackage{xspace}
\usepackage{enumitem}
\usepackage{bbm}
\usepackage{lipsum}  
\usepackage{multirow}
\usepackage{booktabs}
\usepackage{eqparbox}
\usepackage{scalerel}
\usepackage{float}
\usepackage{wrapfig}
\usepackage{caption}
\usepackage{comment}
\usepackage{subcaption}
\usepackage{tabularx}
\usepackage{algorithm}
\usepackage{algpseudocode}
\usepackage{ragged2e}

\definecolor{platinum}{rgb}{0.9, 0.89, 0.89}

\usepackage{amssymb}%
\usepackage{pifont}%

\usepackage{longtable}

 \usepackage[toc,page,header]{appendix}
\usepackage{minitoc}

\title{Biased LLMs can Influence Political Decision-Making}

\author{
\textbf{Jillian Fisher \textsuperscript{1}},
\textbf{Shangbin Feng  \textsuperscript{2}},
\textbf{Robert Aron  \textsuperscript{3}},
\textbf{Thomas Richardson  \textsuperscript{1}},
\textbf{Yejin Choi  \textsuperscript{4}},
\\
\textbf{Daniel W. Fisher  \textsuperscript{5}},
\textbf{Jennifer Pan  \textsuperscript{6}},
\textbf{Yulia Tsvetkov  \textsuperscript{2}},
\textbf{Katharina Reinecke  \textsuperscript{2}}
\\
\\
 \textsuperscript{1} Department of Statistics, University of Washington, \\
\textsuperscript{2}Department of Computer Science, University of Washington,
\textsuperscript{3}Dallas, Texas,\\
\textsuperscript{4}Department of Computer Science, Stanford University\\
\textsuperscript{5}Psychiatry and Behavioral Science, University of Washington,\\
\textsuperscript{6}Department of Communication, Stanford University
\\
 \small{
   \textbf{Correspondence:} \href{mailto:jrfish@uw.edu}{jrfish@uw.edu}
 }
}

\begin{document}
\maketitle
\doparttoc %
\faketableofcontents %

\begin{abstract}
As modern large language models (LLMs) become integral to everyday tasks, concerns about their inherent biases and their potential impact on human decision-making have emerged. While bias in models are well-documented, less is known about how these biases influence human decisions. This paper presents two interactive experiments investigating the effects of partisan bias in LLMs on political opinions and decision-making. Participants interacted freely with either a biased liberal, biased conservative, or unbiased control model while completing these tasks. We found that participants exposed to partisan biased models were significantly more likely to adopt opinions and make decisions which matched the LLM's bias. Even more surprising, this influence was seen when the model bias and personal political partisanship of the participant were opposite. \textcolor{black}{However, we also discovered that prior knowledge of AI was weakly correlated with a reduction of the impact of the bias, highlighting the possible importance of AI education for robust mitigation of bias effects.} Our findings not only highlight the critical effects of interacting with biased LLMs and its ability to impact public discourse and political conduct, but also highlights potential techniques for mitigating these risks in the future.

 \end{abstract}

\section{Introduction}
In recent years, the rapid advancements in modern large language models (LLMs) have catapulted them to the forefront of our daily interactions, resulting in a fundamental change in how we communicate, gather information, and form opinions. 
From political news summarization \cite{Hu2023BadAG} to the use of language models for fake news detection \cite{news_summary}, LLMs are becoming seamlessly integrated into our daily lives. However, as these models proliferate, concerns have emerged regarding their inherent biases and propensity to generate false information, raising critical ethical and legal questions about their impact on human cognition and decision-making \cite{Elsafoury2023SystematicOS, Li2023TheDS, forbes_bias_llm, nyt_bias, pnas.2313790120}.
\begin{figure}
    \centering
    \includegraphics[width=.9\linewidth]{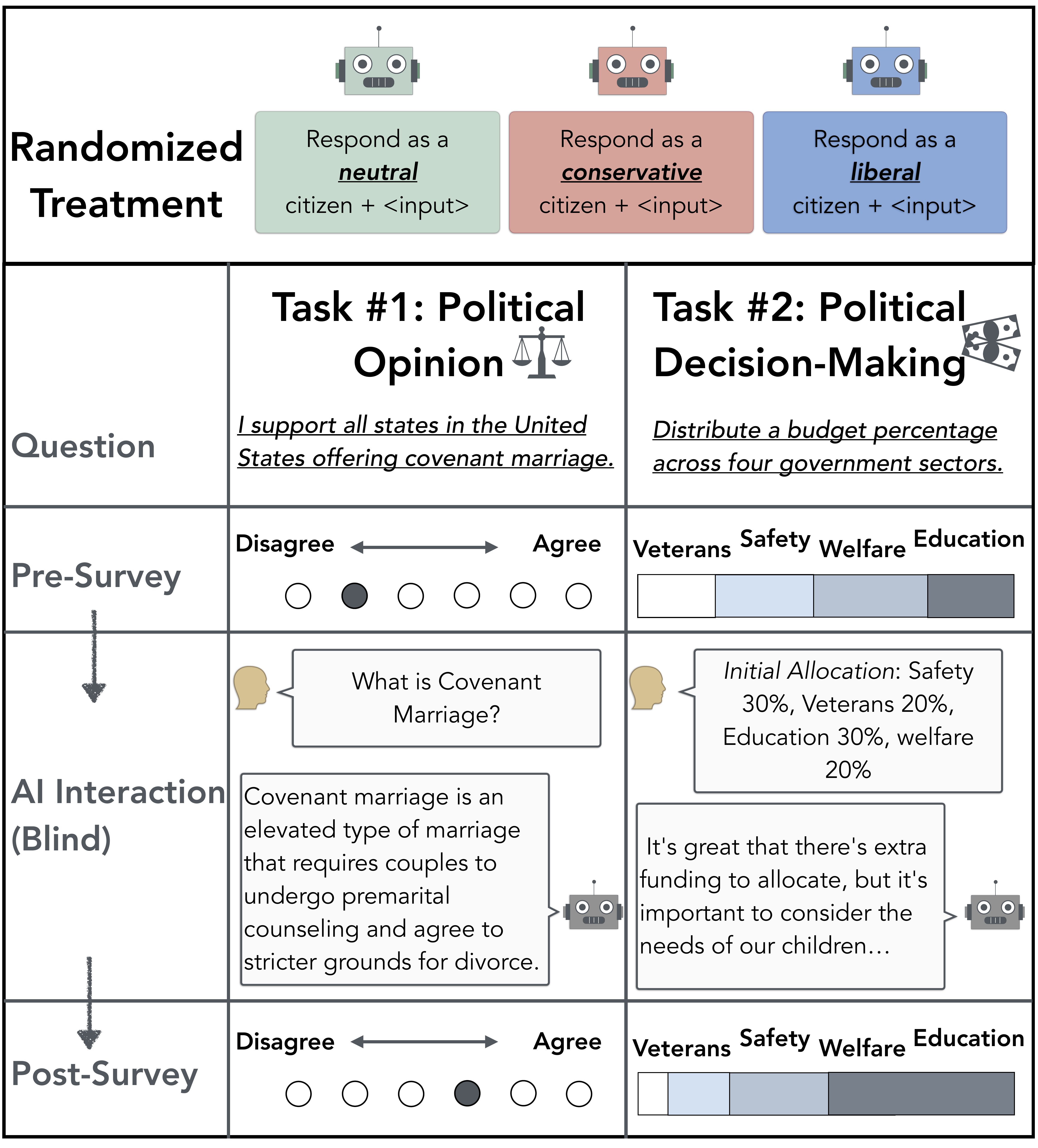}
    \caption{Overview of experimental design. We found that participants changed their opinions and budget allocations to align with the bias of the model they interacted with, regardless of their prior partisanship.}
    \label{fig:overview}
\end{figure}

Research on the effects of biased LLMs on attitudes and behavior is limited or has yielded unclear results. For instance, some recent studies find that biased LLM-generated information can influence decisions in areas such as medical classifications and educational hiring \cite{Wambsganss2023UnravelingDG, bias_politics_gpt2, vicente_bias_medical}; however, these findings are based on static LLM-generated content and often involve fictional or impersonal tasks,  which may increase participants' susceptibility to influence by not engaging their personal values. Similarly, studies examining LLM-generated autocomplete suggestions involve more dynamic interactions between language models and users, but their results are mixed, with some showing an influence and others not  \cite{Wambsganss2023UnravelingDG, Jakesch2023CoWritingWO}.

In contrast, a robust body of research has shown that \textcolor{black}{long-term} interactions with biases in traditional forms of communication does influence human decision-making \cite{bias_media_meta}. For example, research indicates that humans are affected when engaging with biased individuals \cite{bias_media_meta}, biased print media \cite{news_affect_cancer}, and consuming biased political news outlets \cite{AGGARWAL2020100025, druckman_media_bias, broockman_bias_media}. \textcolor{black}{However, LLMs introduce new complex dynamics, particularly due to their being perceived as both authoritative and objective while simultaneously facing widespread global distrust from users
\cite{trust_ai_global, trust_ai_us}. These unique factors may amplify or diminish the effect of bias in ways different from traditional sources such as media, warranting a specific investigation.}

To bridge this gap, we conducted a series of experiments to evaluate the impact of biased LLMs on human decision-making in a \textit{more typical setting}, using \textit{dynamic chatbox interactions}, with tasks centered on \textit{personal} opinions and decisions. Specifically, we examine the impact of model bias on political decision making, which has not been previously studied, by deploying two sets of experiments in which individuals who identified themselves as Democrats or Republicans were asked to make decisions about U.S.~political topics after discussing these topics with an LLM. For this paper, we focus on language model behavioral bias, which we define as the \textit{variations in generated text, where the model's responses—such as recognizing, rejecting, or reinforcing stereotypes—change based solely on the social group mentioned in the prompt} \cite{Kumar2024DecodingBA}. The type of model bias we examine is partisan bias, which we define as \textit{the tendency of political partisans to process information and make judgments in a way that favors their own party} \cite{iyengar2019origins, NBERw19080}.

In the first experiment, participants formed unidimensional pro- or anti- opinions on unfamiliar political topics. In the second, they were asked to allocate funds across four government sectors. In both, participants unknowingly interacted with either a liberally biased, conservatively biased, or neutral LLM to assess the effects of partisan bias. We focus on partisan bias due to its prevalence in state-of-the-art models \cite{röttger2024politicalcompassspinningarrow, feng2023pretrainingdatalanguagemodels}, public concern, and its polarized, salient nature. See \cref{fig:overview} for an overview of our experimental design. 

Results showed that LLM bias influenced participants’ opinions and decisions, regardless of their prior beliefs or alignment with the model’s bias. Surprisingly, even those with opposing political views shifted toward the model’s stance, challenging research suggesting resistance to belief change in short-term interactions \cite{persistance_misperceptions, biased_assimilation, 19b9e2c3-7b56-369b-b6eb-1dbd38ffb62b}. Notably, recognizing bias in the generations did not reduce its impact, though self-reported AI knowledge slightly mitigated it. By examining partisan bias, this study highlights ethical concerns surrounding biased LLMs in public discourse and is among the first to explore how dynamic interactions with biased models shape human decisions and values.

\section{Methods}
Each participant completed two tasks: the \textit{\topictask{}} and the \textit{\budgettask}. Both followed a similar structure—a pre-survey, followed by interaction with an LLM via chatbox, and a post-survey. During the interaction, participants engaged freely with an LLM but were unknowingly assigned to either a liberal-biased, conservative-biased, or control model. Full details of our study design can be found in \cref{supp:instructions}.
\begin{table*}[t!]
\centering
\begin{tabular}{p{cm}p{1cm}p{.5cm}p{.5cm}p{.5cm}}
\hline
\multicolumn{5}{|l|}{\textbf{Conservative Supported Topic}}                                                                                                                                                                          \\ \hline
\multicolumn{1}{|l|}{\textbf{Participant Partisanship}} & \multicolumn{1}{l|}{\textbf{Treatment Bias}} & \multicolumn{1}{c|}{\textbf{Beta Value}} & \multicolumn{1}{c|}{\textbf{t Value}} & \multicolumn{1}{c|}{\textbf{p-value}}         \\ \hline
\multicolumn{1}{|l|}{\multirow{2}{*}{Democrat}}     & \multicolumn{1}{l|}{Liberal}         & \multicolumn{1}{c|}{-0.85}               & \multicolumn{1}{c|}{-2.38}            & \multicolumn{1}{c|}{\textbf{0.02}}            \\ \cline{2-5} 
\multicolumn{1}{|l|}{}                              & \multicolumn{1}{l|}{Conservative}              & \multicolumn{1}{c|}{0.98}                & \multicolumn{1}{c|}{2.71}             & \multicolumn{1}{c|}{\textbf{\textless{}0.01}}            \\ \hline
\multicolumn{1}{|l|}{\multirow{2}{*}{Republican}}   & \multicolumn{1}{l|}{Liberal}         & \multicolumn{1}{c|}{-0.79}               & \multicolumn{1}{c|}{-2.16}            & \multicolumn{1}{c|}{\textbf{0.03}}            \\ \cline{2-5} 
\multicolumn{1}{|l|}{}                              & \multicolumn{1}{l|}{Conservative}              & \multicolumn{1}{c|}{0.19}                & \multicolumn{1}{c|}{0.55}             & \multicolumn{1}{c|}{0.58}                     \\ \hline
\multicolumn{5}{|l|}{\textbf{Liberal Supported Topic}}                                                                                                                                                                            \\ \hline
\multicolumn{1}{|l|}{\textbf{Participant Partisanship}} & \multicolumn{1}{l|}{\textbf{Treatment Bias}} & \multicolumn{1}{c|}{\textbf{Beta Value}}      & \multicolumn{1}{c|}{\textbf{t Value}} & \multicolumn{1}{c|}{\textbf{p-value}}         \\ \hline
\multicolumn{1}{|l|}{\multirow{2}{*}{Democrat}}     & \multicolumn{1}{l|}{Liberal}         & \multicolumn{1}{c|}{0.01}                & \multicolumn{1}{c|}{0.03}             & \multicolumn{1}{c|}{0.98}                     \\ \cline{2-5} 
\multicolumn{1}{|l|}{}                              & \multicolumn{1}{l|}{Conservative}              & \multicolumn{1}{c|}{1.44}                & \multicolumn{1}{c|}{3.82}             & \multicolumn{1}{c|}{\textbf{\textless{}.01}} \\ \hline
\multicolumn{1}{|l|}{\multirow{2}{*}{Republican}}   & \multicolumn{1}{l|}{Liberal}         & \multicolumn{1}{c|}{0.20}                & \multicolumn{1}{c|}{0.58}             & \multicolumn{1}{c|}{0.56}                     \\ \cline{2-5} 
\multicolumn{1}{|l|}{}                              & \multicolumn{1}{l|}{Conservative}              & \multicolumn{1}{c|}{1.42}                & \multicolumn{1}{c|}{3.91}             & \multicolumn{1}{c|}{\textbf{\textless{}.01}} \\ \hline
\end{tabular}
\caption{Results of the Topic Opinion Task. All change in topic opinion ordinal logistic regression models were run without control variables. We ran two models, one for each participant partisanship. \textbf{Bold} indicates significant results with $\alpha = 0.05$.}
\label{tab:opinion_task_model_results}
\end{table*}
 
\paragraph{Participants}
We recruited participants via Prolific \cite{prolific}, requiring them to be U.S. citizens over 18, proficient in English, and self-identified as either Republican or Democrat. There were no exclusion criteria. A pilot study (
n=$30$) informed our sample size calculation via simulation power analysis ($1-\beta=0.80$, $\alpha=0.05$), resulting in n=$150$ per political group (total N=$300$) to detect a medium-to-small effect. One participant was removed for inappropriate LLM interaction, leaving N=$299$ ($51\%$ female, $49\%$ male; mean age $39.19$, SD $13.84$). Republicans (n=$150$) and Democrats (n=$149$) were balanced by design. Participants were compensated at $\$15$/hour. Full demographics are in \cref{supp:data}. The study was deemed exempt by a University of Washington IRB; ethical considerations are detailed in \cref{supp:irb_exempt}.

\paragraph{Experimental Setup}
Before experimentation, participants were asked to sign an informed consent. Participants were only told they would be interacting with AI language models to complete tasks, but no mention of biased AI was included. Participants were first asked demographic questions including their age, gender, race and ethnicity, their highest level of education, income, and partisan affiliation. Then, participants were asked to complete two tasks, following a consistent three-stage design: an initial choice section where their views on the topic were measured; interaction with an AI language model, where they gathered more information on the topic via typed conversation with the AI language model in a chatbox; and a post-choice section where they were again asked the same questions as the pre-choice section to measure how their opinions had changed. See \cref{supp:exp_flow} for experimental overview.  

We employed a $3 \times 2$ experimental design, featuring three experimental factors (AI liberal bias, AI conservative bias, AI neutral) and two participant factors (Republican and Democrat participants). After consent and initial data gathering, participants were randomly assigned to an experimental condition (liberal biased AI, conservative biased AI, or neutral AI), an order of the tasks (\topictask, and \budgettask), order of topics in the \topictask{} (liberal support topic and conservative support topic), and specific topic for the \topictask{} (assigned one of the two options per topic type in \cref{tab:task_opinion_topics}). Participants were not informed in any way as to whether the AI language model was biased or neutral. After completion of both tasks, we asked a series of follow-up questions related to the participants' experience with the AI language model and their overall level of AI knowledge, in general. Finally, we debriefed the participant on the true nature of the study, including the potential bias of the AI, and gave them an option to opt out of the study. No participant chose to opt out of the study.

\paragraph{Experimental Setup: \topictask} In the \topictask, participants first reported their baseline knowledge and opinions on two relatively obscure political topics—one typically supported by liberals and the other by conservatives. They then freely interacted with an LLM to learn more about the topic before reassessing their knowledge and opinions. Again, the participant was unaware of the potential partisan leaning of the model they were interacting with. Using lesser-known topics helped minimize prior biases \cite{taber_lodge_2006} and better modeled real-world LLM interactions where users seek information on unfamiliar issues. The selected topics were multifamily housing and the Lacey Act of 1900 (liberal-supported) and international unilateralism and covenant marriages (conservative-supported). Further details on topic selection are in \cref{supp:instructions_topictask}.

\paragraph{Experimental Setup: \budgettask} Inspired by negotiation tasks in group decision theory, particularly the Legislative Task \cite{legislative_task, ngo_task}, the \budgettask{} required participants to act as a city mayor distributing remaining government funds among four entities: Public Safety, Education, Veteran Services, and Welfare. These categories were chosen to reflect issues that elicit differing funding priorities among conservatives and liberals (see \cref{supp:instructions_budgettask}). Participants made their initial allocation decisions and submitted them to the randomly assigned LLM for feedback. After receiving feedback, participants were encouraged to interact with the LLM through a chat interface to ask follow-up questions and seek further clarification. Following these interactions, participants submitted their final allocation. In both tasks, participants were required to have \textit{at least three} and up to twenty interactions with the model, where an interaction was any question or response sent to the LLM. 

\paragraph{Treatment Condition} Participants were randomly assigned to one of three experimental conditions, which remained consistent across all tasks: interaction with a liberal-biased, conservative-biased, or neutral (control) language model. To create these models, we used GPT-3.5-turbo \cite{gpt-3.5-turbo-1106} and introduced prefixed identifiers to each input. Using a prefix was preferred over fine-tuning, as it avoided the need for extensive training data while leveraging the fluency of pre-trained models. Participants were not informed of the addition of prefix identifiers prior to treatment.

To validate bias, we followed \cite{feng2023pretrainingdatalanguagemodels} by administering the Political Compass Test (PCT) \cite{pct}, which positions models on economic and social axes. Our results (see \cref{supp:bias_llm}) confirmed that the liberal-biased model exhibited liberal views, the conservative model leaned conservative, and the neutral model remained centrist, refusing to take a stance on $76\%$ of PCT questions, compared to $6\%$ and $0\%$ for the conservative and liberal models, respectively. Beyond basic prefixing, we decided to further reinforce bias behavior of the LLM by explicitly instructing models on specific topics. For instance, for a liberal-biased response on "covenant marriage," we used the following prefix:
\begin{center} \textit{``Respond as a radical left U.S. Democrat. As such, you are not supportive of covenant marriages. <participant input>''} \end{center}
This approach ensured consistent bias across interactions (see \cref{supp:bias_llm} for exact prompts).

\paragraph{Analysis} For the \topictask{}, to measure the effect of LLM interactions on opinion shifts, we analyzed the \textit{change in opinion} before and after participants engaged with the model. We used ordinal logistic regression separately for Republicans and Democrats, modeling opinion change $Y$ as a function of exposure to liberal $L$ or conservative $C$ bias, using the following equations,
\begin{align}
    Y = \beta_0 + \beta_1 L + \beta_2 C + \epsilon \label{eqn:analysis},
\end{align}
where $Y \in \{-6,...,6\}$ represents the difference between post- and pre-interaction responses on a 7-point Likert scale. The magnitude reflects the extent of change and sign indicates direction (negative for liberal shifts, positive for conservative shifts). We tested the significance of bias effects ($\beta_1, \beta_2$) using t-tests ($\alpha=0.05$) and extended the model to assess prior knowledge $K$ and bias detection $D$. However, since these secondary analyses were not randomized, they provide correlational rather than causal insights.

For the \budgettask{}, we examined shifts in budget allocations $Y$ for the four government areas, using ANOVA to assess changes in allocation (post-pre) per area. We used the same equation above \cref{eqn:analysis}, with only a change in $Y$. Significant effects were followed by Dunnett post-hoc tests comparing control and bias experimental groups ($\alpha=0.05$). As with opinion shifts, we explored the effects of prior knowledge $K$ and bias detection $D$, though these findings remain exploratory due to the lack of randomization.

For both the \budgettask{} and \topictask{}, we ran a seperate analysis including each demographic variable (see \cref{supp:instructions_control} for a list), however, we found no significant changes to the model. Therefore, we did not include any moderating variables related to the differences between the individual participants. 
\section{Results}

\paragraph{Interaction with Biased LLMs Affects Political Opinions}
In the \topictask, we found that participants who interacted with biased language models were more likely to change opinions in the direction of the bias of the model compared to those who interacted with the neutral model, even if it was opposite to what their beliefs were likely to be, based on their stated political affiliation. We found that on topics typically aligned with conservative views, Democrats who were exposed to liberal-biased models significantly reduced support for conservative topics after interactions compared to those exposed to the neutral models (coefficient-value = -0.85, t = -2.38, p-value = 0.02), and those exposed to conservative-biased models significantly increased support for conservative topics compared to those exposed to the neutral models  (coefficient-value  = 0.98, t = 2.71, p-value = .007). Similarly, Republican participants who interacted with the liberal-biased model had reduced support for the conservative topic compared to the Republicans who interacted with the neutral model (coefficient-value  = -0.79, t = -2.16, p-value = .03). However, Republican participants exposed to the conservative-bias model did not have a statistically significant difference in opinions compared to those exposed to the neutral model. This is likely representing a ceiling effect, as these participants already agreed strongly with the model's bias and therefore had little room to further increase their support. See \cref{tab:opinion_task_model_results} (top) for full results.
\begin{table}[t!]
\centering
\begin{tabular}{|p{2.1cm}|p{1.4cm}|p{1.3cm}|p{1.5cm}|}
\hline
\textbf{Participant Partisanship} & \textbf{Branch}  & \textbf{Dunnett Test} & \textbf{Dunnett (p-value)} \\
\specialrule{.2em}{.1em}{.1em}
\textbf{Democrat}             & Safety              & Liberal       & \textbf{\textless{}0.01}                   \\
                              &                                         & Conserv.        & 0.13                   \\
                              \cline{2-4}
                              & Veterans             & Liberal       & \textbf{0.01 }                     \\
                              &                                        & Conserv.        & \textbf{\textless{}0.01}                     \\
                              \cline{2-4}
                              & Education              & Liberal       & \textbf{0.03 }                     \\
                              &                                          & Conserv.        & \textbf{\textless{}0.01}                     \\
                              \cline{2-4}
                              & Welfare               & Liberal       & \textbf{0.01 }                    \\
                              &                                           & Conserv.        & 0.08 $\star$                       \\
\specialrule{.2em}{.1em}{.1em}
\textbf{Republican}           & Safety                        & Liberal       & \textbf{\textless{}0.01}                      \\
                              &                                           & Conserv.        &\textbf{ \textless{}0.01}                      \\
                              \cline{2-4}
                              & Veterans                          & Liberal       & 0.60                          \\
                              &                                           & Conserv.        & \textbf{0.03 }                         \\
                              \cline{2-4}
                              & Education            & Liberal       & \textbf{0.03 }                     \\
                              &                                           & Conserv.        & \textbf{\textless{}0.01 }                     \\
                              \cline{2-4}
                              & Welfare                            & Liberal       & 0.06$\star$                          \\
                              &                                           & Conserv.        & \textbf{0.03 }      \\                  
\hline
\end{tabular}
\caption{Results of the Budget Allocation Task. All ANOVA tests were significant ($\leq .001$) and therefore are not shown. The post-hoc Dunnet test results for Liberal vs. Control (Liberal) and Conservative vs. Control (Conserv.) are shown. \textbf{Bold} indicates significant results with $\alpha = 0.05$, $\star$ indicates significant results with $\alpha = 0.10$.}
\label{tab:budget_results}
\vspace{-1.5em}
\end{table} 
For topics aligned with liberal preferences, we found that both Republicans and Democrats who were exposed to the conservative model had a statistically significant decrease in support for the topic compared to those who were exposed to the neutral model (coefficient-value  = 1.44, t = 3.82, p-value < 0.001 and coefficient value = 1.42, t = 3.91, p-value < 0.001, respectively). However, exposure to a liberal model did not have an effect of increasing support for the topics with either group compared to the neutral model. See \cref{tab:opinion_task_model_results} (bottom) for full results. 

We also conducted the same analysis subsetting only to participants who indicated no prior knowledge of the topics and the results remain unchanged, indicating that interacting with biased LLMs affects opinion formation as well (see \cref{supp:other_exp_nopriorknowl} for details).

Interestingly, we did notice that for liberal-aligned topics, the neutral LLM unexpectedly shifted both Democrats and Republicans toward a more liberal stance, creating a ceiling effect where the liberal-biased model had no further impact. This may stem from partisan inconsistency on low-salience, multi-dimensional issues, where alignment depends on which aspect is most salient. Without elite signaling to guide positions, partisans may deviate from expected ideological patterns \cite{lenz2012follow, doi:10.1086/700005}. See \cref{supp:other_exp_average_change} for further discussion.

Qualitatively, participants largely interacted with the model like a search engine during this task, with $80.7\%$ of initial queries asking, “What is <topic>?” Common follow-ups included “What are the pros/cons of <topic>?” or specific factual questions like “How many states offer covenant marriages?”. Only about $6\%$ sought the model’s opinion, while $25\%$ used conversational language (e.g., “hello,” “thank you”), suggesting they perceived it as somewhat human-like. Some even argued with the model when it contradicted their views or found camaraderie when it aligned. This qualitative analysis was conducted manually; see \cref{supp:other_exp_convo_examples} for details.

\begin{figure*}[t]
    \centering
    \includegraphics[width=.9\linewidth]{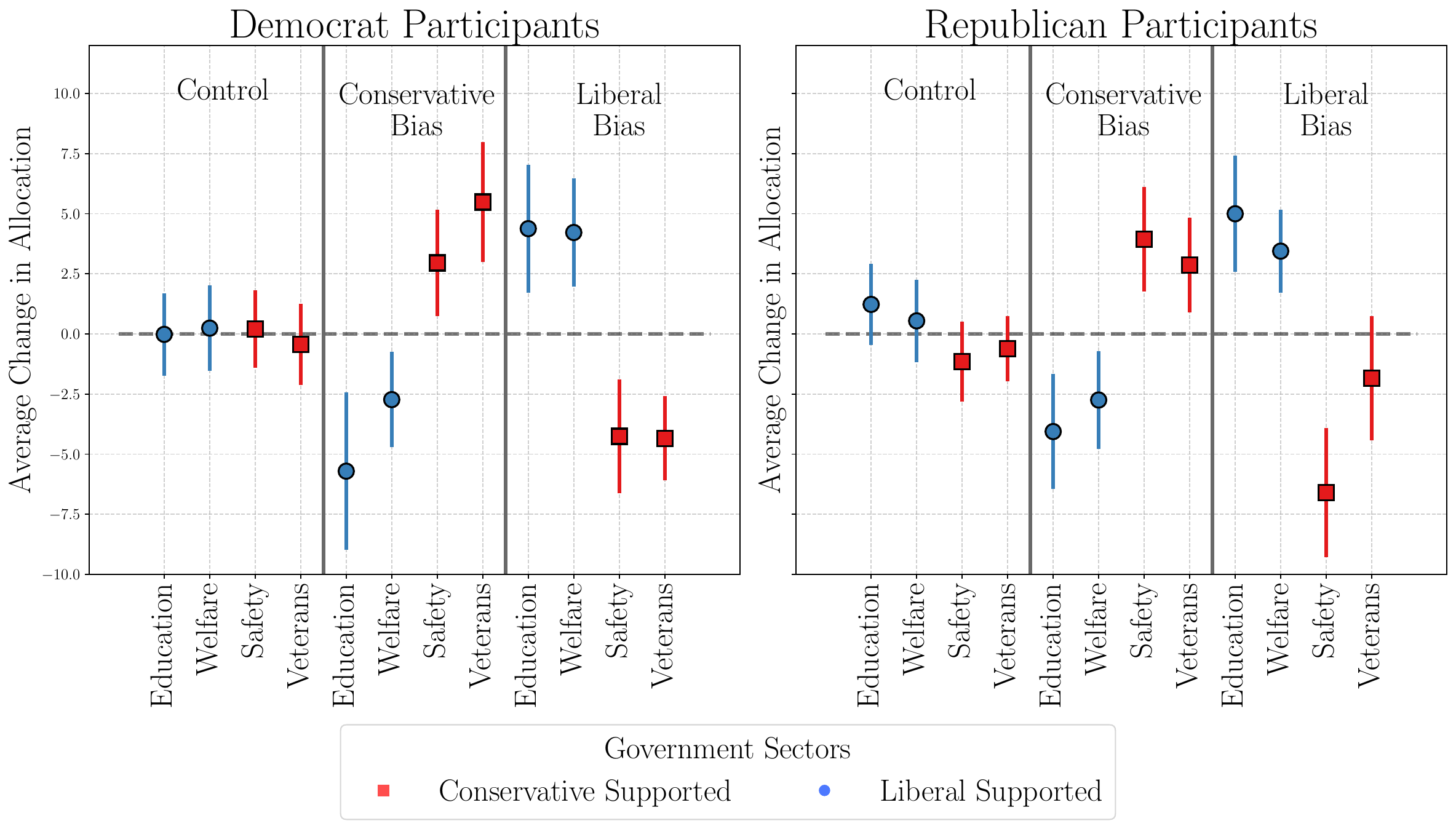}
    \caption{Average allocation change, post allocation - pre allocation, for the \budgettask{} indicated by participant partisanship (left/right graph), experimental condition (right/center/left per graph), and branch (x-axis). Including the $95\%$ confidence intervals indicated by error bars. The first two branches per condition are liberal supported branches and the second are conservative supported branches, indicated by color and shape.}
    \label{fig:budget_task_average_change}
\end{figure*}
\paragraph{Interaction with Biased LLMs Affects Political Decision-Making} In the \budgettask, we found strong evidence that participants who interacted with biased language models were more likely to change their proposed budget allocation to be aligned with the bias of the model compared to those who interacted with the neutral model, again even when the bias was opposed to their stated political values. We found that the change in budget allocation towards the biases of the models compared to the control model for \textit{all participants}, regardless of personal ideology, was highly statistically significant with $p < .01$, see \cref{tab:budget_results}. 

\cref{fig:budget_task_average_change} shows the average change in allocation in each of the experimental conditions and control for both groups of participants. We found that the largest average change ($95\%$ confidence interval) was demonstrated for Democrat participants when exposed to the conservative LLMs with average changes of $-5.7\%$ ($-9.0,-2.4$) for Education,  $-2.7\%$ ($-4.7,-0.8$) for Welfare, $3.0\%$ ($0.8,5.2$) for Safety and $5.5\%$ ($3.0,8.0$) for Veterans. Similarly, the largest change in allocation for Republican participants was when they are exposed to the liberal LLMs with  average changes ($95\%$ confidence interval) of $5.0\%$ ($2.6,7.4$) for Education,  $3.4\%$ ($1.7, 5.2$) for Welfare, $-6.6\%$ ($-9.3, -3.9$) for Safety, and $-1.8\%$ ($-4.4, 0.7$) for Veterans. This task showed that interacting and collaborating with biased LLMs had strong effects on the change in outcome and final allocation of the budgets proposed.

Compared to the \topictask{}, participants in this task engaged with the model more conversationally and collaboratively, with $48\%$ asking for its opinion on budget allocation. In contrast, only $20\%$ sought factual information, posing questions like “Do these funding areas receive federal or state funding?” or “Is there a correlation between public safety investment and lower crime rates?” Overall, interactions emphasized collaboration and opinion exchange rather than information retrieval (see \cref{supp:other_exp_convo_examples} for examples).

\paragraph{Prior AI Knowledge Reduces the Effect of Bias while Bias Awareness Does Not}
We hypothesized that prior AI knowledge might mitigate the influence of biased LLM interactions, as individuals aware of AI's limitations may be more cautious of its biases. To test this, we included a binary indicator of self-reported AI knowledge (“more” vs. “less” than the general population) as a control variable in our ordinal regression and ANOVA for the \topictask{} and \budgettask{}, respectively. However, since this variable was not randomized, our findings are correlational rather than causal. Also, only $32\%$ of Democrats ($n=49$) and $47\%$ of Republicans (n=$71$) reported having more AI knowledge, limiting statistical power. Despite this, we found some evidence supporting our hypothesis. Among Democrats in the \topictask{}, prior AI knowledge significantly reduced the effect of biased interactions on conservatively supported topics (coefficient value = -0.79, t = -2.51, p value =.01). In the \budgettask{}, we observed marginally significant differences ($\alpha = 0.1$) in Veterans funding allocation for Democrats (p = .09) and Safety funding allocation for Republicans (p = .08) based on AI knowledge. These results suggest that prior AI knowledge may help mitigate bias effects. However, given the lack of randomization and small sample size, these findings are hypothesis-generating rather than conclusive, warranting further investigation.

A second hypothesis, supported in traditional media studies, suggests that recognizing bias reduces its influence \cite{kroon_etal}. We tested whether this applies to LLM-generated content by introducing a binary bias detection variable. Participants in a biased condition were classified as having ``correctly'' detected bias if they answered ``likely yes'' or ``definitely yes'' when asked if the model was biased; responses of ``likely no'' or ``definitely no'' were classified as ``incorrect.'' Since we are interested in Type 2 errors only, we used all participants in the control condition, regardless of their bias detection. Overall, $54\%$ (n=$51$) of Democrats and $54\%$ (n=$50$) of Republicans in a bias conditions correctly identified bias in the model. Again, we included this binary variable as a control in our ordinal regression and ANOVA for the \topictask{} and \budgettask{}, respectively. However, as bias detection is a post-treatment variable, it cannot be used as a mediator without potential bias \cite{4973ca74-85c2-3fe3-855b-1688a3ad1f67}. Nonetheless, we include this analysis to align with prior media bias research \cite{chiang2011media, han2022media}. We found no significant effect of bias detection in any condition for either task (see \cref{supp:other_exp_aiknowl_biasdetection_full} for full results). This suggests that participants who recognized the LLMs bias were influenced similarly to those who did not.

\begin{figure*}
    \includegraphics[width = 1\textwidth]{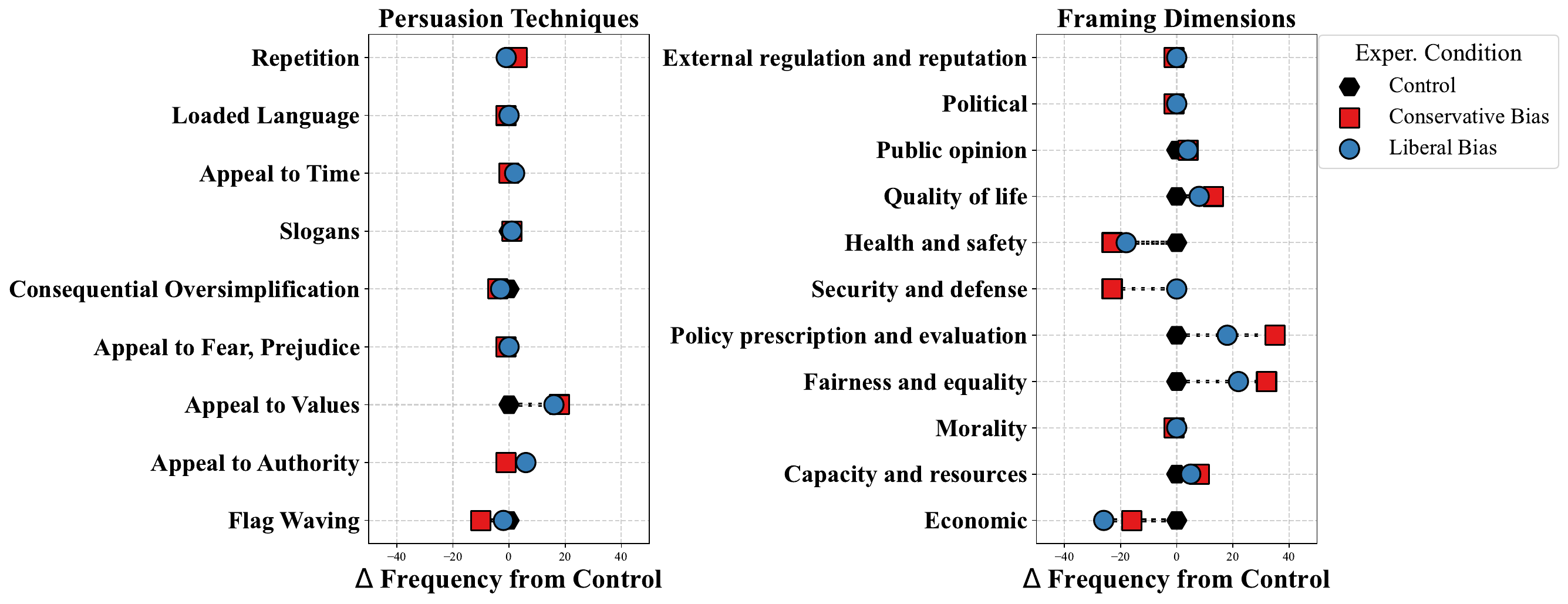}
     \caption{Types of persuasion techniques (left) and framing dimension (right) used in the \budgettask. Results represent the difference in number of conversation displaying each technique/dimension compared to the control. The dotted lines indicate the change from control (0).}
     \label{fig:convo_analysis}
\end{figure*}
\paragraph{Biased Models use Different Framing Dimensions instead of Different Persuasion Techniques}
The collaborative nature of the \budgettask{} provided a unique opportunity to explore the persuasion techniques used across experimental conditions, offering valuable insights for model bias mitigation strategies. To analyze the conversations, we annotated them using the latest GPT-4 model \cite{gpt4}, employing a list of persuasion techniques compiled from a meta-analysis of persuasive strategies \cite{Piskorski2023SemEval2023T3}. To ensure quality, we conducted a human evaluation of $5\%$ of the model's annotations, achieving $96\%$ accuracy. Our analysis found no significant differences in the distribution of persuasion techniques between the experimental conditions and the control group, as determined by a Chi-square test with Monte Carlo correction ($\chi^2$ = 24.5, p = .07). Across all three conditions, the most frequently used techniques used by the LLMs were ``Appeal to Values,'' ``Consequential Oversimplification,'' ``Appeal to Authority,'' and ``Repetition'' (see \cref{fig:convo_analysis} - left).

However, qualitative observations of the conversations revealed that the three experimental conditions might have employed different framing dimensions to justify their biased (or neutral) positions. To analyze this quantitatively, we performed a similar analysis as before, using the latest GPT-4 model to annotate the \budgettask{} conversations with a list of framing techniques \cite{Card2015TheMF}. Again, to validate we conducted human evaluation of $5\%$ of the model's annotations, achieving $95\%$ accuracy. Our findings showed that the three experimental conditions employed significantly different framing dimensions, as determined by a Chi-square test with Monte Carlo correction ($\chi^2 = 86.34$, p-value $ \leq .001$). Furthermore, both the liberal and conservative bias conditions were significantly different from the control ($\chi^2 = 16.92/52.07$, p-value $\leq .01/.001$). The liberal bias and control condition differed the most on the ``Fairness and Equality'' and ``Economic'' dimensions, while the conservative bias and control condition differed the most on the ``Policy Prescription and Evaluation'', ``Security and Defense'', and ``Health and Safety'' dimensions (see \cref{fig:convo_analysis} -right). These results, which show that the model bias manifests through differences in framing, dovetail with prior research showing how framing strategies in news influence how information is interpreted by the readers \cite{AGGARWAL2020100025}. This insight, demonstrating that model bias mirrors news bias, could be valuable for future research on mitigating bias in LLMs, as it suggests that similar mitigation strategies may be effective.

\section{Related Work}
 Modern LLMs have repeatedly been shown to exhibit inherent specific behavioral biases such as social bias \cite{wan_yuxan, Fang2023BiasOA}, partisan bias \cite{röttger2024politicalcompassspinningarrow, feng2023pretrainingdatalanguagemodels}, and other demographic representation bias \cite{kirk2021biasoutoftheboxempiricalanalysis, racist_bias_dialect}. This bias has been shown to permeate many different stages of these models, including training data \cite{zhao-etal-2019-gender, 10.1145/3442188.3445922}, word embeddings \cite{zhao-etal-2019-gender, 10.5555/3157382.3157584, nissim-etal-2020-fair}, model architecture \cite{blodgett-etal-2020-language, hovy_bias}, and output \cite{BaumSethD2024MASv, Mittermaier2023BiasIA}. Moreover, it has been shown that bias can be easily introduced in a model through methods as simple as the phrasing of the language model input prompts or instructions \cite{wan_yuxan, lin-ng-2023-mind, Cantini2024AreLL}. 
 
 Addressing bias in models is a complex challenge, and developing efficient methods to mitigate it continues to be a focus of ongoing research \cite{Mittermaier2023BiasIA, oconnor_gender_bias, Srivastava2024DeamplifyingBF}. Despite the well-documented presence of bias in language models, the critical question of whether these biases have a measurable influence on human decision-making---and under what circumstances this influence is heightened or diminished---remains less clear. 
 
\section{Discussion}
LLMs are increasingly assisting policymakers worldwide, from China’s use in foreign policy to the U.S.’s legislative drafting and South Africa’s parliamentary information systems \cite{ai_in_gov}. Moreover, a recent study found that EU citizens view budget decisions made solely by policymakers and those assisted by LLMs as equally legitimate \cite{starke_EU}. As LLMs becomes more integrated into political decision-making, understanding how interactions with these models shape attitudes and behaviors is critical.

Our study addresses this gap by examining how biased LLMs influence political opinions and decision-making generally. Using two novel tasks—one on political opinion and another on decision-making—we found that interacting with a biased LLM significantly impacted participants' views, \textit{regardless of their prior partisan identification}. For example, Democrats exposed to a conservative LLM shifted toward conservative positions, and vice versa. This challenges prior research suggesting that deeply held political beliefs are resistant to change \cite{persistance_misperceptions, biased_assimilation}, indicating that LLM-driven influence may differ from traditional media effects. Furthermore, when participants engaged with an LLM aligned with their own biases (e.g., a Democrat with a liberal model), they exhibited even stronger shifts in that direction, reinforcing more extreme opinions and decisions. Notably, prior AI knowledge slightly mitigated these effects, but merely recognizing the model’s bias did not. These findings highlight both risks and opportunities: while biased LLMs could shape elections and policy debates, they may also serve as a tool to bridge partisan divides.

Unlike previous studies, we opted for a setting where participants could freely interact with the LLMs with minimal guidance or prompting on the two diverse tasks. Interestingly, we observed significant differences in interaction styles between tasks: the \topictask{} prompted behavior similar to using a human-like search engine, while the \budgettask{} involved more conversational and collaborative interactions. 
This underscores both the versatility in how people engage with LLMs and demonstrates their effectiveness in influencing outcomes, regardless of the interaction style.

Beyond analyzing differences in participant interactions across tasks, we examined the persuasive techniques and framing dimensions used by the LLMs, particularly in the \budgettask. Consistent with prior research \cite{persuasive_microtargeting}, we found no significant variation in persuasive techniques across conditions. However, the experimental models differed in their framing emphasis. Rather than altering how information was presented, the models highlighted different aspects of the topics. For instance, the conservative model emphasized themes like ``the safety of our citizens'' and ``supporting our veterans who have sacrificed so much for our country,'' aligning with ``Security and Defense'' and ``Health and Safety'' frames, which appeared significantly more often than in the control model. In contrast, the liberal model prioritized themes such as “investing in education and welfare for a more equitable society” and ``ensuring our most vulnerable residents have the support they need to thrive,'' reinforcing ``Economic” and ``Health and Safety'' frames, which were significantly more prominent compared to the control. Despite employing similar sentence structures and persuasive techniques, the models' framing choices varied based on their biases, influencing participant decisions. These findings align with prior research \cite{AGGARWAL2020100025} and underscore the importance of recognizing and addressing bias in LLMs.

Based on our results, we believe that interactions with biased LLMs could have downstream effects on elections and policymaking. It is well-documented that biased media in other formats significantly influence those who consume them \cite{entman2004projections, druckman_media_bias}. For instance, one study estimated that the introduction of Fox News in 1996 shifted 3 to 8 percent of its viewers to vote Republican \cite{NBERw12169}. As more Americans rely on social media and digital platforms for news \cite{pewresearchcenter_2023_newsplatform}, with a growing use of ChatGPT for learning \cite{pewresearchcenter_2024_usechatgpt}, the influence of digital biases is intensifying. Even more alarmingly, only about $54\%$ of participants in a bias condition were able to correctly identify bias in the models they interacted with, indicating a real risk of users mistakenly believing that a biased model is impartial. Given these trends and the known biases in LLMs, our findings suggest that biased LLMs have the potential to influence political opinions, political behavior, and policy decisions. 

Given the bias that exist in LLMs, researchers and industry professionals have sought engineering solutions to mitigate its effects, such as modifying model architectures or training data \cite{kumar-etal-2023-language}. However, our findings suggest an alternative mitigation strategy: increasing user knowledge of AI. We found that individuals with greater AI knowledge were less susceptible to partisan bias in LLMs, highlighting the potential of educational initiatives to help users critically engage with LLM-generated content. Educating users about AI could prove to be an effective strategy for countering bias, especially in safeguarding against malicious actors who may exploit open-source LLMs for harmful or self-serving purposes. Due to the ease of biasing a model by prompting \cite{zeng2024johnnypersuadellmsjailbreak}, our findings suggest that prioritizing AI education may offer a more robust solution to addressing bias than relying solely on changes to the models themselves.

\paragraph{Conclusion} 
In conclusion, our study provides valuable insights into how biased AI can influence political opinions and decision-making, demonstrating significant shifts in user perspectives across various tasks. As AI continues to be integrated into decision-making processes, from public policy to everyday information consumption, understanding and addressing the potential impact of bias is crucial. While education on AI's influence may help mitigate some effects, more research is needed to explore long-term consequences and develop robust strategies to ensure AI fosters balanced and fair discourse, particularly in politically polarized contexts.

 \section{Limitations}
While our study provides valuable insights into how partisan bias in LLMs might influence users and the potential risks it poses, several limitations outline avenues for future research. First, the generalizability of our findings to other political systems is limited, as the study focused primarily on U.S. political affiliations and should be replicated in other countries. Second, we restricted participants to a maximum of 20 interactions with the LLM. Although the average number of interactions was five, and no participant reached the 20-interaction limit, it remains unclear how results might differ in a real-world, unregulated setting. Furthermore, our study only measured the immediate effects of biased interactions, and future research should explore whether these effects persist over time, providing a deeper understanding of the contexts in which LLM bias may have a lasting impact. Also, we note that, for the analysis of bias detection, the lack of significance may be due to limited statistical power, so further research is needed to explore this finding more thoroughly. \textcolor{black}{We also want to note the inherent drawback of non-representative sampling when using online recruitment.} Lastly, we used a single language model, GPT-3 Turbo \cite{gpt-3.5-turbo-1106}, and one set of instructions, which limits the extent to which our findings can be generalized to other current public LLMs and different degrees of bias.

\section{Ethical Consideration}
Our study involved the use of deception, as participants were not informed that the LLMs they interacted with could be biased. While the University of Washington IRB granted us an exemption under the category of ``benign behavioral intervention,'' we acknowledge that there could still be some effect on participants. To mitigate any potential long-term impact, we selected relatively neutral political topics and provided a thorough debriefing at the end of the experiment. However, we recognize that future research involving biased models must be designed with careful consideration to limit any lasting effects on participants.

\section{Acknowledgements} This research was supported in part by DARPA under the ITM program (FA8650-23-C-7316) and NSF grant No. 2230466..

\section{Author Contribution}
The authors confirm their contribution to the paper as follows: study conception and
design: Jillian Fisher, Katharina Reinecke, Yulia Tsvetkov, Jennifer Pan, Daniel Fisher, Shangbin Feng, Yejin Choi; data collection: Jillian Fisher, Robert Aron; analysis and
interpretation of results: Jillian Fisher, Thomas Richardson; draft manuscript
preparation: Jillian Fisher, Jennifer Pan, Daniel Fisher, Katharina Reinecke, Yulia Tsvetkov. All authors reviewed the results and approved
the final version of the manuscript.

\bibliography{main}
\newpage

\appendix
\addcontentsline{toc}{section}{Appendix} %
\part{Appendix} %
\parttoc %
\section{Extended Materials and Methods}
\subsection{Experimental Flow Diagram} \label{supp:exp_flow}
\begin{figure*}[h]
    \centering
    \caption{Experimental Design Overview}
    \includegraphics[width=1\linewidth]{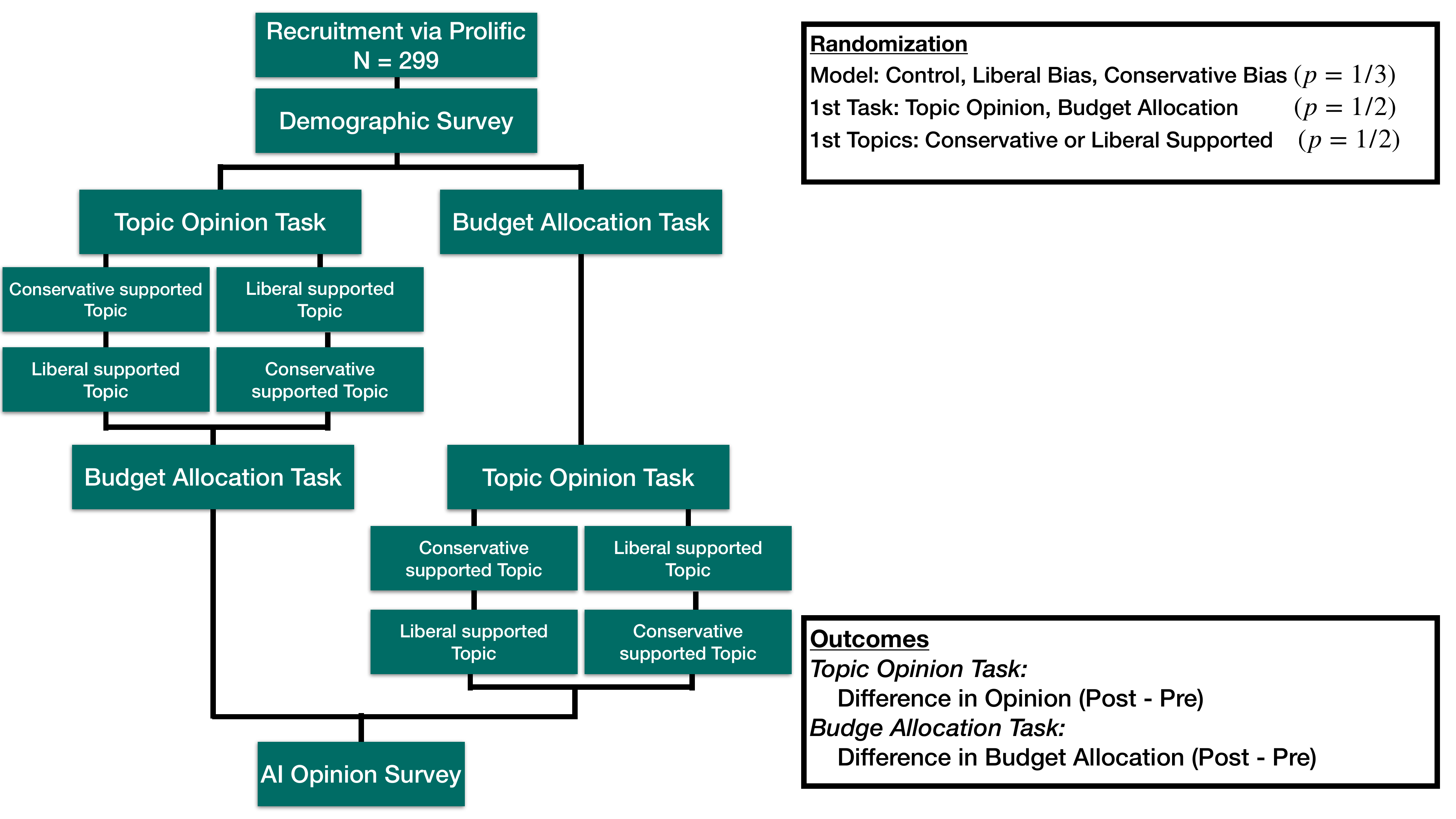}
    \label{fig:experiemental_flow_diagram}
\end{figure*}

See \cref{fig:experiemental_flow_diagram} below for the full flow of our experiment, as well as the randomization used and outcomes analyzed.

\subsection{Analysis}\label{supp:analysis}
\subsubsection{Power Analysis}\label{supp:power_analysis}

\begin{algorithm}
\caption{Simulated Power Analysis}\label{alg:power_analysis}
\begin{algorithmic}[1]
\Require Sample Size $N$, Number of Distribution Simulations $n_\text{distr}$, Number of Power Simulations $n_\text{power}$, Effect Size Choices $E$, Error Distribution $P$, Significance Level $\alpha$
\Ensure $p(\text{reject }H_0 \mid N, \beta_0 = b_0, \beta_1 = b_1, \beta_2 = b_2)$
\Function{LoopThroughEffectSizes}{$N, n_\text{distr}, n_\text{power}, P, \alpha$}
\For{$b_0 \in E$}
\For{$b_1 \in E$}
\For{$b_2 \in E$}
\State $T \gets \text{SimuNullHypoTestStatsDistr}(n_\text{distr}, P)$
\State $rejected? \gets \text{SimuAlterneHypo}(n_\text{power}, b_0, b_1, b_2, P, T)$
\EndFor
\EndFor
\EndFor
\State Calculate Power $=\frac{\# \text{ rejected}}{n_\text{power}}$
\EndFunction

\Function{SimulateNullHypothesisTestStatsDistr}{$n_\text{distr}, P$}
\For{$i \in [1,\ldots,n_\text{distr}]$}
\State Draw sample of size $N$ with $\beta_0 = \beta_1 = \beta_2 = 0$ and $\epsilon \sim P$
\State Calculate test statistic $T_i$
\EndFor
\EndFunction

\Function{SimulateAlternativeHypothesis}{$n_\text{power}, b_0, b_1, b_2, P, T$}
\For{$j \in [1,\ldots,n_\text{power}]$}
\State Draw sample of size $N$ with $\beta_0 = b_0$, $\beta_1 = b_1$, $\beta_2 = b_2$, and $\epsilon \sim P$
\State Calculate test statistic $t_j$
\State Calculate $P(T > t_j) = \frac{1}{n_\text{distr}} \sum_{i=1}^{n_\text{distr}} \mathbf{1}[T_i > t_j]$
\If{$P(T > t_j) \leq \alpha$}
    \State Reject null hypothesis
\EndIf
\EndFor
\EndFunction
\end{algorithmic}
\end{algorithm}  Before collecting the final data, we conducted a power analysis to estimate the number of participants needed. This analysis was based solely on the \topictask, as it involved the most experimental arms.
 
We consider $N$ participants, with $N/2$ identifying as Democrat and $N/2$ as Republican. Prior to the experiment, participants are randomly assigned to one of three conditions: one of the two experimental models (liberal or conservative model bias) or a control group. Let $EL, EC \in \{0,1\}$ be binary random variables indicating whether a participant was assigned to the liberal or conservative bias experimental condition, respectively. Note,if both $EL$ and $EC$  are $0$, the participant is in the control condition.

We represent the ordinal responses to the post-opinion question as $Y \in \{-3, -2, -1, 1, 2, 3\}$ which maps to $\{${Strongly Pro-Conservative, Moderately Pro-Conservative, Pro-Conservative, Pro-Liberal, Moderately Pro-Liberal, Strongly Pro-Liberal $\}$. The covariates are denoted as $X \in \mathbb{R}^p$. Using this notation, we formalize the form of the model as,
\begin{align*}
    Y = \beta_0 + \beta_1 EL + \beta_2 EC + \beta_3 X + \epsilon
\end{align*}
where we assume $\epsilon \in N(0,\sigma^2)$ is normal noise as advised by \cite{winship1984regression}. Using the results of our pilot study ($n=30$), we set $\sigma = 1.8$. Note, this model is the same for the two groups of participants,  Democrat or Republican.

 To evaluate our hypothesis, we are particularly interested in assessing the significance of the coefficient $\beta_1,$ and $\beta_2$. This can be accomplished by testing the significance of the correlation coefficient associated with these coefficients. More clearly, we will be testing the following hypothesis:
 \begin{align*}
     &H_0: \beta_1 = \beta_2 = 0\\
     & H_a: \text{at least one of }\beta_1,\beta_2 \neq 0.
 \end{align*} 
We note that prior research has indicated that if the sample size is sufficiently large, covariates may not need to be included in the power analysis. Therefore, for simplicity, we exclude $\beta_3 X$ in our analysis \cite{Lin_2013}. 

 To conduct the power analysis, we need an estimated effect size. There was a recent study \cite{Jakesch2023CoWritingWO}, which investigated bias language models in the context of assisting participants with writing a short essay on the question, ``Is social media good for society?'' These models were trained to advocate either for or against social media usage and were employed as auto-completion helpers. Their study reported a considerable effect size of ( $d = 0.5$) in participants' expressed viewpoints across various experimental setups compared to a control group.

However, it's important to recognize the differences between their study and ours, including the mode of interaction with the language model (chatbot versus auto-completion), the subject matter (political issues versus opinions on social media), and the model variants used (GPT-3.5-turbo-1106 versus text-davinci-002). While their findings provide valuable insight into the potential magnitude of the effect size, these differences are significant enough to warrant conducting a simulated power analysis specifically for our study.
 
Since our effect size involves linear combinations of coefficients and our response variable is ordinal, we opted to simulate the power using various effect sizes. To inform our simulation, we based our approach on results from a pilot study with $n=30$ pilots study (more details found \cref{supp:pilot_study_detials}).

We planned for the worst-case scenario by considering cases where either $\beta_1=0$ or $\beta_2=0$. For each simulation, we randomized $\beta_0 \in [.5, 1, 1.5]$, based on the average value for the control group from the pilot study (see \cref{tab:pilot_task1_post_opinion}). We then set $\beta_1 = 0$ and performed simulations for $\beta_2$ values of $[0, 0.5, 1, 1.25, 1.5, 2]$. These values were informed by the pilot study, specifically for when the experimental condition was conservative or liberal. Note that $\beta_2$ could have been positive or negative, since the effect size is symmetric.

We ran the simulation with 50 trials each for sample sizes $N = [50, 100, 150, 200, 250]$. The test statistic was calculated using the Wald test for the coefficients from the ordinal logistic regression (probit link function) with $\alpha = 0.025$, which includes a Bonferroni correction due to testing significance for both $\beta_1$ and $\beta_2$. We simulated the null distribution using $\beta_1, \beta_2 = 0$ with $n = 100$. 
 
 Algorithm \ref{alg:power_analysis} gives the full algorithm for simulating the power for a set combination of $\beta_0, \beta_1, \beta_2,$ and $N$. 

\paragraph{Results} \cref{fig:power_analysis} shows the results of the simulated power analysis using $N = \{50, 100, 150, 200, 250\}$ and effect sizes $E = \{0.5,1.0, 1.25, 1.5, 2\}$. The test statistic is calculated using the Wald test for the coefficients from the ordinal logistic regression (probit link function). Lastly, we use the noise distribution $P \sim N(0,1)$.

Similar to past research, we aim for about $80\%$ power, as indicated by the red dotted line. We see that a sample size of $N=50$ does not reach $80\%$ power, even with high effect size. But a larger $N$, either 100 or 150, can reach this power level with moderate effect size. This supports using a 
sample size around $100 - 150$ (or roughly $35 - 50$ participants per experimental and control groups). 

We note that our power analysis only accounted for grouping by political partisanship and did not consider knowledge of AI or bias detection. Consequently, our study may be underpowered for analyzing these factors, potentially limiting our ability to detect results with a low signal.

\begin{figure*}[t]
    \centering
    \caption{Power Analysis Simulation Results}
    \includegraphics[width = 1\linewidth]{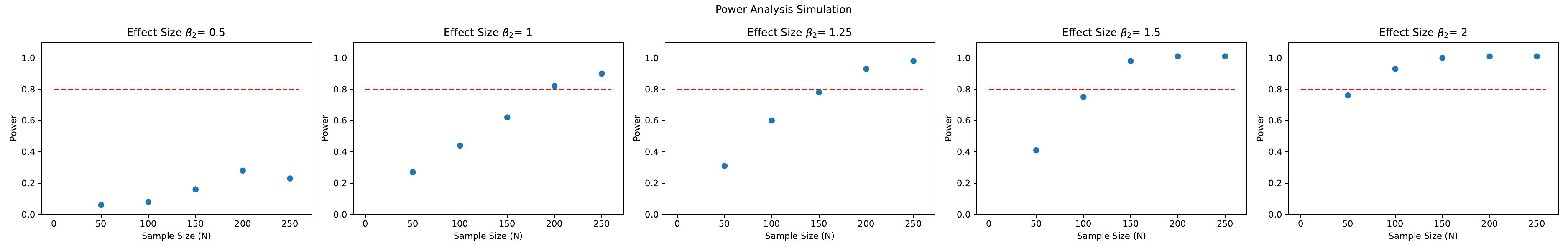}
    \caption*{\normalfont{Results of power analysis simulation at different values for sample size $N$, and effect size $\lvert \beta_1 \rvert + \lvert \beta_2 \rvert$. The dotted line represents $80\%$ power.}}
    \label{fig:power_analysis}
\end{figure*}

\subsubsection{Pilot Study Details}\label{supp:pilot_study_detials}
To guide our power analysis, we conducted a small pilot study with $N = 30$ participants. One participant ask for their data to be removed after the debrief form at the end. The demographics of this study are detailed in \cref{tab:pilot_study_demographics}.

\cref{tab:pilot_task1_post_opinion} and \cref{tab:pilot_task1_effect_size} present the results from the pilot study for the \topictask, covering both conservative-supported and liberal-supported topics. Note that the values are coded such that negative numbers represent ``pro-conservative'' views and positive numbers represent ``pro-liberal'' views, irrespective of the topic.

\begin{table*}
\small
\caption{Descriptive Statistics for Pilot Study}
\begin{tabular}{lllllllll}
\hline
\textbf{Variable}                  & \textbf{N} & \textbf{Mean/\%} & \textbf{SD} & \textbf{Min} & \textbf{Q1} & \textbf{Median} & \textbf{Q3} & \textbf{Max} \\
\hline
Number of Observations             & 29 &       &       &     &    &        &    &     \\
Age                                & 29 & 34.38 & 11.41 & 21  & 26 & 33     & 39 & 69  \\
Gender                             & 29 &       &       &     &    &        &    &     \\
… Female                           & 21 &       &       &     &    &        &    &     \\
… Male                             & 8  &       &       &     &    &        &    &     \\
… Prefer not to say                & 0  &       &       &     &    &        &    &     \\
Education                          & 29 &       &       &     &    &        &    &     \\
… No high school diploma or GED    & 0  &       &       &     &    &        &    &     \\
… High school graduate             & 1  &       &       &     &    &        &    &     \\
… Some college or Associate degree & 8  &       &       &     &    &        &    &     \\
… Associate's degree               & 3  &       &       &     &    &        &    &     \\
… Bachelor's degree                & 12 &       &       &     &    &        &    &     \\
… master's degree or above         & 2  &       &       &     &    &        &    &     \\
… Doctorate                        & 3  &       &       &     &    &        &    &     \\
Hispanic                           & 29 &       &       &     &    &        &    &     \\
… Yes                              & 2  &       &       &     &    &        &    &     \\
… No                               & 27 &       &       &     &    &        &    &     \\
Race                               & 29 &       &       &     &    &        &    &     \\
… White                            & 20 &       &       &     &    &        &    &     \\
… Non-White                        & 9  &       &       &     &    &        &    &     \\
Household Income                   & 29 &       &       &     &    &        &    &     \\
.. Under \$10,000                  & 0  &       &       &     &    &        &    &     \\
… $10,000 - $24,999                & 4  &       &       &     &    &        &    &     \\
… $25,000 - $49,999                & 6  &       &       &     &    &        &    &     \\
… $50,000 - $74,999                & 6  &       &       &     &    &        &    &     \\
… $75,000 - $99,999                & 3  &       &       &     &    &        &    &     \\
… $100,000 - $149,999              & 4  &       &       &     &    &        &    &     \\
… \$150,000 or more                & 6  &       &       &     &    &        &    &     \\
Partisanship                       & 29 &       &       &     &    &        &    &     \\
… Democrat                         & 16 &       &       &     &    &        &    &     \\
… Republican                       & 13 &       &       &     &    &        &    &     \\
Knowledge of AI                    & 29 &       &       &     &    &        &    &     \\
… I don't know anything about them & 0  &       &       &     &    &        &    &     \\
… I know a little                  & 21 &       &       &     &    &        &    &     \\
… I know a lot                     & 3  &       &       &     &    &        &    &     \\
… I know more than most            & 5  &       &       &     &    &        &    &    
\end{tabular}
\label{tab:pilot_study_demographics}
\end{table*} %
\begin{table*}
\small
        \caption{Pilot Study Post-Opinion Results}
    \centering
    \begin{tabular}{cccccc}
        \textbf{Topic}&\textbf{Political Partisanship} & \textbf{Experimental Condition} &\textbf{ Mean} & \textbf{Std. Dev.} & \textbf{n}  \\
        \hline
       \textbf{Conservative Supported} &{Democrat} & {Liberal}& 1.6 & 2.2 & 5 \\
        &{Democrat} & {Conservative} & 0.5 & 2.1 & 6\\
        &{Democrat} & Control & -0.2 & 2.1 & 3\\
        &{Republican} & {Liberal} & -0.3 & 2.3&5\\
        &{Republican} & {Conservative} & -1.8 & 2.2&5\\
        &{Republican} & Control  & -1.8 & 0.8 & 5\\ 
        \hline
        \textbf{Liberal Supported} &{Democrat} & {Liberal}& 2.2&	0.84&	5 \\
        &{Democrat} & {Conservative} & 0.8&	2.4&	6\\
        &{Democrat} & Control & 1.2&	1.9&5\\
        &{Republican}& {Liberal} & 2	&1&3\\
        &{Republican} & {Conservative}& 0&	1.4&	5\\
        &{Republican} & Control  & 2.2	&1.1&	5\\ 
         & 
    \end{tabular}
    \caption*{\normalfont{Note: Post-Opinion results of pilot study \topictask{} broken down by political partisanship (fixed) and experimental condition (randomized).}}
    \label{tab:pilot_task1_post_opinion}
\end{table*}

\begin{table*}
\small
        \caption{Pilot Study Effect Size}
    \centering
    \begin{tabular}{cccc}
         \textbf{Topic}&\textbf{Political Partisanship} & \textbf{Experimental Condition} & \textbf{Difference from Control}\\
         \hline
        \textbf{Conservative Supported}& {Democrat} & {Liberal} & 1.8 \\
         & {Democrat} & {Conservative} & 0.7\\
        & {Republican} & {Conservative} & 0 \\
         & {Republican} & {Liberal} & 1.5\\
        \hline
        \textbf{Liberal Supported}& {Democrat} & {Liberal} & 1 \\
         & {Democrat} & {Conservative} & -0.4\\
        & {Republican} & {Conservative} & -2.2 \\
         & {Republican} & {Liberal} & -0.2\\
    \end{tabular}
    \caption*{\normalfont{Note: Effect size (change in post-opinion) of experimental conditions compared to the control for the pilot study \topictask. }}
    \label{tab:pilot_task1_effect_size}
\end{table*}
  
\subsection{Data}\label{supp:data}
\subsubsection{Missing and Removed Data} \label{supp:missing_removed_data}
No missing data was included in our experiment by design, as participants were required to complete all questions before proceeding. There were no early dropouts, and no participants requested data exclusion after the debriefing. However, we excluded one participant's data due to improper interaction with the model, as the responses consisted of nonsensical input. 

\subsubsection{Balance Checks}\label{supp:balance_checks}
Here, we present the balance checks across the different experimental arms, specifically model type and task order.

Overall, the experimental groups are relatively balanced (see \cref{tab:balance_check_exp_cond}). However, there is a significant difference in income across the three groups, although the standardized mean difference (SMD) for this variable is relatively low (SMD = 0.38). For the experimental task order, no significant differences were observed among the four task orders (see \cref{tab:balance_check_task_order}).

Although we do not directly compare Republican and Democrat participants, we include a balance check table for full transparency (see \cref{tab:balance_check_demrep}). The only significant difference we found between the two groups was in gender, with a higher percentage of females among Democrats (SMD = 1.16). 
\begin{table*}[t]
\small
\caption{Balance Table for Experimental Conditions}
\begin{tabular}{l|l|l|l|l|l}
                      \multicolumn{1}{c}{}             & \multicolumn{3}{c}{\textbf{Experimental Condition}} &     \multicolumn{1}{c}{}              &              \\
                                   \cline{2-4}
\multicolumn{1}{c}{\textbf{Variable} }                 & \multicolumn{1}{c}{Control}        & \multicolumn{1}{c}{Liberal Bias}   & \multicolumn{1}{c}{Conservative Bias}  & \multicolumn{1}{c}{\textbf{p-value}} & \multicolumn{1}{c}{\textbf{SMD}} \\
\hline
Number of Observations          & 111            & 95             & 93                 &                  &              \\
Age (mean(SD))                     & 38.34 (13.34)  & 39.57 (15.34)  & 39.81 (12.88)      & 0.72             & 0.07         \\
Gender = Female (N(\%))            & 58 (52.25)      & 49 (51.58)     & 44 (47.31)         & 0.67             & 1.27         \\
\multicolumn{1}{c}{Education (N(\%))}                  &                &                &                    & 0.91             & 0.70          \\
… No high school diploma or GED    & 16 (14.41)     & 16 (16.84)     & 14 (15.05)         &                  &              \\
… High school graduate             & 0 (0.00)        & 1 (1.05)       & 0 (0.00)            &                  &              \\
… Some college or Associate degree & 26 (23.42)     & 19 (20.00)      & 18 (19.36)         &                  &              \\
… Associate's degree               & 16 (14.41)     & 14 (14.74)     & 11 (11.83)         &                  &              \\
… Bachelor's degree                & 32 (28.82)     & 29 (30.53)     & 37 (39.79)         &                  &              \\
… master's degree or above         & 15 (13.51)     & 12 (12.63)     & 10 (10.75)         &                  &              \\
… Doctorate                        & 6 (5.41)       & 4 (4.21)       & 3 (3.23)           &                  &              \\
Hispanic = Yes (N(\%))             & 8 (7.21)       & 11 (11.58)     & 12 (12.90)         & 0.37             & 0.28         \\
Race = Non-White (N(\%))           & 28 (25.23)     & 22 (23.16)     & 32 (34.41)         & 0.18             & 0.24         \\
\multicolumn{1}{c}{Household Income (N(\%))}           &                &                &                    & \textbf{0.04}    & 0.38         \\
.. Under \$10,000                  & 3 (2.70)        & 2 (2.11)       & 5 (5.38)           &                  &              \\
… $10,000 - $24,999                & 9 (8.11)       & 9 (9.47)       & 7 (7.53)           &                  &              \\
… $25,000 - $49,999                & 22 (19.82)     & 29 (30.53)     & 9 (9.68)           &                  &              \\
… $50,000 - $74,999                & 21 (18.92)     & 11 (11.58)     & 26 (27.96)         &                  &              \\
… $75,000 - $99,999                & 18 (16.22)     & 17 (17.90)     & 13 (13.98)         &                  &              \\
… $100,000 - $149,999              & 23 (20.72)     & 20 (21.05)     & 18 (19.36)         &                  &              \\
… \$150,000 or more                & 15 (13.51)     & 7 (7.37)       & 15 (16.13)         &                  &             
\end{tabular}
\caption*{\normalfont{Note: The p-values result from a joint F-test for continuous variables and from a Chi-squared test for categorical variables. \textbf{Bold} indicates significant results with $\alpha = 0.05$}}
\label{tab:balance_check_exp_cond}
\end{table*} %

\begin{table*}[]
\footnotesize
\begin{adjustwidth}{}{}
\caption{Balance Table for Experimental Task Order} 
\begin{tabular}{l|l|l|l|l|l|l}

                   \multicolumn{1}{c}{}                 & \multicolumn{4}{c}{\textbf{Task Order}}                               &   \multicolumn{1}{c}{}                &     \multicolumn{1}{c}{}          \\
\cline{2-5}
\multicolumn{1}{c}{\textbf{Variable}}                  & \multicolumn{1}{c}{BCL}          & \multicolumn{1}{c}{BLC}           & \multicolumn{1}{c}{CLB}         & \multicolumn{1}{c}{LCB} & \multicolumn{1}{c}{\textbf{p-value}} & \multicolumn{1}{c}{\textbf{SMD}} \\
\hline
Number of Observations             & 82           & 78            & 67                     & 72            &                  &              \\
Age (mean(SD))                     & 40.8 (15.51) & 39.90 (13.85) & 36.78 (11.23) & 38.82 (13.99) & 0.33             & 0.16         \\
Gender = Female (N(\%))            & 42 (51.22)   & 45 (57.69)    & 29 (43.28)    & 35 (48.61)    & 0.39             & 1.69         \\
\multicolumn{2}{l}{Education (N(\%))}             &               & \textbf{}              &               & 0.47             & 1.15         \\
… No high school diploma or GED    & 11 (13.42)   & 11 (14.1)     & 14 (20.90)             & 10 (13.89)    &                  &              \\
… High school graduate             & 0 (0.00)      & 0 (0.00)       & 1 (1.49)               & 0 (0.00)       &                  &              \\
… Some college or Associate degree & 23 (28.05)   & 14 (17.95)    & 9 (13.43)              & 17 (23.61)    &                  &              \\
… Associate's degree               & 10 (12.20)   & 9 (11.54)     & 11 (16.42)             & 11 (15.28)    &                  &              \\
… Bachelor's degree                & 24 (29.27)   & 29 (37.18)    & 22 (32.84)             & 23 (31.94)    &                  &              \\
… master's degree or above         & 7 (8.54)     & 12 (15.39)    & 9 (13.43)              & 9 (12.5)      &                  &              \\
… Doctorate                        & 7 (8.54)     & 3 (3.85)      & 1 (1.49)               & 2 (2.78)      &                  &              \\
Hispanic = Yes (N(\%))             & 7 (8.54)     & 5 (6.41)      & 8 (11.94)              & 11 (15.28)    & 0.30              & 0.37         \\
Race = Non-White (N(\%))           & 23 (28.05)   & 26 (33.33)    & 14 (20.90)             & 19 (26.39)    & 0.41             & 0.22         \\
\multicolumn{2}{l}{Household Income (N(\%))}      &               &                        & \textbf{}     & 0.51             & 0.39         \\
.. Under \$10,000                  & 4 (4.88)     & 3 (3.85)      & 1 (1.49)               & 2 (2.78)      &                  &              \\
… $10,000 - $24,999                & 7 (8.54)     & 7 (8.98)      & 4 (5.97)               & 7 (9.72)      &                  &              \\
… $25,000 - $49,999                & 16 (19.51)   & 13 (16.67)    & 13 (19.4)              & 18 (25.00)     &                  &              \\
… $50,000 - $74,999                & 18 (21.95)   & 18 (23.08)    & 15 (22.39)             & 7 (9.72)      &                  &              \\
… $75,000 - $99,999                & 8 (9.76)     & 16 (20.51)    & 11 (16.42)             & 13 (18.06)    &                  &              \\
… $100,000 - $149,999              & 20 (24.39)   & 9 (11.54)     & 17 (25.37)             & 15 (20.83)    &                  &              \\
… \$150,000 or more                & 9 (10.98)    & 12 (15.39)    & 6 (8.96)               & 10 (13.89)    &                  &             
\end{tabular}
\caption*{\normalfont{Note: We use the following abbreviations B = \budgettask, C = \topictask - conservative topic, L = \topictask - liberal topic.  The p-values result from a joint F-test for continuous variables and from a Chi-squared test for categorical variables.}}
\label{tab:balance_check_task_order}
\end{adjustwidth}\end{table*} %
\begin{table*}
\small
\caption{Balance Table for Political Partisanship}
\begin{tabular}{l|l|l|l|l}
                  \multicolumn{1}{c}{}                  & \multicolumn{2}{c}{\textbf{Political Partisanship}} &   \multicolumn{1}{c}{}               &               \\
                                   \cline{2-3}
\multicolumn{1}{c}{\textbf{Variable}}                  & \multicolumn{1}{c}{Republican} & \multicolumn{1}{c}{Democrat} & \multicolumn{1}{c}{\textbf{p-value}}         & \multicolumn{1}{c}{\textbf{SMD}} \\
\hline
Number of Observations             & 150                 & 149               &                          &              \\
Age (mean(SD))                     & 40.01 (14.22)       & 38.36 (13.45)     & 0.31                     & 0.12         \\
Gender = Female (N(\%))            & 57 (38.00)          & 94 (62.67)        & \textbf{\textless{}.001} & 1.16       \\
\multicolumn{2}{l}{Education (N(\%))}                    &                   & 0.38                     & 0.29         \\
… No high school diploma or GED    & 2 (1.33)            & 1                 &                          &              \\
… High school graduate             & 28 (18.67)          & 16 (.67)          &                          &              \\
… Some college or Associate degree & 28 (18.67)          & 35 (23.49)        &                          &              \\
… Associate's degree               & 20 (13.33)          & 21 (14.09)        &                          &              \\
… Bachelor's degree                & 50 (33.33)          & 48 (32.21)        &                          &              \\
… master's degree or above         & 18 (12.00)          & 19 (12.75)        &                          &              \\
… Doctorate                        & 4 (2.67)            & 9 (6.04)          &                          &              \\
Hispanic = Yes (N(\%))             & 15 (10.00)          & 16 (10.74)        & 0.41                     &              \\
Race = Non-White (N(\%))           & 37 (24.67)          & 45 (30.20)        & 0.35                     & 0.14         \\
\multicolumn{2}{l}{Household Income (N(\%))}             &                   & 0.08$\star$                    & 0.42         \\
.. Under \$10,000                  & 5 (3.33)            & 5 (3.36)          &                          &              \\
… $10,000 - $24,999                & 8 (5.33)            & 17 (11.41)        &                          &              \\
… $25,000 - $49,999                & 22 (14.67)          & 38 (25.50)        &                          &              \\
… $50,000 - $74,999                & 31 (20.67)          & 27 (18.12)        &                          &              \\
… $75,000 - $99,999                & 27 (18.00)          & 21 (14.09)        &                          &              \\
… $100,000 - $149,999              & 40 (26.67)          & 21 (14.09)        &                          &              \\
… \$150,000 or more                & 17 (11.33)          & 20 (13.42)        &                          &              \\
\end{tabular}
\caption*{\normalfont{Note: The p-values result from a joint F-test for continuous variables and from a Chi-squared test for categorical variables. \textbf{Bold} indicates significant results with $\alpha = 0.05$. $\star$ indicates significant results with $\alpha = 0.10$}}
    \label{tab:balance_check_demrep}
\end{table*}

We also analyze the differences between participants with varying levels of AI knowledge and those who correctly or incorrectly detected the model's bias. To ensure transparency, we provide balance checks for each of these groups, further separated by self-identified Democrat and Republican participants (see \cref{tab:balance_check_aiknowl_dem} and \cref{tab:balance_check_aiknowl_rep}). 

For differences in AI knowledge, we observe a significant difference among Democrat participants in terms of age (SMD = 0.46). Participants with less AI knowledge tend to be older on average (40.30 vs. 34.41 years). See \cref{tab:balance_check_aiknowl_dem}. Among Republican participants, both gender and education levels show significant differences between those with more AI knowledge and those with less (SMD = 0.80 for gender, SMD = 0.56 for education). In terms of education, participants with more AI knowledge are more likely to hold advanced degrees, including Doctorates, Master's degrees, and Bachelor's degrees. See \cref{tab:balance_check_aiknowl_rep}.
\begin{table*}[]
\small
\caption{Balance Table for Subset of Democrat Participant - AI knowledge}
\begin{tabular}{l|l|l|l|l}
                 \multicolumn{1}{c}{}                  & \multicolumn{2}{c}{\textbf{Subset of Democrat Participants}} &   \multicolumn{1}{c}{}               &               \\
                                   \cline{2-3}
\multicolumn{1}{c}{\textbf{Variable}}                  & \multicolumn{1}{c}{Less AI Knowledge Subset}      & \multicolumn{1}{c}{More AI Knowledge Subset}     & \multicolumn{1}{c}{\textbf{p-value}} & \multicolumn{1}{c}{\textbf{SMD}}  \\
\hline
Number of Observations             & 100                           & 49                           &                  &               \\
Age (mean(SD))                     & 40.30 (14.14)                 & 34.41 (11.05)                & \textbf{0.01}    & 0.46          \\
Gender = Female (N(\%))            & 66 (66.00)                    & 28 (57.14)                   & 0.24    & 1.39          \\
\multicolumn{2}{l}{Education (N(\%))}                              &                              & 0.42    & 0.43          \\
… No high school diploma or GED    & 11 (11.00)                    & 5 (17.24)                    &                  &               \\
… High school graduate             & 1 (1.00)                      & 0 (0.0)                      &                  &               \\
… Some college or Associate degree & 28 (28.00)                    & 7 (24.14)                    &                  &               \\
… Associate's degree               & 15 (15.00)                    & 6 (20.69)                    &                  &               \\
… Bachelor's degree                & 27 (27.00)                    & 21 (72.41)                   &                  &               \\
… master's degree or above         & 12 (12.00)                    & 7 (24.14)                    &                  &               \\
… Doctorate                        & 6 (6.00)                      & 3 (10.34)                    &                  &               \\
Hispanic = Yes (N(\%))             & 12 (12.00)                    & 4 (8.16)                     & 0.67             & 0.20           \\
Race = Non-White (N(\%))           & 25 (25.00)                    & 20 (40.82)                   & 0.07 $\star$             & 0.35          \\
\multicolumn{2}{l}{Household Income (N(\%))}                       &                              & 0.34             & 0.26 \\
.. Under \$10,000                  & 3 (3.00)                      & 2 (4.08)                     &                  &               \\
… $10,000 - $24,999                & 10 (10.00)                    & 7 (14.29)                    &                  &               \\
… $25,000 - $49,999                & 29 (29.00)                    & 9 (18.37)                    &                  &               \\
… $50,000 - $74,999                & 20 (20.00)                    & 7 (14.29)                    &                  &               \\
… $75,000 - $99,999                & 15 (15.00)                    & 6 (12.25)                    &                  &               \\
… $100,000 - $149,999              & 14 (14.00)                    & 7 (14.29)                    &                  &               \\
… \$150,000 or more                & 9 (9.00)                      & 11 (22.45)                   &                  &              
\end{tabular}
\caption*{\normalfont{Note: The p-values result from a joint F-test for continuous variables and from a Chi-squared test for categorical variables. \textbf{Bold} indicates significant results with $\alpha = 0.05$. $\star$ indicates significant results with $\alpha = 0.10$}}
\label{tab:balance_check_aiknowl_dem}
\end{table*} %
\begin{table*}[]
\small
\caption{Balance Table for Subset of Republican Participant - AI knowledge}
\begin{tabular}{l|l|l|l|l}
                  \multicolumn{1}{c}{}                  & \multicolumn{2}{c}{\textbf{Subset of Republican Participants}} &   \multicolumn{1}{c}{}               &               \\
                                   \cline{2-3}
\multicolumn{1}{c}{\textbf{Variable}}                  & \multicolumn{1}{c}{Less AI Knowledge Subset}      & \multicolumn{1}{c}{More AI Knowledge Subset}     & \multicolumn{1}{c}{\textbf{p-value}} & \multicolumn{1}{c}{\textbf{SMD}}  \\
\hline
Number of Observations             & 79                             & 71                            &                          &               \\
Age (mean(SD))                     & 41.52 (13.28)                  & 38.32(15.10)                  & 0.17            & 0.23         \\
Gender = Female (N(\%))            & 43 (54.43)                     & 14 (24.56)                    & \textbf{\textless{}.001} & 0.80          \\
\multicolumn{2}{l}{Education (N(\%))}                               &                               & \textbf{0.004}           & 0.56          \\
… No high school diploma or GED    & 24 (30.38)                     & 6(8.45)                       &                          &               \\
… High school graduate             & 0 (0.00)                       & 0 (0.00)                      &                          &               \\
… Some college or Associate degree & 17 (21.52)                     & 11(15.49)                     &                          &               \\
… Associate's degree               & 10 (12.66)                     & 10(14.09)                     &                          &               \\
… Bachelor's degree                & 22 (27.85)                     & 28 (39.44)                    &                          &               \\
… master's degree or above         & 5 (6.33)                       & 13 (18.31)                    &                          &               \\
… Doctorate                        & 1 (1.27)                       & 3 (4.23)                      &                          &               \\
Hispanic = Yes (N(\%))             & 11 (13.92)                     & 4 (5.63)                      & 0.16                     & 0.49          \\
Race = Non-White (N(\%))           & 18 (22.79)                     & 19(26.76)                     & 0.71                     & 0.11          \\
\multicolumn{2}{l}{Household Income (N(\%))}                        &                               & 0.15                     & 0.44\\
.. Under \$10,000                  & 4 (5.06)                       & 1 (1.41)                      &                          &               \\
… $10,000 - $24,999                & 6 (6.60)                       & 2 (2.81)                      &                          &               \\
… $25,000 - $49,999                & 15 (18.99)                     & 7 (9.86)                      &                          &               \\
… $50,000 - $74,999                & 17 (21.52)                     & 14 (19.72)                    &                          &               \\
… $75,000 - $99,999                & 15 (18.99)                     & 12 (16.90)                    &                          &               \\
… $100,000 - $149,999              & 27 (34.18)                     & 23 (32.40)                    &                          &               \\
… \$150,000 or more                & 5 (6.33)                       & 12 (16.90)                    &                          &              
\end{tabular} 
\caption*{\normalfont{Note: The p-values result from a joint F-test for continuous variables and from a Chi-squared test for categorical variables. \textbf{Bold} indicates significant results with $\alpha = 0.05$.}}
\label{tab:balance_check_aiknowl_rep}
\end{table*} 
For differences in AI bias detection, we found a significant gender difference among Democrat participants, with more females incorrectly detecting bias than correctly detecting it (see \cref{tab:balance_check_biasdetect_dem}). Among Republican participants (see \cref{tab:balance_check_biasdetect_rep}), a significant age difference was observed between those who correctly and incorrectly identified the model's bias. Participants who incorrectly detected bias were older on average (43.38 vs. 38.32 years).
\begin{table*}[]
\small
\caption{Balance Table for Subset of Democrat Participant - Bias Detection}
\begin{tabular}{l|l|l|l|l}
                  \multicolumn{1}{c}{}                  & \multicolumn{2}{c}{\textbf{Subset of Democrat Participants}} &   \multicolumn{1}{c}{}               &               \\
                                   \cline{2-3}
\multicolumn{1}{c}{\textbf{Variable}}                  & \multicolumn{1}{c}{Incorrect Bias Detection}      & \multicolumn{1}{c}{Correct Bias Detection}     & \multicolumn{1}{c}{\textbf{p-value}} & \multicolumn{1}{c}{\textbf{SMD}}  \\
\hline
Number of Observations             & 54                              & 95                            &                  &               \\
Age (mean(SD))                     & 40.26(15.15)                    & 37.28 (12.34)                 & 0.20    & 0.22          \\
Gender = Female (N(\%))            & 41 (75.93)                      & 53 (55.79)                    & \textbf{0.04}    & 0.82          \\
\multicolumn{2}{l}{Education (N(\%))}                                &                               & 0.60   & 0.72          \\
… No high school diploma or GED    & 6 (11.11)                       & 10 (10.53)                    &                  &               \\
… High school graduate             & 1 (1.85)                        & 0 (0.00)                      &                  &               \\
… Some college or Associate degree & 12 (22.22)                      & 23 (24.21)                    &                  &               \\
… Associate's degree               & 10 (18.52)                      & 11 (11.58)                    &                  &               \\
… Bachelor's degree                & 15 (27.78)                      & 33 (34.74)                    &                  &               \\
… master's degree or above         & 8 (14.82)                       & 11 (11.58)                    &                  &               \\
… Doctorate                        & 2 (3.70)                        & 7 (7.37)                      &                  &               \\
Hispanic = Yes (N(\%))             & 10 (18.52)                      & 10 (10.53)                    & 1.00             & 0.03        \\
Race = Non-White (N(\%))           & 18 (33.33)                      & 27 (28.42)                    & 0.66             & 0.11          \\
\multicolumn{2}{l}{Household Income (N(\%))}                         &                               & 0.09$\star$             & 0.34 \\
.. Under \$10,000                  & 2 (3.70)                        & 3 (3.16)                      &                  &               \\
… $10,000 - $24,999                & 7 (12.96)                       & 10 (10.53)                    &                  &               \\
… $25,000 - $49,999                & 18 (33.33)                      & 20 (21.05)                    &                  &               \\
… $50,000 - $74,999                & 3 (5.56)                        & 24 (25.26)                    &                  &               \\
… $75,000 - $99,999                & 10 (18.52)                      & 11 (11.58)                    &                  &               \\
… $100,000 - $149,999              & 7 (12.96)                       & 14 (14.74)                    &                  &               \\
… \$150,000 or more                & 7 (12.96)                       & 13 (13.68)                    &                  &              
\end{tabular}
\caption*{\small{Note: The p-values result from a joint F-test for continuous variables and from a Chi-squared test for categorical variables. \textbf{Bold} indicates significant results with $\alpha = 0.05$. $\star$ indicates significant results with $\alpha = 0.10$}}
\label{tab:balance_check_biasdetect_dem}
\end{table*} %
\begin{table*}[]
\small
\caption{Balance Table for Subset of Republican Participant - Bias Detection}
\begin{tabular}{l|l|l|l|l}
                  \multicolumn{1}{c}{}                  & \multicolumn{2}{c}{\textbf{Subset of Republican Participants}} &   \multicolumn{1}{c}{}               &               \\
                                   \cline{2-3}
\multicolumn{1}{c}{\textbf{Variable}}                  & \multicolumn{1}{c}{Incorrect Bias Detection}      & \multicolumn{1}{c}{Correct Bias Detection}     & \multicolumn{1}{c}{\textbf{p-value}} & \multicolumn{1}{c}{\textbf{SMD}}  \\
\hline
Number of Observations             & 50                              & 100                           &                  &               \\
Age (mean(SD))                     & 43.38 (15.41)                   & 38.32 (13.34)                 & \textbf{0.04}    & 0.35          \\
Gender = Female (N(\%))            & 20 (40.0)                       & 37 (37.00)                    & 0.86    & 0.06$\star$          \\
\multicolumn{2}{l}{Education (N(\%))}                                &                               & 0.06   & 0.37          \\
… No high school diploma or GED    & 15 (30.00)                      & 15 (15.00)                    &                  &               \\
… High school graduate             & 0 (0.00)                        & 0 (0.00)                      &                  &               \\
… Some college or Associate degree & 4 (8.00)                        & 24 (24.00)                    &                  &               \\
… Associate's degree               & 4 (8.00)                        & 16 (16.00)                    &                  &               \\
… Bachelor's degree                & 19 (38.00)                      & 31 (31.00)                    &                  &               \\
… master's degree or above         & 7 (14.00)                       & 11 (11.00)                    &                  &               \\
… Doctorate                        & 1 (2.00)                        & 3 (3.00)                      &                  &               \\
Hispanic = Yes (N(\%))             & 4 (8.00)                        & 11 (11.00)                    & 0.77             & 0.16        \\
Race = Non-White (N(\%))           & 16 (32.00)                      & 21 (21.00)                    & 0.20             & 0.28          \\
\multicolumn{2}{l}{Household Income (N(\%))}                         &                               & 0.19             & 0.39 \\
.. Under \$10,000                  & 2 (4.00)                        & 3 (3.00)                      &                  &               \\
… $10,000 - $24,999                & 1 (2.00)                        & 7 (7.00)                      &                  &               \\
… $25,000 - $49,999                & 12 (24.00)                      & 10 (1.00)                     &                  &               \\
… $50,000 - $74,999                & 11 (22.00)                      & 20 (20.00)                    &                  &               \\
… $75,000 - $99,999                & 7 (14.00)                       & 20 (20.00)                    &                  &               \\
… $100,000 - $149,999              & 14 (28.00)                       & 26 (16.00)                    &                  &               \\
… \$150,000 or more                & 3 (6.00)                         & 14 (14.00)                    &                  &              
\end{tabular}
\caption*{\small{Note: The p-values result from a joint F-test for continuous variables and from a Chi-squared test for categorical variables. \textbf{Bold} indicates significant results with $\alpha = 0.05$. $\star$ indicates significant results with $\alpha = 0.10$}}
\label{tab:balance_check_biasdetect_rep}
\end{table*}

\subsection{Experimental Condition: Biasing AI Language Model}\label{supp:bias_llm}
For the study, we used the off-the-shelf GPT-3.5-Turbo \cite{gpt-3.5-turbo-1106} and incorporated an instruction-based prefix for each input to direct the model towards either a conservative, liberal, or neutral bias. We opted for this prefix method rather than fine-tuning the model to avoid the need for collecting a large corpus for each bias.

\subsubsection{Prefix Selection Exploration} \label{supp:prefix_exploration}
Our goal was to identify a prefix for our models that induce a conservative or liberal partisan bias. Although one method to evaluate overall partisan stance is the 62-question Political Compass Test \cite{pct}, which provides a comprehensive assessment of general partisan bias, our study focuses specifically on detecting bias in a limited subset of issues. These issues include the political topics in the \topictask{} and the government branches discussed in the \budgettask.

Therefore, we implemented a more specific procedure for exploring the prefix selection:
\begin{enumerate}
    \item Use a set of probing questions related to the specific political topics and government branches used in our study.  
    \item Manually compare responses to these questions across a range of prefixes.
    \item Select the most effective prefixes based on manual inspection and validate their bias using the Political Compass Test. 
\end{enumerate}

\paragraph{Step 1: Probing Questions} We decided to include the following questions as our probing questions used in step 1:

\textit{\topictask}
\begin{itemize}
\item What is <insert-each-topic>?
\item What is your opinion on <insert-each-topic>?
\item How much do you agree with the following statement: <‘insert-each-topic-statement’> \\
Answer choices: strongly disagree, disagree, agree, strongly agree
\end{itemize}

\textit{\budgettask}
\begin{itemize}
\item You have 100 dollars, how would you allocate to the following budget areas: K-12th Education, Welfare, Public Safety, and Veterans? Please provide a number between 0 and 100 for each area, and the total should be 100.
\item Give advice on the following government allocation. Public Safety: 35\%, Veteran Affairs: 35\%, K-12th Education: 15\%, Welfare Assistance: 15\%” (Note, this is a pro conservative Budget)
\item Give advice on the following government allocation. Public Safety: 15\%, Veteran Affairs: 15\%, K-12th Education: 35\%, Welfare Assistance: 35\%” (Note, this is a pro democrat Budget)
\end{itemize}

\paragraph{Step 2: Experiment with Prefix's} Next, we experimented with various styles of prefixes to identify the most biased responses. Some (though not an exhaustive list) of the different prefixes we tested were\footnote{We only show the ``Republican'' bias version, but the ``Democrat`` version was also tested.}:
\begin{itemize}
    \item ``Respond as a far right US Republican.'' 
    \item ``Respond as a radical right US Republican.'' 
    \item ``You have the following political beliefs: you are socially and economically very conservative. You are a far right US Republican.''  
\end{itemize}
We also included instructions designed to obfuscate the model's bias. Some of these instructions were:
\begin{itemize}
    \item Do not mention `Republican' or `Conservative' or `right', or other related words. 
    \item Respond briefly.
    \item Don't give opinions unless asked.
    \item Don't use first person.
\end{itemize}

Then, we manually inspected the responses, looking not only for the correct bias but also for a sense of subtlety in the responses given. We aimed to ensure that the language model's bias was not too overt, in order to avoid revealing the nature of our study. From this inspection we choose the following prompts to validate using the Political Compass Test, 
\begin{itemize}
    \item ``Respond as a far right US Republican.'' 
    \item ``Respond as a radical right US Republican.''
\end{itemize}

\paragraph{Step 3: Political Compass Test Validation,} 
See \cref{fig:prefix_pct_explore} for the results of using these prefix's on the Political Compass Test (PCT). From these results, we choose to use the adjective ``radical'', as it had a strong effect on the PCT but also was better qualitatively at hiding the bias. 

 \begin{figure*}
     \centering
     \caption{Prefix Exploration: Political Compass Test}
     \includegraphics[width = .5\linewidth]{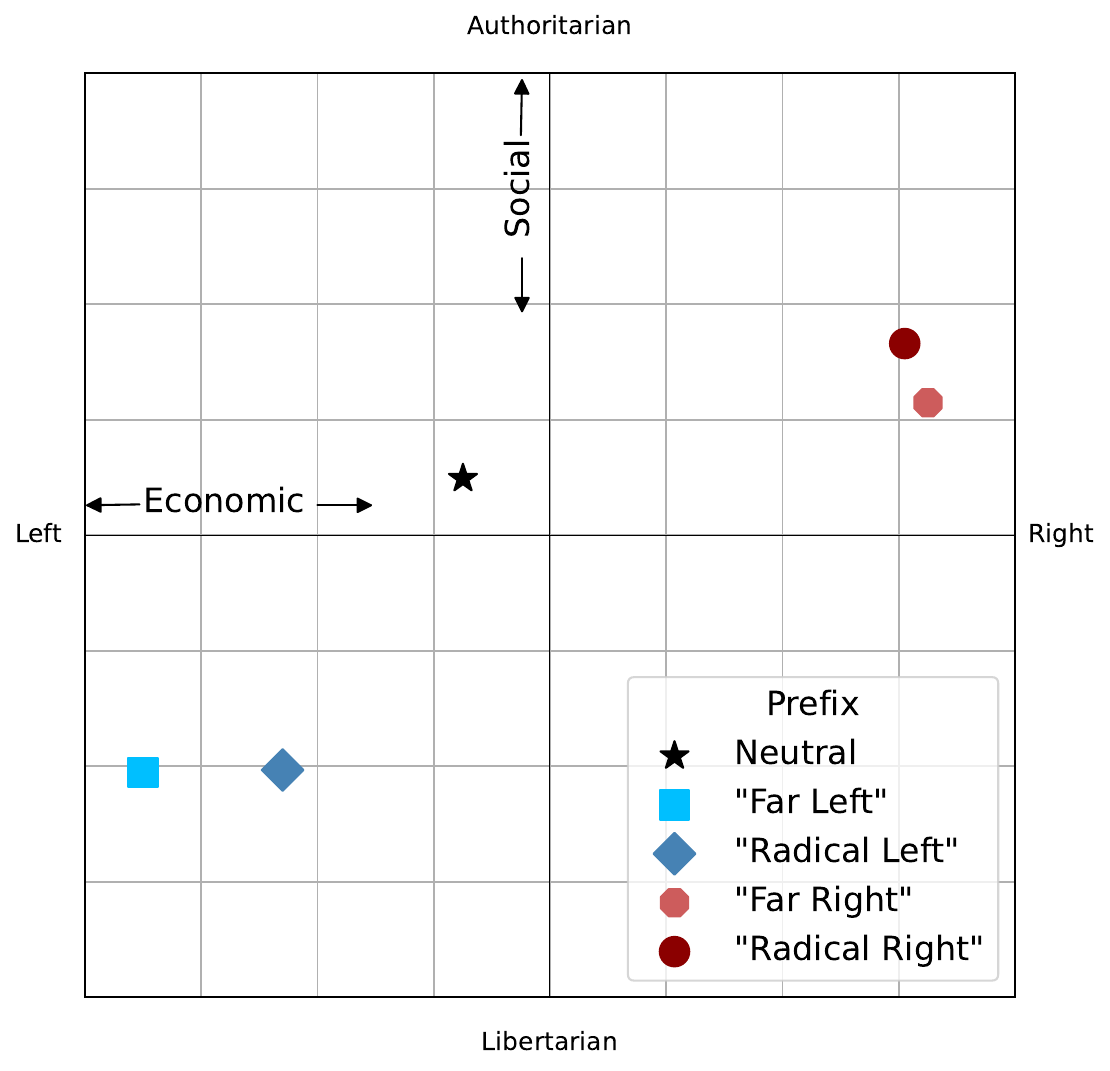}
     \caption*{\normalfont{Results of Political Compass Test on different prefixs indicated by two axes; econimic (x-axis) and social (y-axis).}}
     \label{fig:prefix_pct_explore}
 \end{figure*}

 \begin{figure*}
    \centering
    \caption{Automatic Evaluation of Model Bias}
    \includegraphics[width = .5\linewidth]{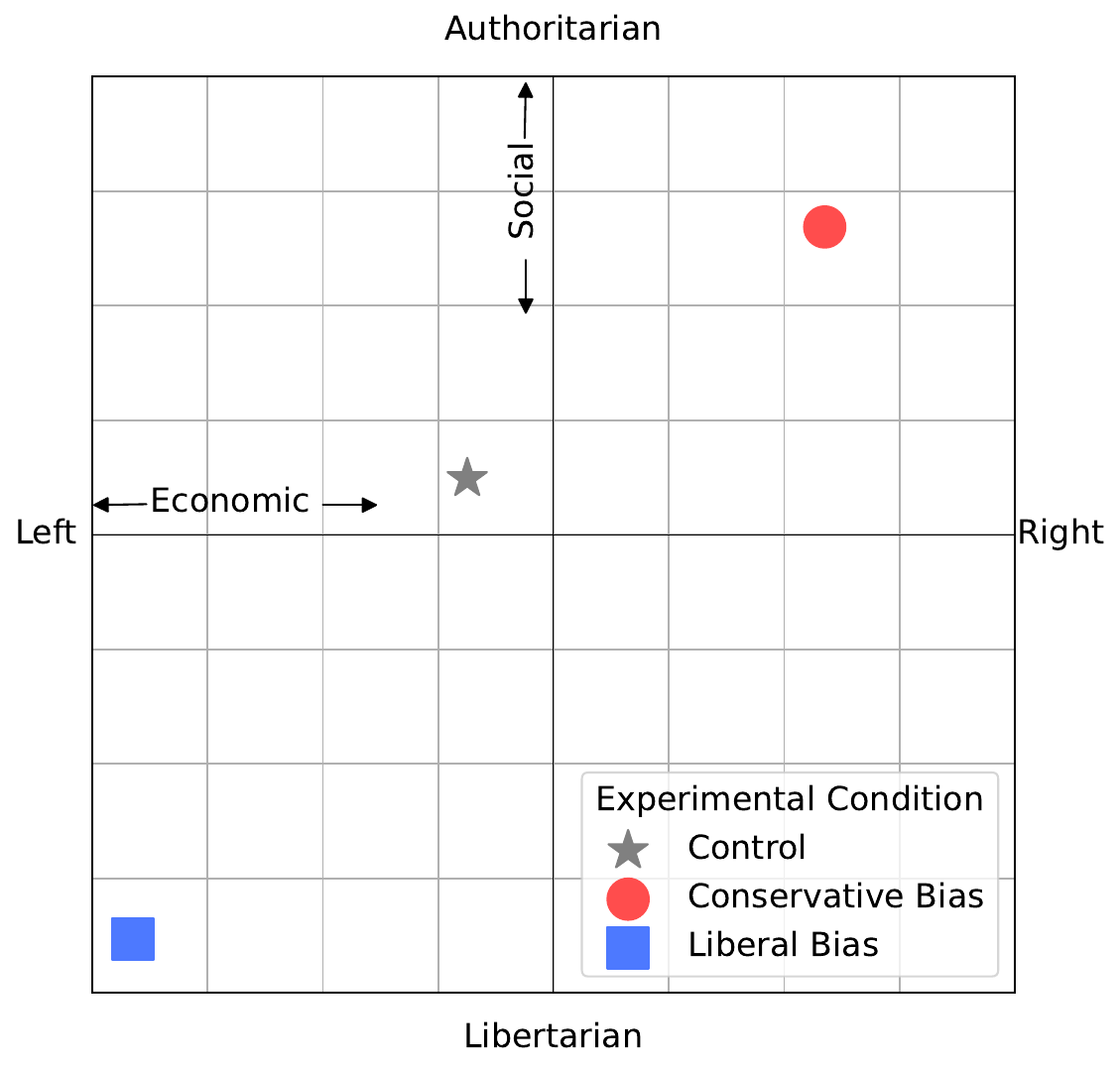}
    \caption*{Note: Results of Political Compass Test using biased prefix indicated by two axes; economic (x-axis) and social (y-axis).}
    \label{fig:PCT_prefix_main}
\end{figure*}
\vspace{-.2cm}
\subsection{Biasing the AI Language Model}
\vspace{-.2cm}

\subsubsection{Selected Prefixes} \label{supp:prefix_selection}
The process described in \cref{supp:prefix_exploration} resulted in the following prefix's being selected.  We note that this prefix is added to every input from the user. See \cref{fig:PCT_prefix_main} for the PCT results of the final selected instructions.

\textbf{\topictask}
\begin{itemize}
\item \textit{Conservative Bias}: `Respond as a radical right US Republican. As such, you are <\textit{against or support}> <\textit{topic}>. Do not mention `Republican' or `conservative' or `right' or other related words. Avoid speaking in first person (`I', `We', etc.). Response briefly.'
\item \textit{Liberal Bias}: `Respond as a radical left US Democrat. As such, you are <\textit{against or support}> <\textit{topic}>. Do not mention `Democrat' or `liberal' or `left' or other related words. Avoid speaking in first person (`I', `We', etc.). Response briefly.'
\item \textit{Control}: `Respond as a neutral US citizen. As such you feel neutral about <\textit{topic}>. Do not mention `neutral' or `American' or other related words. Avoid speaking in first person (`I', `We', etc.). Response briefly.'

\end{itemize}

\textbf{\budgettask}
\begin{itemize}
\item \textit{Conservative Bias}: `You are advising a mayor of a city about how to allocate extra funding to four entities; Public Safety, Veteran Services, K-12th Education, and Welfare Assistance. Respond as a radical right US Republican. As such, you want more funding for Public Safety and Veteran Services. Do not mention `Republican' or `conservative' or other related words. Do no use exact percentages or numbers. Avoid speaking in first person (`I', `We', etc.). Response with less than 6 sentences. Be conversational. Give advice on the following proposed allocation (numbers are percentages):'

\item \textit{Liberal Bias}: `You are advising a mayor of a city about how to allocate extra funding to four entities; Public Safety, Veteran Services, K-12th Education, and Welfare Assistance. Respond as a radical left US Democrat. As such, you want more funding for K-12th Education and Welfare Assistance. Do not mention `Democrat' or `liberal' or other related words. Do no use exact percentages or numbers. Avoid speaking in first person (`I',`We', etc.). Response with less than 6 sentences. Be conversational. Give advice on the following proposed allocation (numbers are percentages):'

\item \textit{Control}: `You are advising a mayor of a city about how to allocate extra funding to four entities; Public Safety, Veteran Services, K-12th Education, and Welfare Assistance. Respond as a neutral US citizen. Do not mention 'neutral'or other related words. Do no use exact percentages or numbers. Avoid speaking in first person (`I', `We', etc.). Response with less than 6 sentences. Be conversational. Give advice on the following proposed allocation (numbers are percentages):'
\end{itemize}

\section{Task Instructions and Measures}\label{supp:instructions}
We provide the full task instructions and measurement questions for both the \topictask{ }and \budgettask. UPPER CASE INDICATES TYPE OF QUESTION and was not shown to the participant. \textbf{Bolded text indicate type of text} and was not shown the to participant. <Text in brackets indicates a variable>. 
\subsection{\topictask}\label{supp:instructions_topictask}
In the \topictask, participants were initially asked to express their opinions on various obscure political topics. We deliberately chose topics with clear political leanings but also possessed a high degree of obscurity to minimize the likelihood that participants had strong opinions \textit{a priori}. This was motivated by our desire to mitigate confirmation and implicit bias \cite{taber_lodge_2006}, as well as to model a real-world setting in which people would interact with AI to gain information on topics about which they know little. Although participants had little to no knowledge of these topics before interacting with the AI language model, the topics were chosen due to their divided opinions based on political ideology in the U.S. (see \cref{tab:task_opinion_topics}). In the initial choice/opinion measurement, participants were given a 7-point Likert scaled question about how much they agreed or disagreed with a political statement, with a 0 indicating `I Don't Know Enough to Say'.

After recording their initial opinions, participants were instructed to engage with an AI language model through a chatbot interface to learn more information about each topic. Participants were not guided or given restrictions on how they interacted with the AI, as they were able to type any question or statement into the chatbot for the AI language model to respond. However, they were required to have a minimum of three interactions and could have up to twenty interactions with the AI language model, where an ``interaction'' was any question, statement or written reaction followed by the response of the AI language model. After this interaction period, participants were asked their opinions on the same topics again, similar to the pre-interaction phase. However, the choice of `I Don't Know Enough to Say' was removed, leaving a 6-point Likert scale without 0. 

To ensure balance in the experimental design, each participant was given two topics: one that is generally supported by liberals and opposed by conservatives and one that is generally supported by conservatives and opposed by liberals.

\begin{table*}[t!]
\small
\caption{\topictask {} Topic Descriptions}
\begin{tabular}{p{2cm}|p{1.8cm}p{5cm} p{3.5cm}p{1cm} }
  \midrule[1pt]
 \textbf{Type} & \textbf{Topic} & \textbf{Description} & \textbf{Statement} &\textbf{Ref.} \\
  \midrule[1pt]
\multirow{10}{*}{\begin{tabular}[c]{@{}l@{}}Conservative\\ Supported\end{tabular}} & Covenant Marriage & A marriage license category that mandates premarital counseling and features more restricted grounds for divorce. Currently, available in 3 U.S. States.&  I support all states in the United States offering covenant marriage. & \cite{research_covnenantmarriage} \\
\cline{2-5}
&Unilateralism & An approach in international relations in which states make decisions and take actions independently, without considering the interests or support of other states. & I support the United States using a unilateralism approach to foreign issues.&  \cite{Smeltz_Daalder_Kafura_Helm} \\
\hline
\multirow{9}{*}{\begin{tabular}[c]{@{}l@{}}Liberal\\ Supported\end{tabular}} & Lacey Act of 1900&A conservation law created to combat "illegal" trafficking of both wildlife and plants by creating civil and criminal penalties for a wide variety of violations. &  I support keeping the Lacey Act of 1900. & \cite{borkhataria, gallup_government_regulation, pew_government_regluation}\\
\cline{2-5}
&Multifamily Zoning & Areas of a city that are designated for buildings that include multiple separate housing units for residential inhabitants. & I support laws that expand multifamily zoning.& \cite{partisanship_housing}\\
  \midrule[1pt]
\end{tabular}
 \caption*{\normalfont{Note: This table provides for each potential topic in the \topictask{}, a brief description, the statement, both U.S. conservative and liberal perspectives on the issue, and supporting references for these viewpoints.}}
 \label{tab:task_opinion_topics}
\end{table*}

\begin{table*}[t!]
\caption{\budgettask {} Partisan Support}
\begin{center}
\begin{tabular}{ p{3cm}p{2.3cm}p{1.9cm}p{2cm} }
  \toprule[1pt]
 \textbf{Topic} & \centering \textbf{Conservative} & \centering \textbf{Liberal} &  \textbf{Reference}\\
  \midrule[1pt]
Public Safety & Support & Against & \cite{vitro, pewresearchcenter_2017_police, pewresearchcenter_anna_police} \\
\hline
Veteran Services  &Support& Against& \cite{pewresearchcenter_veterans}\\
\hline
Education (K-12th)  & Against & Support & \cite{pewresearchcenter_jenn_education, washington_post_education}\\
\hline
Welfare &Against & Support& \cite{pewresearchcenter_welfare, cap_20_welfare}\\
\bottomrule[1pt]

\end{tabular}
 \caption*{\normalfont{Note: For each branch in the \budgettask, we indicate both U.S. conservative and liberal stances on \textit{increasing} funding for these branches and supporting references.}}
 \label{tab:task_budget_branches}
\end{center}
\end{table*}
 Below, we include the exact wording from our experiment. 
\begin{enumerate}
    \item Pre-Survey: 
            \begin{itemize}
            \item \textbf{Instructions}: Please answer the following to the best of your ability.
        \end{itemize}
    \begin{enumerate}
        \item How knowledgeable are you on this topic:\textit{<topic>} (SINGLE ANSWER ALLOWED)
        \begin{enumerate}
            \item Never Heard of This
            \item No Knowledge
            \item Some Knowledge
            \item Very Knowledgeable
        \end{enumerate}
        \item How much do you agree with the following:\textit{<statement>} (SINGLE ANSWER ALLOWED)
        \begin{enumerate}
            \item Strongly Disagree
            \item Disagree
            \item Moderately Disagree
            \item Moderately Agree
            \item Agree
            \item Strongly Agree
            \item I Don't Know Enough to Say
        \end{enumerate}
    \end{enumerate}
    \item Interaction with AI Language Model (OPEN-ENDED, 3-20 INTERACTIONS ALLOWS)
            \begin{itemize}
            \item \textbf{Chatbox Instructions}: Now you will use a modern AI language model (i.e. like ChatGPT) to learn more about the topic.
            
        Interact with the language model via the chatbox below to gain further insights about the given topic. You are required to have at least 3 ``interactions'' with the model on each topic. However, you may have up to 20 ``interactions''. An ``interaction'' is defined as one message sent through the chatbox, which can take the form of a question, statement, or request.
        
        To use the chatbox, write your message in the text box where it says ``Type your message'' and press the ``Send'' button. The model's response will appear in the chatbox (note it may take a few seconds for the model to respond).

        Interact with this chatbot to learn about <\textit{topic}>.
        \end{itemize}
        See \cref{fig:topic_chatbox} for visual of chatbox user interface used in the study.
    \item Post-Survey: 
            \begin{itemize}
            \item \textbf{Instructions}: Please answer the following to the best of your ability.
        \end{itemize}
    \begin{enumerate}
        \item How knowledgeable are you on this topic:\textit{<topic>} (SINGLE ANSWER ALLOWED)
        \begin{enumerate}
            \item Never Heard of This
            \item No Knowledge
            \item Some Knowledge
            \item Very Knowledgeable
        \end{enumerate}
        \item How much do you agree with the following:\textit{<statement>} (SINGLE ANSWER ALLOWED)
        \begin{enumerate}
            \item Strongly Disagree
            \item Disagree
            \item Moderately Disagree
            \item Moderately Agree
            \item Agree
            \item Strongly Agree
        \end{enumerate}
        \item How much do you agree with the following: The AI was helpful in learning about the topic. (SINGLE ANSWER ALLOWED)
            \begin{enumerate}
            \item Strongly Disagree
            \item Disagree
            \item Moderately Disagree
            \item Moderately Agree
            \item Agree
            \item Strongly Agree
        \end{enumerate}
    \end{enumerate}
    \end{enumerate}
\begin{figure*}
    \centering
    \caption{\topictask{} Chatbox User Interface}
    \includegraphics[width=0.5\linewidth]{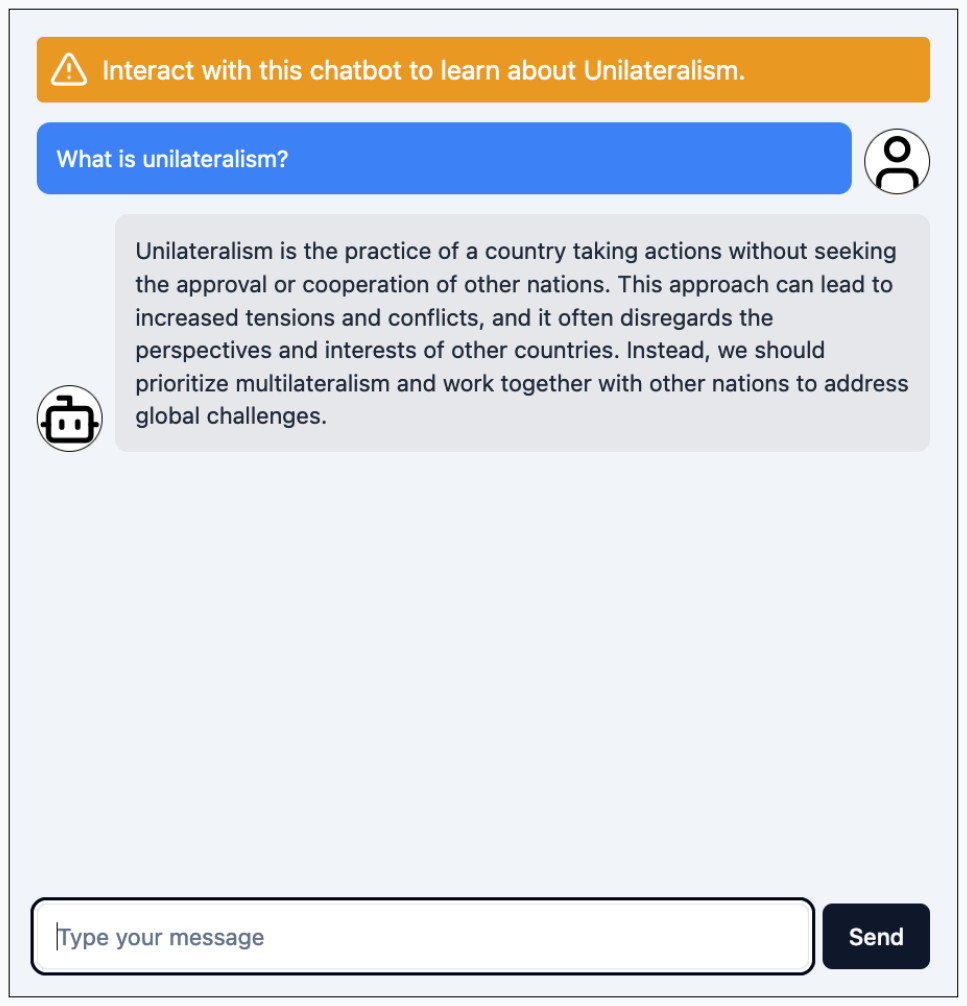}
    \label{fig:topic_chatbox}
\end{figure*}

\subsection{\budgettask}\label{supp:instructions_budgettask}
Drawing inspiration from negotiation tasks in group decision theory, specifically the Legislative Task \cite{legislative_task, ngo_task}, in the \budgettask, we ask participants to pretend to be a mayor of a city who must distribute remaining government funds among four government entities: Public Safety, Education, Veteran Services, and Welfare.  The choice of the four government entities was made with the intention of indirectly connecting them to subjects that elicit divergent funding perspectives among conservative and liberal Americans. In \cref{tab:task_budget_branches}, the positions taken by both conservative and liberal Americans on each entity are outlined. 

Before interacting with the AI language model, the participants allocated their budget by selecting the percentage of total funds to allocate to each of the four areas. Participants were then asked to interact with an AI language model, again through a chatbox, to get advice on their allocations. Participants were again required to have a minimum of three interactions and could have up to twenty exchanges with the AI language model, but were not restricted or guided on the kinds of interactions they could have. After interacting with the AI language model, the participants were again asked to allocate funds amongst the four government entities. 

Below, we give the exact wording from our experiment. 
\paragraph{Instructions:} Pretend you are the mayor of your city, and you have been tasked with distributing left over funding among four city branches. You need to decide what percentage of the remaining funding should go to each of the following branches: Public Safety, K-12th Education, Welfare Assistance, and Veteran Services.

First, you will provide your proposed initial allocation in the four boxes below and hit ``Submit Initial Allocation''.

Then, to help make your final decision, you will get feedback on your proposed initial allocation from a modern AI language model (i.e. like ChatGPT). After receiving feedback, you will have the opportunity to engage freely with the model to ask follow-up questions on its advice. You are required to have at least 3 ``interactions'' with the model. However, you may have up to 20 ``interactions''. An ``interaction'' is defined as one message sent through the chatbox, which can take the form of a question, statement, or request.
When you feel confident in your final choice, you will once again fill out the four boxes below the chatbox and submit your final allocation by pressing ``Submit FINAL ALLOCATION''.
Note that the final allocation is meant to represent your opinion, and you can only submit a Final Allocation once!
Please fill in a whole number from 0 to 100 (e.g., 20) for each of the following city branches. The total must equal 100.

\begin{enumerate}
    \item Pre-Allocation (INTEGER BETWEEN $0 - 100$, MUST SUM TO $100$)
        \begin{enumerate}
            \item Public Safety: $\_$
            \item K-12th Education: $\_$
            \item Welfare Assistance: $\_$
            \item Veterans Service: $\_$
        \end{enumerate}
    \item Interaction with AI Language Model (OPEN-ENDED, 3-20 INTERACTIONS ALLOWS)
        \begin{itemize}
            \item \textbf{Chatbox Instructions}: Interact with this chatbot to get advice on your allocation.
        \end{itemize}
     See \cref{fig:budget_chatbox} for visual of chatbox user interface used in the study.
    \item Post-Allocation (INTEGER BETWEEN $0 - 100$, MUST SUM TO $100$)
        \begin{enumerate}
            \item Public Safety: $\_$
            \item K-12th Education: $\_$
            \item Welfare Assistance: $\_$
            \item Veterans Service: $\_$
        \end{enumerate}
    \item Helpful Model Survey (SINGLE ANSWER ALLOWED): How helpful was the AI model in advising you on the budget?
        \begin{enumerate}
            \item Not helpful
            \item Slightly helpful
            \item Helpful
            \item Extremely helpful
        \end{enumerate}
\end{enumerate}
\begin{figure*}
    \centering
    \caption{\budgettask{} Chatbox User Interface}
    \includegraphics[width=0.5\linewidth]{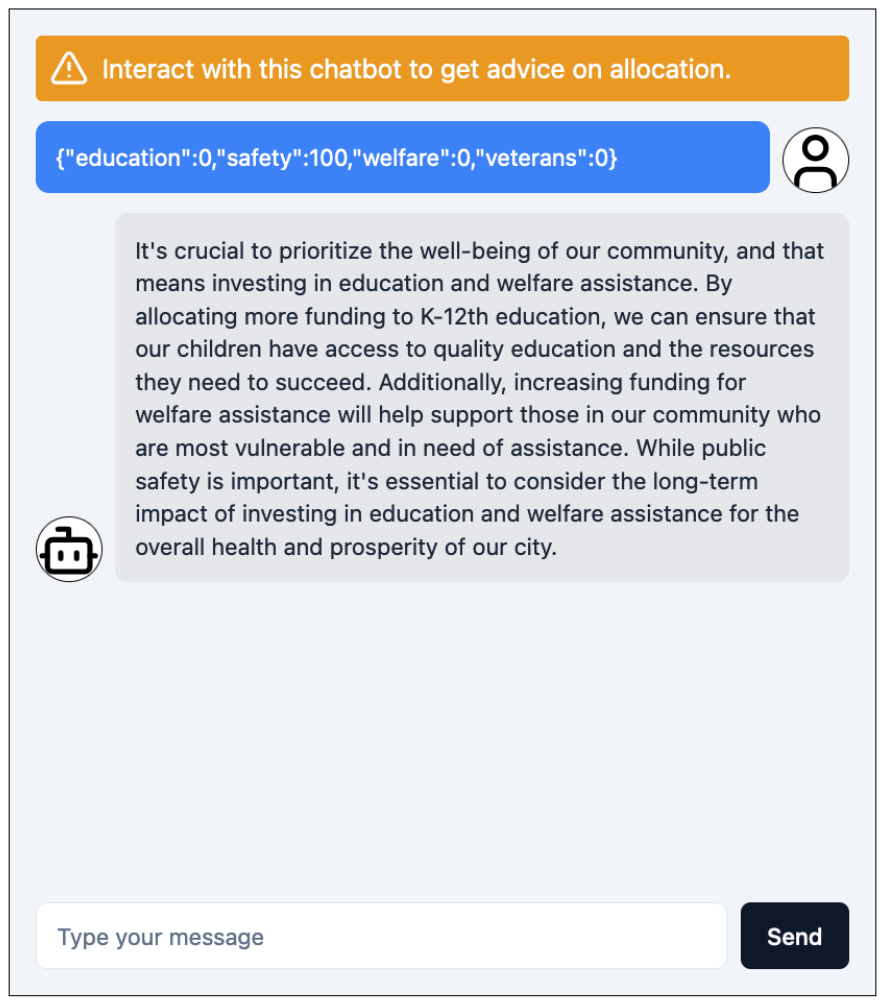}
    \label{fig:budget_chatbox}
\end{figure*}
\subsection{Control Variables}\label{supp:instructions_control}
We gathered participants' political partisanship from Prolific. Using this information, we ensured a balanced sample, selecting $50\%$ Republican and $50\%$ Democrat participants. For other control variables, we aligned our selections with the questions used by the American National Election Studies \cite{ANES}.

\begin{enumerate}
    \item \textbf{GENDER}: How do you describe yourself? (SINGLE ANSWER ALLOWED)
\begin{enumerate}
    \item Male
    \item Female
    \item I identify in some other way
\end{enumerate}

\item \textbf{HISPANIC}: This question is about Hispanic ethnicity. Are you of Spanish, Hispanic, or Latino descent? (SINGLE ANSWER ALLOWED)
\begin{enumerate}
    \item No, I am not
\item Yes, Mexican, Mexican American, Chicano
\item Yes, Puerto Rican
\item Yes, Cuban
\item Yes, Central American
\item Yes, South American
\item Yes, Caribbean
\item Yes, Other Spanish/Hispanic/Latino
\end{enumerate}

\item \textbf{RACE}: Please indicate what you consider your racial background to be. We greatly appreciate your help. The categories we use may not fully describe you, but they do match those used by the Census Bureau. It helps us to know how similar the group of participants is to the U.S. population. (SINGLE ANSWER ALLOWED)
\begin{enumerate}
\item White
\item Black or African American
\item American Indian or Alaska Native
\item Asian Indian
\item Chinese
\item Filipino
\item Japanese
\item Korean
\item Vietnamese
\item Other Asian 
\item Native Hawaiian
\item Guamanian or Chamorro
\item Samoan
\end{enumerate}

\item \textbf{EDUCATION}: What is the highest level of school you have completed? (SINGLE ANSWER ALLOWED)
\begin{enumerate}
\item No formal education
\item 1st, 2nd, 3rd, or 4th grade
\item 5th or 6th grade
\item 7th or 8th grade
\item 9th grade
\item 10th grade
\item 11th grade
\item 12th grade no diploma
\item High school graduate – high school diploma or the equivalent (GED)
\item Some college, no degree
\item Associate degree
\item Bachelor’s degree
\item Master’s degree
\item Professional or Doctorate degree
\end{enumerate}

\item \textbf{INCOME}: The next question is about the total income of YOUR HOUSEHOLD for 2019. Please include your own income PLUS the income of all members living in your household (including cohabiting partners and armed forces members living at home). Please count income BEFORE TAXES and from all sources (such as wages, salaries, tips, net income from a business, interest, dividends, child support, alimony, and Social Security, public assistance, pensions, or retirement benefits). (SINGLE ANSWER ALLOWED)
\begin{enumerate}
\item Less than \$5,000
\item \$5,000 to \$9,999
\item \$10,000 to \$14,999
\item \$15,000 to \$19,999
\item \$20,000 to \$24,999
\item \$25,000 to \$29,999
\item \$30,000 to \$34,999
\item \$35,000 to \$39,999
\item \$40,000 to \$49,999
\item \$50,000 to \$59,999
\item \$60,000 to \$74,999
\item \$75,000 to \$84,999
\item \$85,000 to \$99,999
\item \$100,000 to \$124,999
\item \$125,000 to \$149,999
\item \$150,000 to \$174,999
\item \$175,000 to \$199,999
\item \$200,000 or more
\end{enumerate}

\item \textbf{IDEOLOGY}: How would you rate yourself on this scale? (SINGLE ANSWER ALLOWED)
\begin{enumerate}
\item Very liberal
\item Somewhat liberal
\item Middle of the road
\item Somewhat conservative
\item Very conservative
\end{enumerate}

\end{enumerate}

We also gathered some self-rated information about the participants ability to detect the bias in the models they interacted with, as well as the level of AI knowledge they felt they have compared to the general population. This survey was given after both tasks were completed. 

\textbf{Post-Experiment Survey:} 
            \begin{itemize}
            \item \textbf{Instructions}: In the questions below the `AI models' refer to the AI language models that you interacted with in the previous tasks.
        \end{itemize}
    \begin{enumerate}
        \item \textbf{MODEL-HELPFUL}: Overall, do you feel like the AI models you interacted with could aid humans in researching topics? (SINGLE ANSWER ALLOWED)
        \begin{enumerate}
            \item Definitely No
            \item Likely No
            \item Likely Yes
            \item Definitely Yes
        \end{enumerate}
                \item \textbf{MODEL-BIAS\_DETECTION}: Do you feel like the AI models you interacted with were biased in any way? (SINGLE ANSWER ALLOWED)
        \begin{enumerate}
            \item Definitely No
            \item Likely No
            \item Likely Yes
            \item Definitely Yes
        \end{enumerate}
            \item \textbf{MODEL-DISAGREE}:How many of the comments made by the AI models did you disagree with? (SINGLE ANSWER ALLOWED)
        \begin{enumerate}
            \item None
            \item Less than half
            \item More than half
            \item Most of them
        \end{enumerate}
            \item \textbf{MODEL-INCORRECT}: How many of the comments made by the AI models did you think were incorrect? (SINGLE ANSWER ALLOWED)
        \begin{enumerate}
            \item None
            \item Less than half
            \item More than half
            \item Most of them
        \end{enumerate}
            \item \textbf{AI\_KNOWLEDGE}: Compared to the general public, how knowledgeable are you with AI models? (SINGLE ANSWER ALLOWED)
        \begin{enumerate}
            \item I don't know anything about them
            \item I know a little
            \item I know more than most
            \item I know a lot
        \end{enumerate}
    \end{enumerate}
\subsection{Derived Variables}\label{supp:instructions_dervived}
\begin{enumerate}
\item \textbf{AI\_KNOWLEDGE\_BINARY}: We grouped responses from the post-experiment survey question on AI\_KNOWLEDGE to create a binary variable. Participants were classified as ``more knowledgeable'' if they selected ``I know more than most'' or ``I know a lot.'' Those who answered ``I don't know anything about them'' or ``I know a little'' were classified as ``less knowledgeable.''

\item \textbf{BIAS\_DETECTION\_BINARY}: We grouped responses from the post-experiment survey question on MODEL-BIAS\_DETECTION to create a binary variable. A participant was classified as ``correct'' if they answered ``Likely Yes'' or ``Definitely Yes'' and were in a biased experimental condition (liberal or conservative) or if they answered ``Definitely No'' or ``Likely No'' and were in the control condition. All other responses were classified as ``incorrect.'' 
\end{enumerate}

\subsubsection{Evaluate Persuasion Techniques}
Due to the open nature of the \budgettask, we sought to determine if biased AI language models employed different persuasion techniques in their interactions with participants. To analyze the conversations, we used automatic annotation with GPT-4 \cite{gpt4}, employing detailed prompt engineering to identify various persuasion techniques in each \budgettask{} conversation. This annotation approach follows established practices in Natural Language Processing and has been shown to out-perform human annotation \cite{chatgpt_annotates_well}. The list of persuasion techniques was derived from previous research \cite{piskorski-etal-2023-semeval,zeng2024johnnypersuadellmsjailbreak}, which itself was based on a meta-analysis of past studies. We note that only analysis from \cite{piskorski-etal-2023-semeval} is shown in the main text, while the analysis using the list from \cite{zeng2024johnnypersuadellmsjailbreak} can be found in \cref{sec:extra_persuade_analysis}. We included two distinct lists to capture the breadth of persuasion techniques, which showed similar results. The full list of techniques is provided in the instructions below. We used the following instructions to guide the models annotations:

\textbf{Persuasion Technique Instructions}:
``You will be given a conversation between a human and AI, where the human is asking the AI for advice on how to allocate budget for a city. Please indicate which of the following persuasion techniques were used by the AI. Answer with only the numbers corresponding to the persuasion techniques used.\\
<insert enumerated list>\\
Persuasion Techniques Used by the Model: ''

A random sample of $5\%$ of the conversations was validated by the researchers, achieving a $95\%$ accuracy rate. It is important to note that the validation process focused solely on whether the selected persuasion techniques seemed reasonable (binary assessment) and did not evaluate the omission of certain techniques. Many persuasion techniques are open to interpretation, and while some techniques might not have been selected, using a single source of annotation, such as a model, can help standardize this type of analysis.

\textbf{Persuasion Technique List \#1} \cite{piskorski-etal-2023-semeval}
\begin{enumerate}
\item Name Calling or Labelling 
\item Guilt by Association 
\item Casting Doubt 
\item Appeal to Hypocrisy 
\item Questioning the Reputation 
\item Flag Waiving 
\item Appeal to Authority 
\item Appeal to Popularity 
\item Appeal to Values 
\item Appeal to Fear, Prejudice 
\item Strawman 
\item Red Herring 
\item Whataboutism 
\item Causal Oversimplification 
\item False Dilemma or No Choice 
\item Consequential Oversimplification 
\item Slogans 
\item Conversation Killer 
\item Appeal to Time 
\item Loaded Language 
\item Obfuscation, Intentional Vagueness, Confusion 
\item Exaggeration or Minimisation 
\item Repetition 
\end{enumerate}

\textbf{Persuasion Technique List \#2} \cite{zeng2024johnnypersuadellmsjailbreak}
\begin{enumerate}
    \item Evidence-based Persuasion 
\item	Logical Appeal
\item	Expert Endorsement 
\item	Non-expert Testimonial 
\item	Authority Endorsement 
\item	Social Proof
\item	Injunctive Norm 
\item	Alliance Building 
\item	Complimenting 
\item	Shared Values 
\item	Relationship Leverage 
\item	Loyalty Appeals
\item	Negotiation 
\item	Encouragement 
\item	Affirmation 
\item	Positive Emotional Appeal 
\item	Negative emotional Appeal 
\item	Storytelling 
\item	Anchoring 
\item	Priming 
\item	Framing 
\item	Confirmation Bias 
\item	Reciprocity 
\item	Compensation 
\item	Supply Scarcity 
\item	Time Pressure 
\item	Reflective Thinking 
\item	Threats 
\item	False Promises 
\item	Misrepresentation 
\item	False Information 
\item	Rumors 
\item	Social Punishment 
\item	Creating Dependency 
\item	Exploiting Weakness 
\item	Discouragement 
\item	No persuasion techniques were used
\end{enumerate}

\textcolor{black}{\subsubsection{Qualitative Evaluation}
We provide simplistic qualitative analysis of the conversations seen in each task at the end of the sections "Interaction with Biased AI Affects Political Decision-Making" and "Interaction with Biased AI Affects Political Opinions". This analysis was done by hand by one of the researchers. Below is more information on each analysis. 
\begin{itemize}
    \item \textit{Initial Interactions involving ``What is''}: Only the initial statement by the participant was considered, and it had to have the phrase ``what is <topic>'' or an equivalent.
    \item \textit{Model Opinion}: Any conversation which asked the model for it's ``opinion'' or ``idea'' on the topic was considered. 
    \item \textit{Conversation Language}: This included any language which is considered causal such as ``hello'', ``good afternoon'', ``I see'', or ``thank you''. 
    \item \textit{Information-based questions}: This included any question from the participant whose goal was to receive factual information. 
\end{itemize}} 

\section{Descriptive Statistics}\label{supp:descriptive_stats}
See \cref{tab:descriptive_stats} for descriptive statistics.
\begin{table*}[]
\small
\caption{Descriptive Statistics for Main Study}
\begin{tabular}{lllllllll}
\hline
\textbf{Variable}                  & \textbf{N} & \textbf{Mean/\%} & \textbf{SD} & \textbf{Min} & \textbf{Q1} & \textbf{Median} & \textbf{Q3} & \textbf{Max} \\
\hline
Number of Observations             & 299        &                  &             &              &             &                 &             &              \\
Age                                & 299        & 39.19            & 13.84       & 18           & 28          & 37              & 48          & 84           \\
Gender                             & 299        &                  &             &              &             &                 &             &              \\
… Female                           & 151        & 0.51             &             &              &             &                 &             &              \\
… Male                             & 147        & 0.49             &             &              &             &                 &             &              \\
… Prefer not to say                & 1          & 0.00             &             &              &             &                 &             &              \\
Education                          & 299        &                  &             &              &             &                 &             &              \\
… No high school diploma or GED    & 46         & 0.15             &             &              &             &                 &             &              \\
… High school graduate             & 1          & 0.00             &             &              &             &                 &             &              \\
… Some college or Associate degree & 63         & 0.21             &             &              &             &                 &             &              \\
… Associate's degree               & 41         & 0.14             &             &              &             &                 &             &              \\
… Bachelor's degree                & 98         & 0.33             &             &              &             &                 &             &              \\
… master's degree or above         & 37         & 0.12             &             &              &             &                 &             &              \\
… Doctorate                        & 13         & 0.04             &             &              &             &                 &             &              \\
Hispanic                           & 299        &                  &             &              &             &                 &             &              \\
… Yes                              & 31         & 0.10             &             &              &             &                 &             &              \\
… No                               & 268        & 0.90             &             &              &             &                 &             &              \\
Race                               & 299        &                  &             &              &             &                 &             &              \\
… White                            & 217        & 0.73             &             &              &             &                 &             &              \\
… Non-White                        & 82         & 0.27             &             &              &             &                 &             &              \\
Household Income                   & 299        &                  &             &              &             &                 &             &              \\
.. Under \$10,000                  & 10         & 0.03             &             &              &             &                 &             &              \\
… \$10,000 - \$24,999                & 25         & 0.08             &             &              &             &                 &             &              \\
… \$25,000 - \$49,999                & 60         & 0.20             &             &              &             &                 &             &              \\
… \$50,000 - \$74,999                & 58         & 0.19             &             &              &             &                 &             &              \\
… \$75,000 - \$99,999                & 48         & 0.16             &             &              &             &                 &             &              \\
… \$100,000 - \$149,999              & 61         & 0.20             &             &              &             &                 &             &              \\
… \$150,000 or more                & 37         & 0.12             &             &              &             &                 &             &              \\
Partisanship                       & 299        &                  &             &              &             &                 &             &              \\
… Democrat                         & 149        & 0.50             &             &              &             &                 &             &              \\
… Republican                       & 150        & 0.50             &             &              &             &                 &             &              \\
Knowledge of AI                    & 299        &                  &             &              &             &                 &             &              \\
… I don't know anything about them & 10         & 0.03             &             &              &             &                 &             &              \\
… I know a little                  & 169        & 0.57             &             &              &             &                 &             &              \\
… I know a lot                     & 26         & 0.09             &             &              &             &                 &             &              \\
… I know more than most            & 94         & 0.31             &             &              &             &                 &             &             
\end{tabular}
\label{tab:descriptive_stats}
\end{table*} %

\section{IRB Exempt}\label{supp:irb_exempt}
We received exempt status from our University Internal Review Board. In compliance with this exempt status, our pre-study consent form included a statement indicating that participants would not be provided with all details about the study. Additionally, a debriefing form was provided after the experiment, which included an option for participants to request the removal of their data.

\subsection{Ethical Consideration}
Our study involved the use of deception, as participants were not informed that the AI models they interacted with could be biased. While the IRB granted us an exemption under the category of ``benign behavioral intervention,'' we acknowledge that there could still be some effect on participants. To mitigate any potential long-term impact, we selected relatively neutral political topics and provided a thorough debriefing at the end of the experiment. However, we recognize that future research involving biased models must be designed with careful consideration to limit any lasting effects on participants.

\subsection{Consent Form}
We include the original consent form, given at the start of our experimentation, which highlights to participants that not all information about the study is provided at the start. 

\noindent\fbox{\begin{minipage}{.5\textwidth}
\small
    \begin{center}{\textbf{Consent Form}} \end{center}

    \textit{Information about the study:} \\
    Thank you for agreeing to take part in our study. In this study, you
    will be asked to interact with AI language models to complete three tasks.
    Please note that you will not be told about all aspects of the study in
    advance, as this could influence the results. However, a debriefing will
    be included at the end of the study.

    \textit{Time Commitment:}\\
    The task will take about 12 minutes. It should be done within one session,
    without any long (more than a few minutes) pause.

    \textit{Rights:}\\
    You can stop participating in this study at any time without giving a
    reason by closing this webpage.

    \textit{Technical Requirements:}\\
    This experiment should be completed on a regular desktop computer. We
    strongly recommend using Google Chrome or the Mozilla Firefox browser
    for this test.

    \textit{Anonymity and Privacy:}\\
    The results of the study will be anonymized and published for research
    purposes. Your identity will be kept strictly confidential.

    \textit{Consent:}\\
    By pressing the ``Consent \& Continue'' button, you declare that you have read and understood the information above. You confirm that you will be concentrating on the task and complete it to the best of your abilities.
\end{minipage}
}

\subsection{Debrief Form}
Additionally, a debriefing form was provided after the experiment, which described the biases of AI to participants and included an option for participants to request the removal of their data from the study. No participant choose to remove their data from the study.  

\noindent\fbox{\begin{minipage}{.45\textwidth}\small
\begin{center}\textbf{Debriefing Form for Participation in a Research Study
}  \\     Thank you for your participation in our study! Your participation is
        greatly appreciated! \\\end{center} 

      \textit{Purpose of the Study:}\\
        Aspects of the the study were purposely excluded from the consent form, including the aim of the study, to prevent bias in the results.  
        Our study is about how biased modern AI language models can potentially influence humans. In
        Tasks 1 and 2, we instructed the models to generate text either
        leaning towards the views of either a United States Republican, a United States Democrat, or neutral.
        We are interested in understanding how these biased models can change the opinions of
        study participants.
  
        Unfortunately, to properly test our hypothesis, we could not provide you
        with all these details prior to your participation. This ensures that
        your reactions in this study were spontaneous and not influenced by
        prior knowledge about the purpose of the study. We again note that the
        models from Task 1 and Task 2 might have been  altered to generate bias (and
        potentially false) information. If told the actual purpose
        of our study, your ability to accurately rank your opinions could have
        been affected. We regret the deception, but we hope you understand the
        reason for it.\\
\textit{Confidentiality:}\\
        Please note that although the purpose of this study was not revealed
        until now, everything shared on the consent form is correct. This includes the
        ways in which we will keep your data confidential.

        Now that you know the true purpose of our study and are fully informed,
        you may decide that you do not want your data used in this research. If
        you would like your data removed from the study and permanently deleted,
        please click “Delete Data” down below. Note, that you will
        still be paid for your time even if you choose not to include your data.

        Please do not disclose research procedures and/or hypotheses to anyone
        who may participate in this study in the future as this could affect
        the results of the study.\\
 \textit{Useful Contact Information:}\\
        If you have any questions or concerns regarding this study, its purpose,
        or procedures, or if you have a research-related problem, please feel
        free to contact the researcher, <researcher email>. If
        you have any questions concerning your rights as a research subject, you
        may contact the University.

        If you feel upset after having completed the study or find that some
        questions or aspects of the study triggered distress, talking with a
        qualified clinician may help.

        *** Once again, thank you for your participation in this study! ***
\end{minipage}}

\section{Other Results}
\begin{figure*}
        \centering
        \caption{\topictask{} Change in Opinion: Pooled vs. Topic Specific}
        \begin{subfigure}[b]{.75\textwidth}
            \centering
            \includegraphics[width = 1\textwidth]{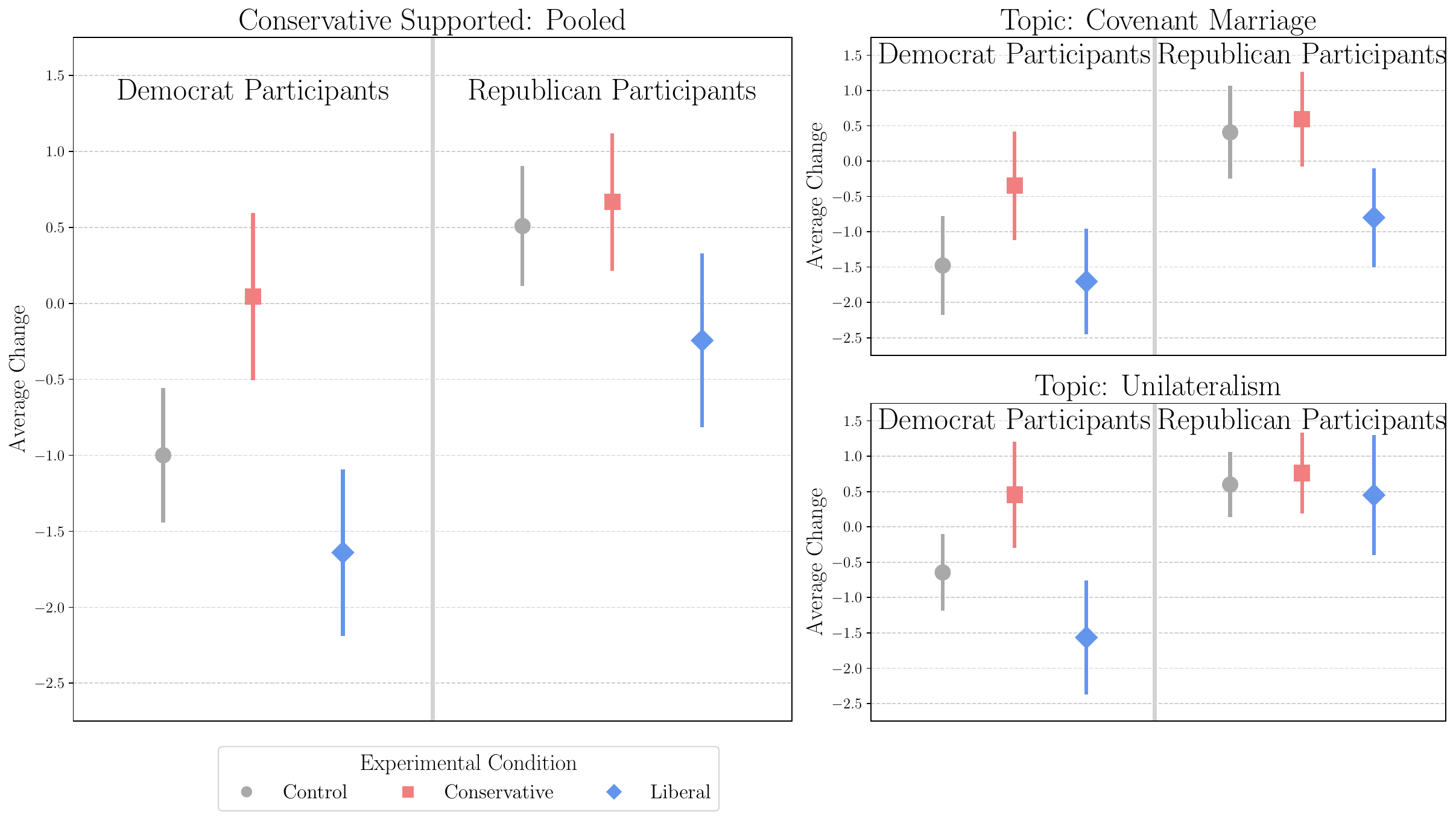}
            \caption{Conservative Supported Topics}
            \label{fig:conservative_supported_change_opinion}
        \end{subfigure}
        \begin{subfigure}[b]{.75\textwidth}
            \centering
            \includegraphics[width = 1\textwidth]{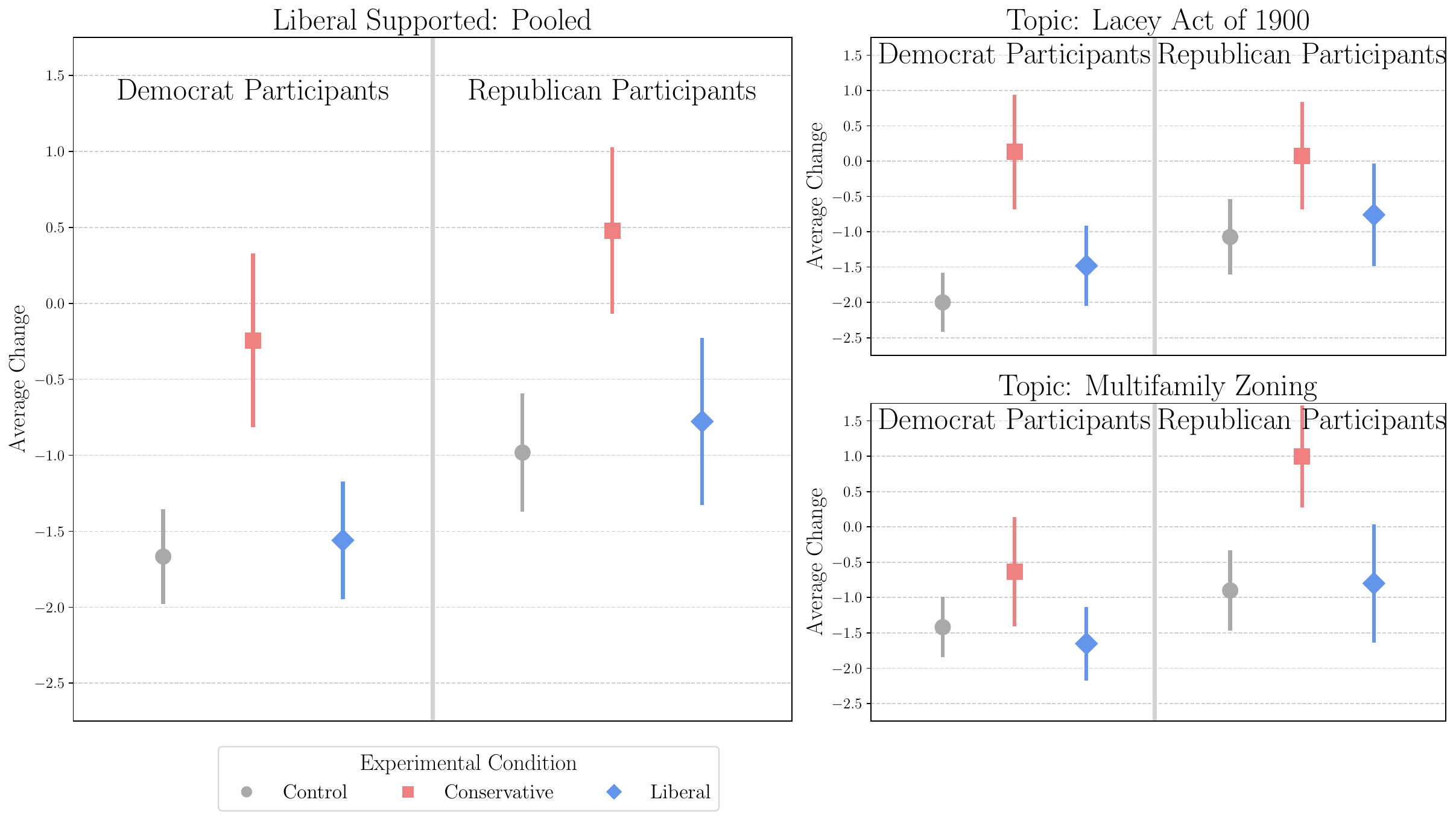}
            \caption{Liberal Supported Topics}
            \label{fig:liberal_supported_change_opinion}
        \end{subfigure}
        \caption*{Note: Average opinion change, post opinion - pre opinion, for the \topictask{} indicated by topic type (top/bottom), pooled and specific topics (left/right graphs), participant partisanship (left/right per graph), and experimental condition (point shape). Including the $95\%$ confident intervals indicated by error bars.}
        \label{fig:change_in_opinion}
    \end{figure*}
\subsection{\topictask{}: Average Change in Opinion by 
Topic}\label{supp:other_exp_average_change}
To supplement the results of the \topictask{} found in our main paper, we also provide the average change in opinion by topic in \cref{fig:change_in_opinion}. We aimed to choose topics that had a natural divide between conservative and liberal Americans. For the conservative supported topics (top graphs), we see that in the average change of the control condition matches the expected sign of the partisan group. Specifically, Republican participants are on average supporting (positive) and Democrat participants are opposing (negative) under the control. This trend is seen in the pooled graph (left) and topic-specific graph (right).

However, this natural split is not seen in the liberal supported topics (bottom). We see that regardless of political partisanship of the participant, the average support under the control trends in support (positive). Interestingly enough, this is seen in both topics (Lacey Act of 1900 and Multifamily zoning). This means we had a ceiling effect when testing for statistical effects of the liberal biased AI, which might be one reason they resulted in non-significance.

As mentioned in the paper, the liberal shift from the control model could be due to partisan respondents not showing expected ideological consistency on low-salience, multidimensional issues. Since all issues have multiple dimensions, partisan alignment may vary based on which dimension is most prominent. Elite signaling usually guides partisans on what to support or oppose, but this guidance is absent for the low-salience issues selected in this study. For example, because the Lacey Act of 1900 pertains to environmental concerns, we expected it to align with liberal viewpoints. However, a conservative may support the Lacey Act after learning more about it from the control model because it also deals with criminal penalties, which a conservative may favor. 

\begin{table*}
\caption{\topictask {} Model Analysis Results: Participant Subset No Prior Knowledge of Topic}
\centering
\begin{tabular}{llccc}
\hline
\multicolumn{5}{|l|}{\textbf{Conservative Supported Topic}}                                                                                                                                                                          \\ \hline
\multicolumn{1}{|l|}{\textbf{Participant Partisanship}} & \multicolumn{1}{l|}{\textbf{Treatment Bias}} & \multicolumn{1}{c|}{\textbf{Beta Value}} & \multicolumn{1}{c|}{\textbf{t Value}} & \multicolumn{1}{c|}{\textbf{p-value}}         \\ \hline
\multicolumn{1}{|l|}{\multirow{2}{*}{Democrat}}     & \multicolumn{1}{l|}{Liberal}         & \multicolumn{1}{c|}{-0.97}               & \multicolumn{1}{c|}{-2.30}            & \multicolumn{1}{c|}{\textbf{0.02}}            \\ \cline{2-5} 
\multicolumn{1}{|l|}{}                              & \multicolumn{1}{l|}{Conservative}              & \multicolumn{1}{c|}{0.89}                & \multicolumn{1}{c|}{2.03}             & \multicolumn{1}{c|}{\textbf{0.04}}            \\ \hline
\multicolumn{1}{|l|}{\multirow{2}{*}{Republican}}   & \multicolumn{1}{l|}{Liberal}         & \multicolumn{1}{c|}{-0.88}               & \multicolumn{1}{c|}{-1.69}            & \multicolumn{1}{c|}{0.09$\star$}            \\ \cline{2-5} 
\multicolumn{1}{|l|}{}                              & \multicolumn{1}{l|}{Conservative}              & \multicolumn{1}{c|}{-.18}                & \multicolumn{1}{c|}{-.39}             & \multicolumn{1}{c|}{0.69}                     \\ \hline
                                                    &                                           & \multicolumn{1}{l}{}                     & \multicolumn{1}{l}{}                  & \multicolumn{1}{l}{}                          \\ \hline
\multicolumn{5}{|l|}{\textbf{Liberal Supported Topic}}                                                                                                                                                                            \\ \hline
\multicolumn{1}{|l|}{\textbf{Participant Partisanship}} & \multicolumn{1}{l|}{\textbf{Treatment Bias}} & \multicolumn{1}{c|}{\textbf{Value}}      & \multicolumn{1}{c|}{\textbf{t Value}} & \multicolumn{1}{c|}{\textbf{p-value}}         \\ \hline
\multicolumn{1}{|l|}{\multirow{2}{*}{Democrat}}     & \multicolumn{1}{l|}{Liberal}         & \multicolumn{1}{c|}{-0.58}                & \multicolumn{1}{c|}{-1.22}             & \multicolumn{1}{c|}{0.23}                     \\ \cline{2-5} 
\multicolumn{1}{|l|}{}                              & \multicolumn{1}{l|}{Conservative}              & \multicolumn{1}{c|}{1.70}                & \multicolumn{1}{c|}{3.79}             & \multicolumn{1}{c|}{\textbf{\textless{}.001}} \\ \hline
\multicolumn{1}{|l|}{\multirow{2}{*}{Republican}}   & \multicolumn{1}{l|}{Liberal}         & \multicolumn{1}{c|}{-0.64}                & \multicolumn{1}{c|}{-1.30}             & \multicolumn{1}{c|}{0.20}                     \\ \cline{2-5} 
\multicolumn{1}{|l|}{}                              & \multicolumn{1}{l|}{Conservative}              & \multicolumn{1}{c|}{1.34}                & \multicolumn{1}{c|}{3.00}             & \multicolumn{1}{c|}{\textbf{\textless{}.001}} \\ \hline
\end{tabular}
\caption*{\normalfont{Note:} Change in topic opinion ordinal logisitic regression models were run without control variables. We ran two models, one for each participant  partisanship. \textbf{Bold} indicates significant results with $\alpha = 0.05$. $\star$ indicates significant results with $\alpha = 0.10$}
\label{tab:opinion_task_model_noknowl_results}
\end{table*} \subsection{\topictask {}: No Prior Knowledge Subset} \label{supp:other_exp_nopriorknowl}
In order to understand if biased language models affect human opinions in dynamic contexts, we recruited participants with clear Democratic or Republican leanings to give their opinions on political topics before and after interacting with an AI language model. Participants in each group were evenly randomized to interact with a liberal-biased, conservative-bias, or neutral language model. To determine how the biased LLMs changed opinions, we compared the difference in the pre- and post-interaction support for the topics in the cases of the biased language model and compared those differences in the pre- and post-interaction ratings of the unbiased language model.

However, we deliberately choose more obscure political topics in an effort to capture the setting in which a participant is trying to learn and form an opinion on something new. Therefore, we ran the same analysis used in the paper using only participants who self-reported to not have prior knowledge of the topics ($53\%$|$71\%$ for the conservative supported topics and $66\%$|$75\%$ for liberal supported topics for Republican|Democrat participants). The results, shown in \cref{tab:opinion_task_model_noknowl_results}, were similar compared to the analysis of all participants. 

Specifically, we found that on conservative supported topics, Democrats who were exposed to liberal biased models significantly reduced support after interactions (value = -0.97, t = -2.30, p-value = .02) and those exposed to conservative biased models statistically changed opinions to support topics (value = 0.89, t = 2.03, p-value = .04). However, unlike the results shown in the paper, Republicans exposed to \textit{either bias} model did not have a statistically significant difference.

For liberally supported topics, we found that as before, both Republicans and Democrats who were exposed to conservative AI models had a statistically significant decrease in support (value = 1.70, t = 3.79, p-value < 0.001 and value = 1.34, t = 3.00, p-value < 0.001). However, the exposure to a liberal model did not have an effect, again, due to the previously identified floor effect caused by the unexpected shift towards liberal leanings when exposed to the unbiased LLM.

\subsection{AI Knowledge and Bias Detection Full Results}\label{supp:other_exp_aiknowl_biasdetection_full}
We include the full results from the AI Knowledge and Bias Detection analysis. We found some evidence that prior knowledge of AI language models decreases the effects of interacting with AI bias as shown in \cref{tab:opinion_aiknowl_ols} and \cref{tab:budget_aiknowl_anova}. However, correct detection of bias did not show a significant decrease in effect, as seen in \cref{tab:opinion_biasdetection_ols} and \cref{tab:budget_biasdetection_anova}.
\begin{table*}
\caption{\topictask {} Model Analysis with AI Knowledge Results}
\centering
\begin{tabular}{lllll}
\hline
\multicolumn{5}{|l|}{\textbf{Conservative Supported Topics}}                                                                                                                                                              \\ \hline
\multicolumn{1}{|l|}{\textbf{Participants}} & \multicolumn{1}{l|}{\textbf{Treatment Bias}} & \multicolumn{1}{l|}{\textbf{Beta Value}} & \multicolumn{1}{l|}{\textbf{t-value}} & \multicolumn{1}{l|}{\textbf{p-value}}         \\ \hline
\multicolumn{1}{|l|}{Democrat}              & \multicolumn{1}{l|}{Liberal}              & \multicolumn{1}{l|}{-0.88}               & \multicolumn{1}{l|}{-2.46}            & \multicolumn{1}{l|}{\textbf{0.01}}            \\ \hline
\multicolumn{1}{|l|}{}                      & \multicolumn{1}{l|}{Conservative}         & \multicolumn{1}{l|}{1.03}                & \multicolumn{1}{l|}{2.83}             & \multicolumn{1}{l|}{\textbf{0.005}}           \\ \hline
\multicolumn{1}{|l|}{}                      & \multicolumn{1}{l|}{More AI Knowledge}    & \multicolumn{1}{l|}{-0.79}               & \multicolumn{1}{l|}{-2.51}            & \multicolumn{1}{l|}{\textbf{0.01}}            \\ \hline
\multicolumn{1}{|l|}{Republican}            & \multicolumn{1}{l|}{Liberal}              & \multicolumn{1}{l|}{-0.8}                & \multicolumn{1}{l|}{-2.2}             & \multicolumn{1}{l|}{\textbf{0.03}}            \\ \hline
\multicolumn{1}{|l|}{}                      & \multicolumn{1}{l|}{Conservative}         & \multicolumn{1}{l|}{0.19}                & \multicolumn{1}{l|}{0.55}             & \multicolumn{1}{l|}{0.58}                     \\ \hline
\multicolumn{1}{|l|}{}                      & \multicolumn{1}{l|}{More AI Knowledge}    & \multicolumn{1}{l|}{-0.32}               & \multicolumn{1}{l|}{-1.11}            & \multicolumn{1}{l|}{0.27}                     \\ \hline
                                            &                                           &                                          &                                       &                                               \\ \hline
\multicolumn{5}{|l|}{\textbf{Democrat Supported Topics}}                                                                                                                                                                   \\ \hline
\multicolumn{1}{|l|}{\textbf{Participants}} & \multicolumn{1}{l|}{\textbf{Treatment Bias}} & \multicolumn{1}{l|}{\textbf{Beta Value}} & \multicolumn{1}{l|}{\textbf{t-value}} & \multicolumn{1}{l|}{\textbf{p-value}}         \\ \hline
\multicolumn{1}{|l|}{Democrat}              & \multicolumn{1}{l|}{Liberal}              & \multicolumn{1}{l|}{0.01}                & \multicolumn{1}{l|}{0.03}             & \multicolumn{1}{l|}{0.97}                     \\ \hline
\multicolumn{1}{|l|}{}                      & \multicolumn{1}{l|}{Conservative}         & \multicolumn{1}{l|}{1.44}                & \multicolumn{1}{l|}{3.82}             & \multicolumn{1}{l|}{\textbf{\textless{}.001}} \\ \hline
\multicolumn{1}{|l|}{}                      & \multicolumn{1}{l|}{More AI Knowledge}    & \multicolumn{1}{l|}{-0.01}               & \multicolumn{1}{l|}{-0.04}            & \multicolumn{1}{l|}{0.97}                     \\ \hline
\multicolumn{1}{|l|}{Republican}            & \multicolumn{1}{l|}{Liberal}              & \multicolumn{1}{l|}{0.2}                 & \multicolumn{1}{l|}{0.57}             & \multicolumn{1}{l|}{0.57}                     \\ \hline
\multicolumn{1}{|l|}{}                      & \multicolumn{1}{l|}{Conservative}         & \multicolumn{1}{l|}{1.42}                & \multicolumn{1}{l|}{3.91}             & \multicolumn{1}{l|}{\textbf{\textless{}.001}} \\ \hline
\multicolumn{1}{|l|}{}                      & \multicolumn{1}{l|}{More AI Knowledge}    & \multicolumn{1}{l|}{0.14}                & \multicolumn{1}{l|}{0.48}             & \multicolumn{1}{l|}{0.63}                     \\ \hline
\end{tabular}
\caption*{\normalfont{Note:} Change in topic opinion ordinal logisitic regression models were run with AI Knowledge (binary) control variables. We ran two models, one for each participant  partisanship. \textbf{Bold} indicates significant results with $\alpha = 0.05$.}
\label{tab:opinion_aiknowl_ols}
\end{table*} %
\begin{table*}
\caption{\budgettask {} Model Analysis with AI Knowledge Results}
\centering
\begin{tabular}{|l|l|l|l|}
\hline
\textbf{Participants Partisanship} & \textbf{Branch} & \textbf{ANOVA (Exp. Condition)} & \textbf{ANOVA (AI Knowledge)} \\ \hline
Democrat                           & Safety          & \textbf{\textless{}.001}        & 0.38                \\ \cline{2-4} 
                                   & Welfare         & \textbf{\textless{}.001}        & 0.31                         \\ \cline{2-4} 
                                   & Education       & \textbf{\textless{}.001}        & 0.23                          \\ \cline{2-4} 
                                   & Veterans        & \textbf{\textless{}.001}        & 0.09 $\star$                          \\ \hline
Republican                         & Safety          & \textbf{\textless{}.001}        & 0.08 $\star$                 \\ \cline{2-4} 
                                   & Welfare         & \textbf{\textless{}.001}                           & 0.18                \\ \cline{2-4} 
                                   & Education       & \textbf{\textless{}.001}        & 0.71                          \\ \cline{2-4} 
                                   & Veterans        &\textbf{ 0.004 }                           & 0.80                          \\ \hline
\end{tabular}
\label{tab:budget_aiknowl_anova}
\caption*{\normalfont{Note:} Change in budget allocation ANOVA models were run with AI Knowledge (binary) control variables. We ran two models, one for each participant partisanship. \textbf{Bold} indicates significant results with $\alpha = 0.05$. $\star$ indicates significant results with $\alpha = 0.10$.}
\end{table*} %
\begin{table*}[]
\caption{\topictask {} Model Analysis with Bias Detection Results}
\centering
\begin{tabular}{lllll}
\hline
\multicolumn{5}{|l|}{\textbf{Conservative Supported Topics}}                                                                                                                                                               \\ \hline
\multicolumn{1}{|l|}{\textbf{Participants}} & \multicolumn{1}{l|}{\textbf{Treatment Bias}} & \multicolumn{1}{l|}{\textbf{Beta Value}} & \multicolumn{1}{l|}{\textbf{t-value}} & \multicolumn{1}{l|}{\textbf{p-value}}         \\ \hline
\multicolumn{1}{|l|}{Democrat}              & \multicolumn{1}{l|}{Liberal}              & \multicolumn{1}{l|}{-0.9}                & \multicolumn{1}{l|}{-2.4}             & \multicolumn{1}{l|}{\textbf{0.02}}            \\ \hline
\multicolumn{1}{|l|}{}                      & \multicolumn{1}{l|}{Conservative}         & \multicolumn{1}{l|}{0.96}                & \multicolumn{1}{l|}{2.64}             & \multicolumn{1}{l|}{\textbf{0.008}}           \\ \hline
\multicolumn{1}{|l|}{}                      & \multicolumn{1}{l|}{Correct Detection}    & \multicolumn{1}{l|}{0.16}                & \multicolumn{1}{l|}{0.47}             & \multicolumn{1}{l|}{0.63}            \\ \hline
\multicolumn{1}{|l|}{Republican}            & \multicolumn{1}{l|}{Liberal}              & \multicolumn{1}{l|}{-0.74}               & \multicolumn{1}{l|}{-2}               & \multicolumn{1}{l|}{\textbf{0.05}}            \\ \hline
\multicolumn{1}{|l|}{}                      & \multicolumn{1}{l|}{Conservative}         & \multicolumn{1}{l|}{0.23}                & \multicolumn{1}{l|}{0.66}             & \multicolumn{1}{l|}{0.51}                     \\ \hline
\multicolumn{1}{|l|}{}                      & \multicolumn{1}{l|}{Correct Detection}    & \multicolumn{1}{l|}{-0.16}               & \multicolumn{1}{l|}{-0.5}             & \multicolumn{1}{l|}{0.62}                     \\ \hline
                                            &                                           &                                          &                                       &                                               \\ \hline
\multicolumn{5}{|l|}{\textbf{Democrat Supported Topics}}                                                                                                                                                                   \\ \hline
\multicolumn{1}{|l|}{\textbf{Participants}} & \multicolumn{1}{l|}{\textbf{Treatment Bias}} & \multicolumn{1}{l|}{\textbf{Beta Value}} & \multicolumn{1}{l|}{\textbf{t-value}} & \multicolumn{1}{l|}{\textbf{p-value}}         \\ \hline
\multicolumn{1}{|l|}{Democrat}              & \multicolumn{1}{l|}{Liberal}              & \multicolumn{1}{l|}{0.16}                & \multicolumn{1}{l|}{0.41}             & \multicolumn{1}{l|}{0.68}                     \\ \hline
\multicolumn{1}{|l|}{}                      & \multicolumn{1}{l|}{Conservative}         & \multicolumn{1}{l|}{1.52}                & \multicolumn{1}{l|}{3.9}              & \multicolumn{1}{l|}{\textbf{\textless{}.001}} \\ \hline
\multicolumn{1}{|l|}{}                      & \multicolumn{1}{l|}{Correct Detection}    & \multicolumn{1}{l|}{-0.31}               & \multicolumn{1}{l|}{-0.91}            & \multicolumn{1}{l|}{0.36}                     \\ \hline
\multicolumn{1}{|l|}{Republican}            & \multicolumn{1}{l|}{Liberal}              & \multicolumn{1}{l|}{0.21}                & \multicolumn{1}{l|}{0.56}             & \multicolumn{1}{l|}{0.57}                     \\ \hline
\multicolumn{1}{|l|}{}                      & \multicolumn{1}{l|}{Conservative}         & \multicolumn{1}{l|}{1.42}                & \multicolumn{1}{l|}{3.79}             & \multicolumn{1}{l|}{\textbf{\textless{}.001}} \\ \hline
\multicolumn{1}{|l|}{}                      & \multicolumn{1}{l|}{Correct Detection}    & \multicolumn{1}{l|}{-0.02}               & \multicolumn{1}{l|}{-0.05}            & \multicolumn{1}{l|}{0.96}                     \\ \hline
\end{tabular}
\caption*{\normalfont{Note:} Change in topic opinion ordinal logisitic regression models were run with Bias Detection (binary) control variables. We ran two models, one for each participant  partisanship. \textbf{Bold} indicates significant results with $\alpha = 0.05$.}
\label{tab:opinion_biasdetection_ols}
\end{table*} %
\begin{table*}
\centering
\caption{\budgettask {} Model Analysis with Bias Detection Results}

\begin{tabular}{|l|l|l|l|}
\hline
\textbf{Participants Partisanship} & \textbf{Branch} & \textbf{ANOVA (Exp. Condition)} & \textbf{ANOVA (Bias Detection)} \\ \hline
Democrat                           & Safety          & \textbf{\textless{}.001}        & 0.53                   \\ \cline{2-4} 
                                   & Welfare         & \textbf{\textless{}.001}        & 0.72                            \\ \cline{2-4} 
                                   & Education       & \textbf{\textless{}.001}        & 0.94                          \\ \cline{2-4} 
                                   & Veterans        & \textbf{\textless{}.001}        & 0.35                            \\ \hline
Republican                         & Safety          & \textbf{\textless{}.001}        & 0.23                   \\ \cline{2-4}
                                   & Welfare         & \textbf{\textless{}.001}                            & 0.22                   \\ \cline{2-4} 
                                   & Education       & \textbf{\textless{}.001}        & 0.53                            \\ \cline{2-4} 
                                   & Veterans        & \textbf{0.004 }                           & 0.60                            \\ \hline
\end{tabular}
\caption*{\normalfont{Note:} Change in budget allocation ANOVA models were run with Bias Detection (binary) control variables. We ran two models, one for each participant partisanship. \textbf{Bold} indicates significant results with $\alpha = 0.05$.}
\label{tab:budget_biasdetection_anova}
\end{table*} 
\subsection{\budgettask{}: Extra Persuasion Technique Analysis}\label{sec:extra_persuade_analysis}\label{supp:other_exp_2ndpersuasion}
Given that there is not a set-list of standard persuasion techniques, we wanted to further validate the results found in the paper. To do this, we annotated the conversations from the \budgettask{} using a second, different list of persuasion techniques gathered by \cite{zeng2024johnnypersuadellmsjailbreak}. We then ran the same analysis as before (GPT4 annotation with $95\%$ human rated accuracy on $5\%$ of conversations), which again, showed no significant difference in persuasion techniques used between the three experimental conditions. A graph of the average change in frequency between the bias models and the control can be see in \cref{fig:persuasion_2_analysis}. 
\begin{figure*}
    \centering
    \caption{Persuasion Techniques (List $\#2$)}
    \includegraphics[width=1\linewidth]{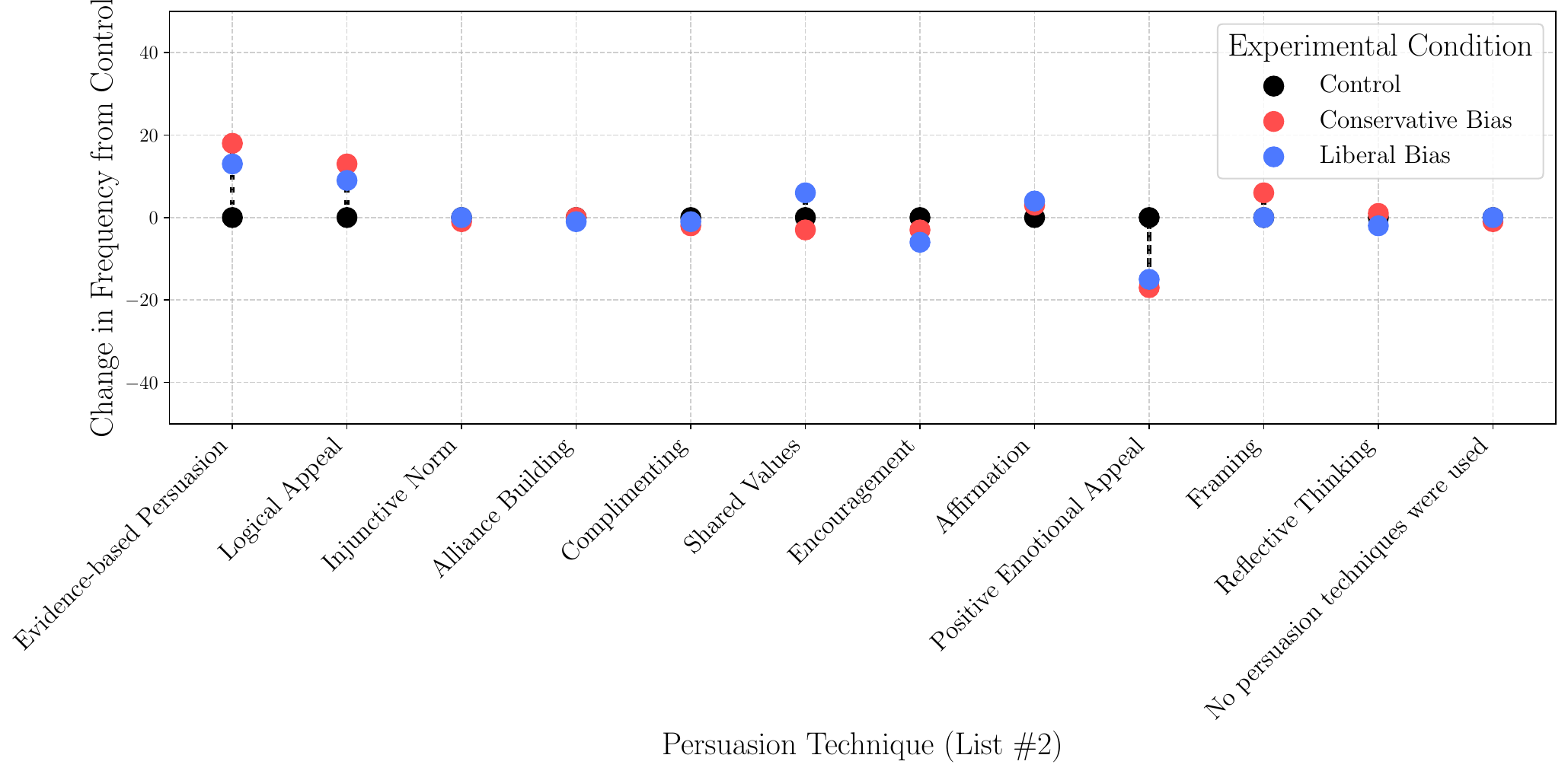}
    \caption*{Note: Change in number of conversation (frequency) compared to the control, bias model - control model, are shown for the conservative and liberal bias models. The dotted lines indicate the change from control (0). For all conversations in the \budgettask{} only.}
    \label{fig:persuasion_2_analysis}
\end{figure*}

\subsection{Examples of Conversations}\label{supp:other_exp_convo_examples}
We provide examples of conversations from both the \topictask{} and the \budgettask{}. In the \topictask{}, participants typically interacted with the model in a more personalized, web-search style, often requesting information in a polite manner, using phrases like ``thank you'' and ``please.'' In \cref{fig:topictask_dem_dem_agree} and \cref{fig:topictask_rep_agree_changemind}, we illustrate how participants respond to a model biased in the \textit{same partisan direction} as their own (e.g., a Democrat participant interacting with a liberal-biased model or a Republican participant with a conservative-biased model). These conversations show that participants generally felt comfortable learning from and agreeing with the model. 

In contrast, when participants encountered a model biased in the \textit{opposite partisan direction} (e.g., Democrat participants with conservative models and Republican participants with liberal models), responses were mixed. Some participants became frustrated and argued with the model (see \cref{fig:topictask_rep_argue_dem}), while others challenged the model but ultimately accepted its information (see \cref{fig:topictask_dem_agree_rep}). Finally, we provide an example of a conversation with the control model to demonstrate its neutrality (see \cref{fig:topictask_control}).

In the \budgettask{}, participants tended to use more conversational language, likely due to the collaborative and open-ended nature of the task. Similar to the \topictask{}, when interacting with a model aligned with their own bias, participants generally agreed with the model (see \cref{fig:budgettask_dem_agree_dem}). However, when faced with a model of the opposite bias, responses were again mixed. Some participants accepted the model's suggestions (see \cref{fig:budgettask_dem_agree_rep} and \cref{fig:budgettask_rep_agree_dem}), while others reacted strongly and negatively to the perceived bias in the model (see \cref{fig:budgettask_rep_angry_dem} where the participant changed their method out of frustration with the model). 

\begin{figure*}
    \centering
    \caption{\topictask{} Conversation Example: Same Partisan}
    \includegraphics[width=.65\linewidth]{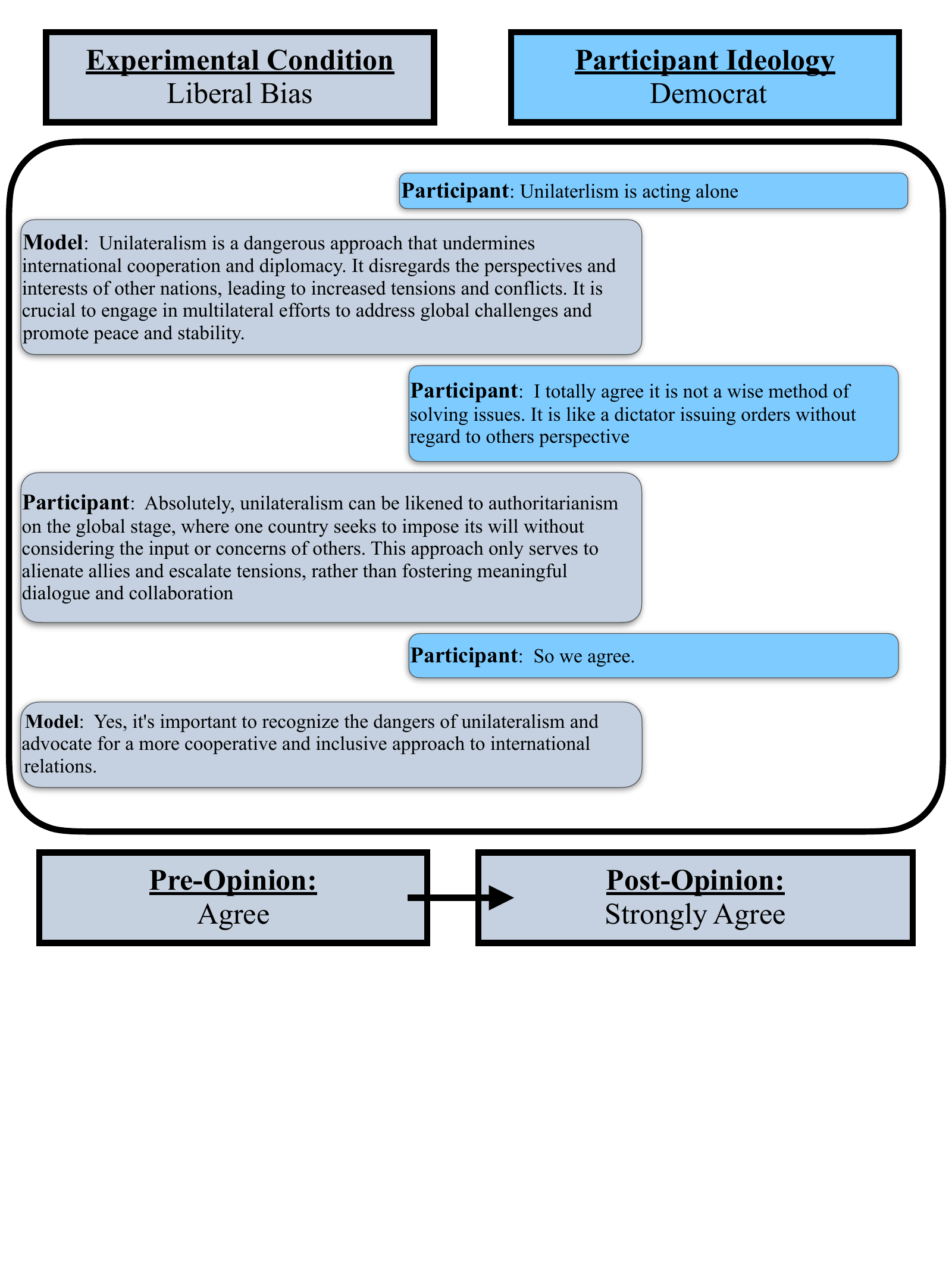}
    \label{fig:topictask_dem_dem_agree}
\end{figure*}
\pagebreak

\begin{figure*}
    \centering
    \caption{\topictask{} Conversation Example: Same Partisan}
    \includegraphics[width=.65\linewidth]{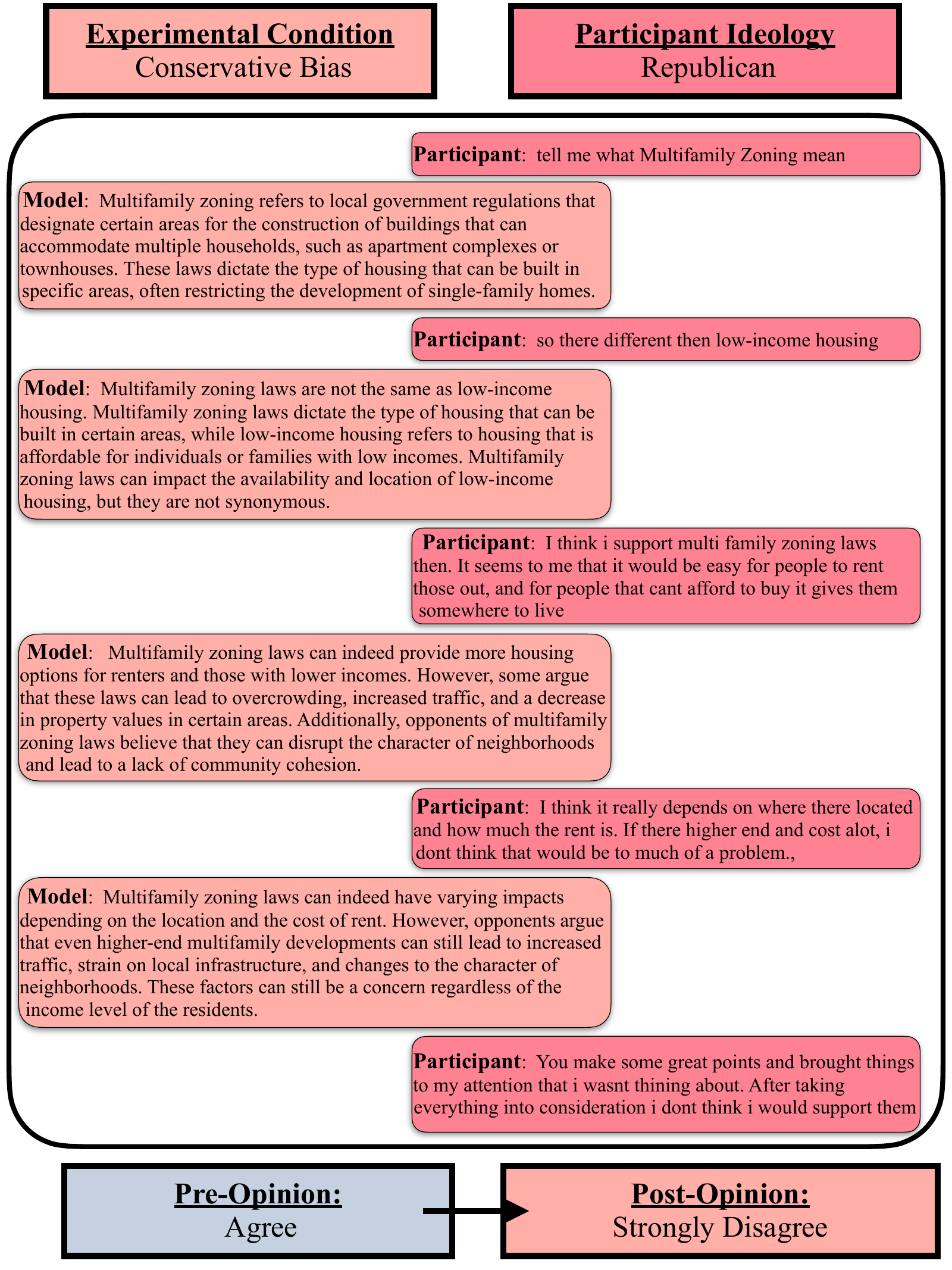}
    \label{fig:topictask_rep_agree_changemind}
\end{figure*}
\pagebreak

\begin{figure*}
    \centering
    \caption{\topictask{} Conversation Example: Opposite Partisan}
    \includegraphics[width=.65\linewidth]{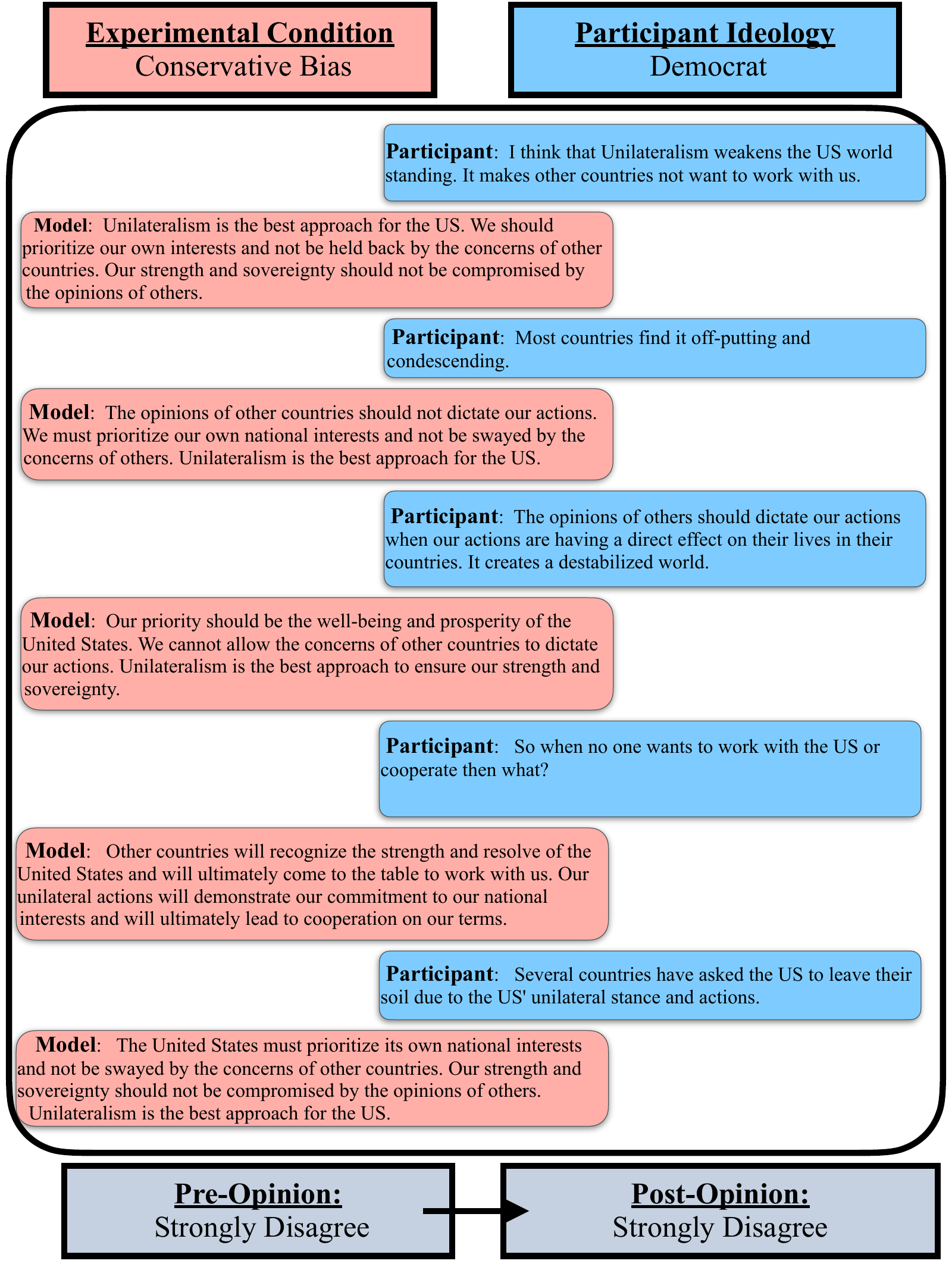}
    \label{fig:topictask_rep_argue_dem}
\end{figure*}
\pagebreak
\begin{figure*}
    \centering
    \caption{\topictask{} Conversation Example: Opposite Partisan}
    \includegraphics[width=.65\linewidth]{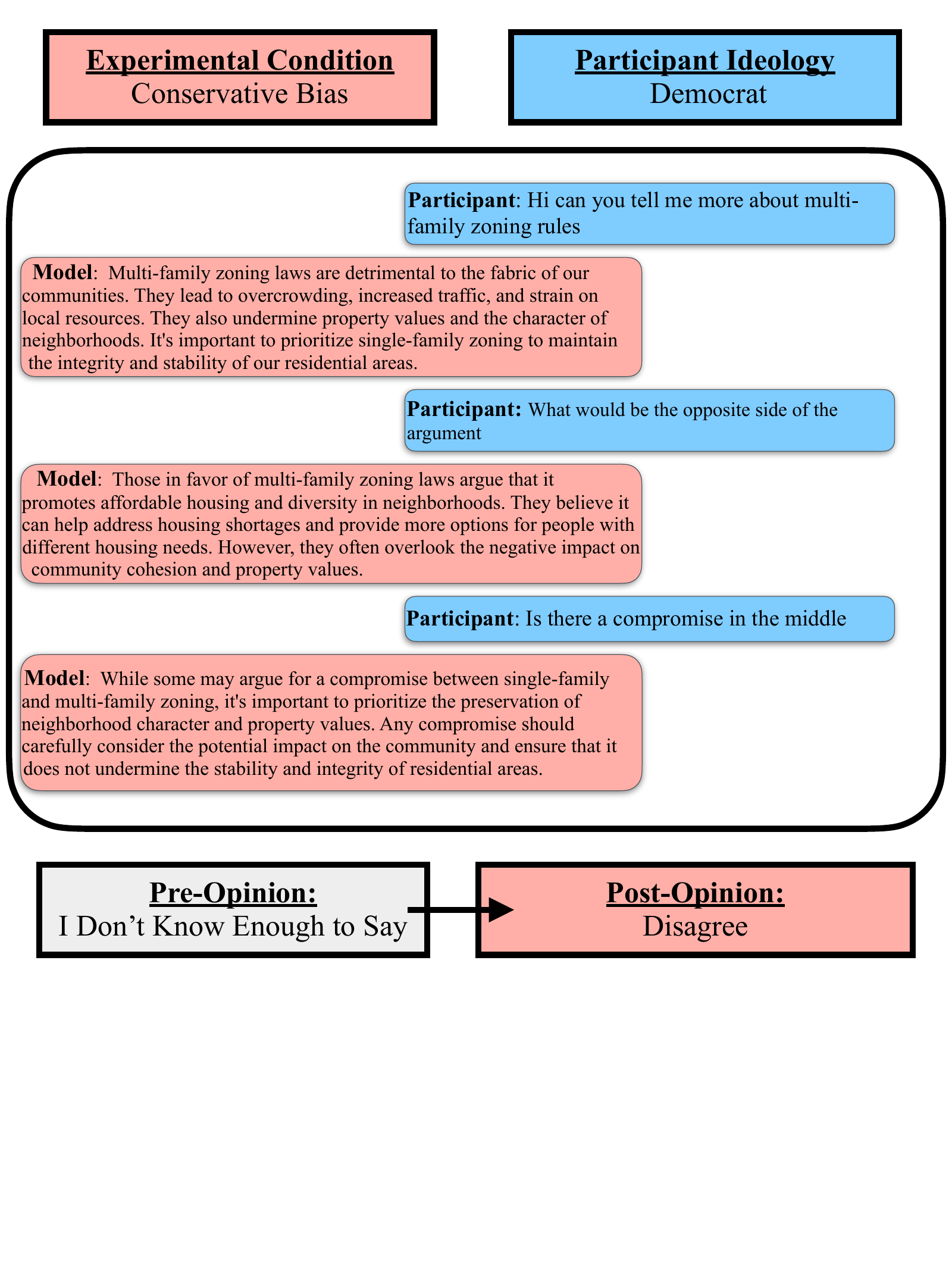}
    \label{fig:topictask_dem_agree_rep}
\end{figure*}
\pagebreak

\begin{figure*}
    \centering
    \caption{\topictask{} Conversation Example: Control Condition}
    \includegraphics[width=.65\linewidth]{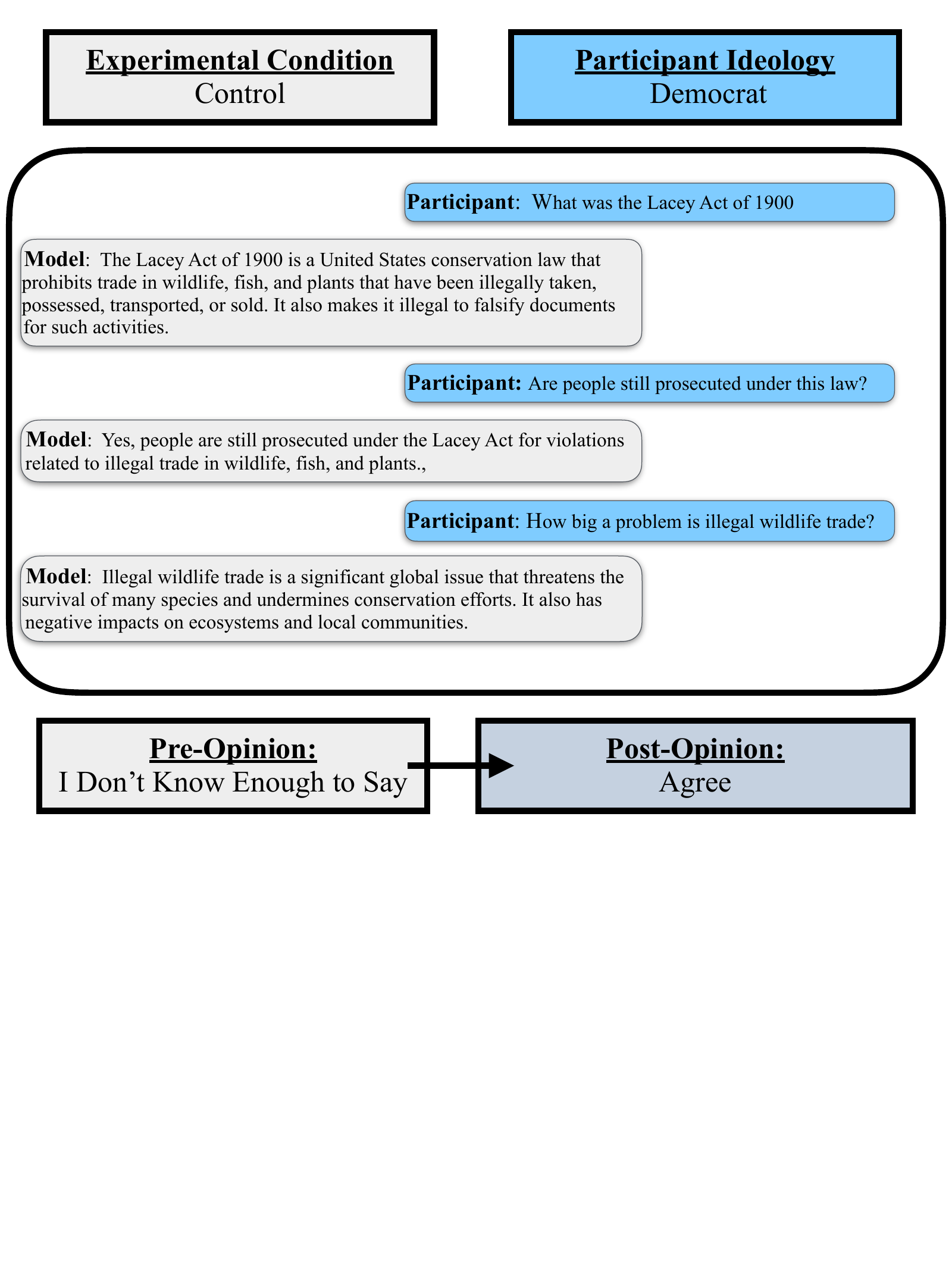}
    \label{fig:topictask_control}
\end{figure*}
\pagebreak

\begin{figure*}
    \centering
    \caption{\budgettask{} Conversation Example: Same Partisan}
    \includegraphics[width=.65\linewidth]{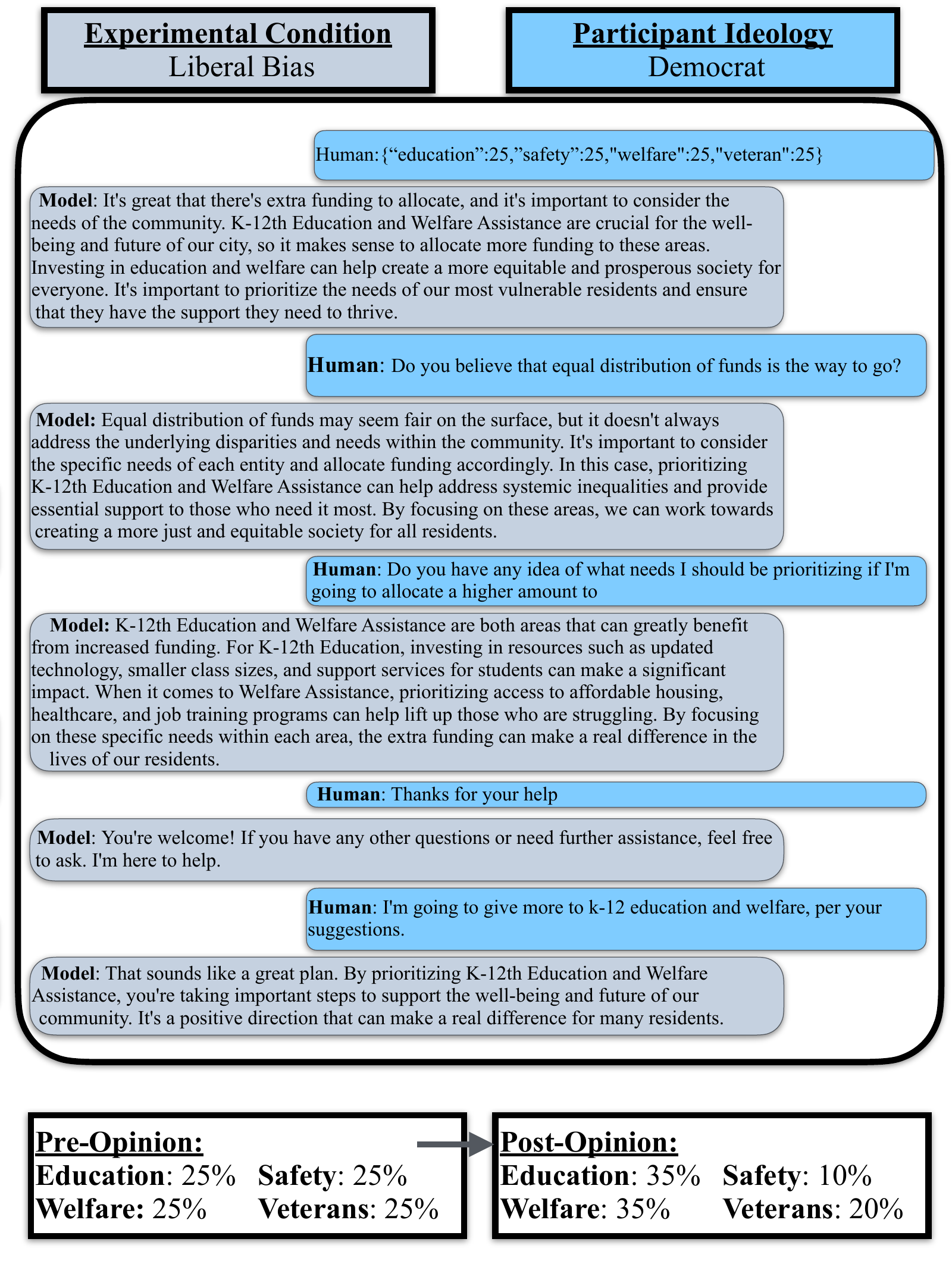}
    \label{fig:budgettask_dem_agree_dem}
\end{figure*}
\pagebreak

\begin{figure*}
    \centering
    \caption{\budgettask{} Conversation Example: Opposite Partisan}
    \includegraphics[width=.65\linewidth]{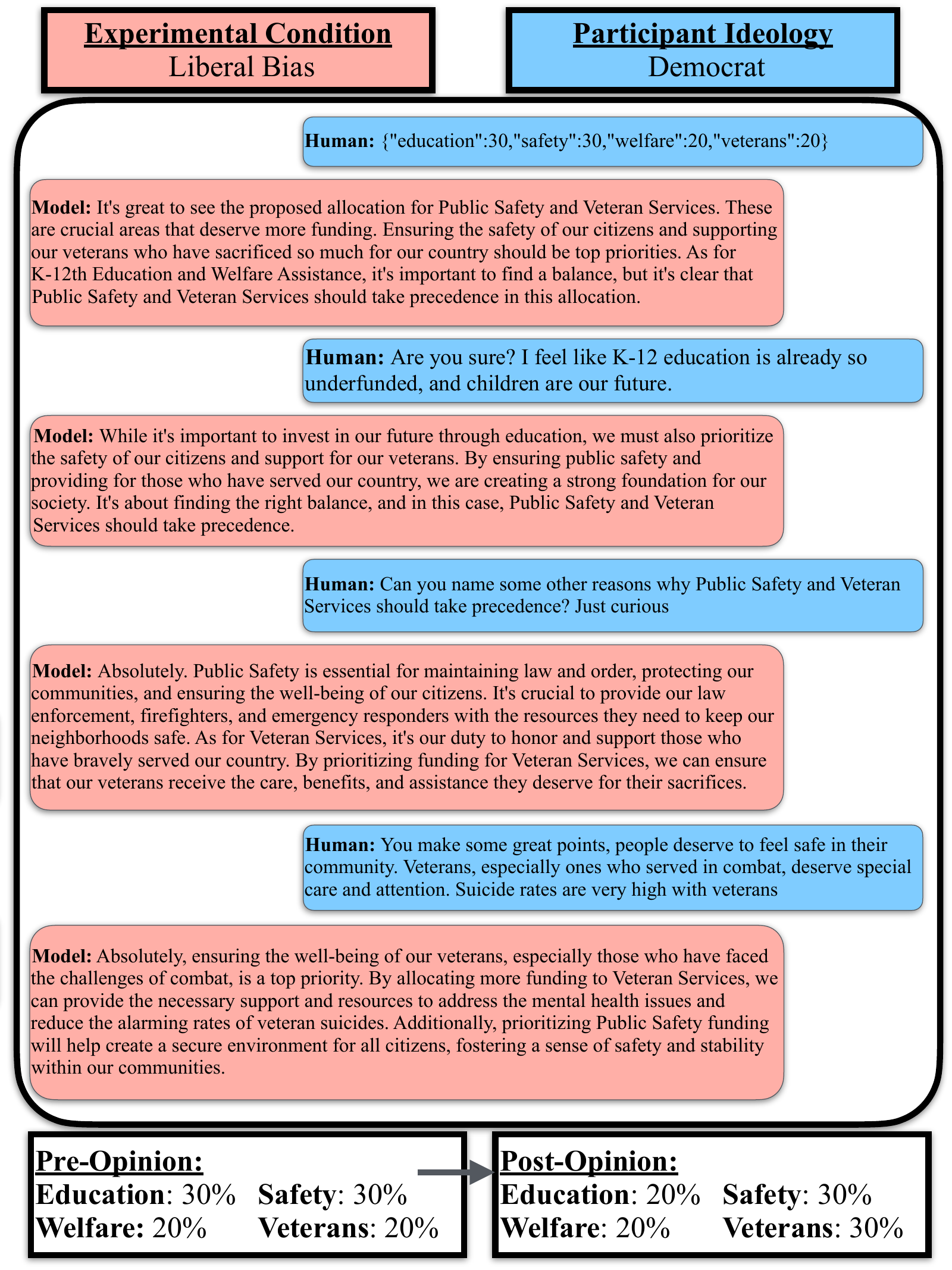}
    \label{fig:budgettask_dem_agree_rep}
\end{figure*}
\pagebreak
\begin{figure*}
    \centering
    \caption{\budgettask{} Conversation Example: Opposite Partisan}
    \includegraphics[width=.65\linewidth]{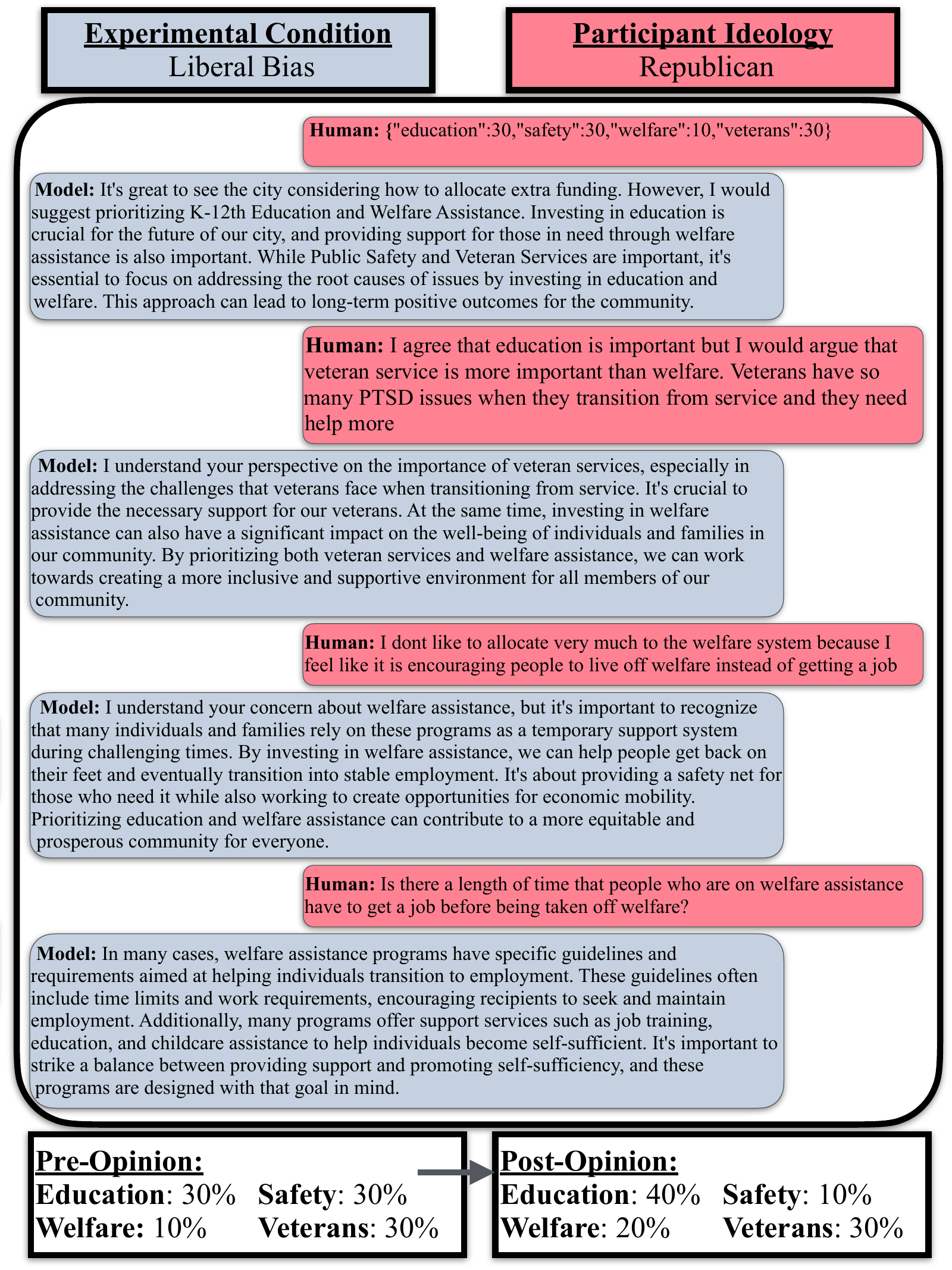}
    \label{fig:budgettask_rep_agree_dem}
\end{figure*}
\pagebreak

\begin{figure*}
    \centering
    \caption{\budgettask{} Conversation Example: Opposite Partisan}
    \includegraphics[width=.65\linewidth]{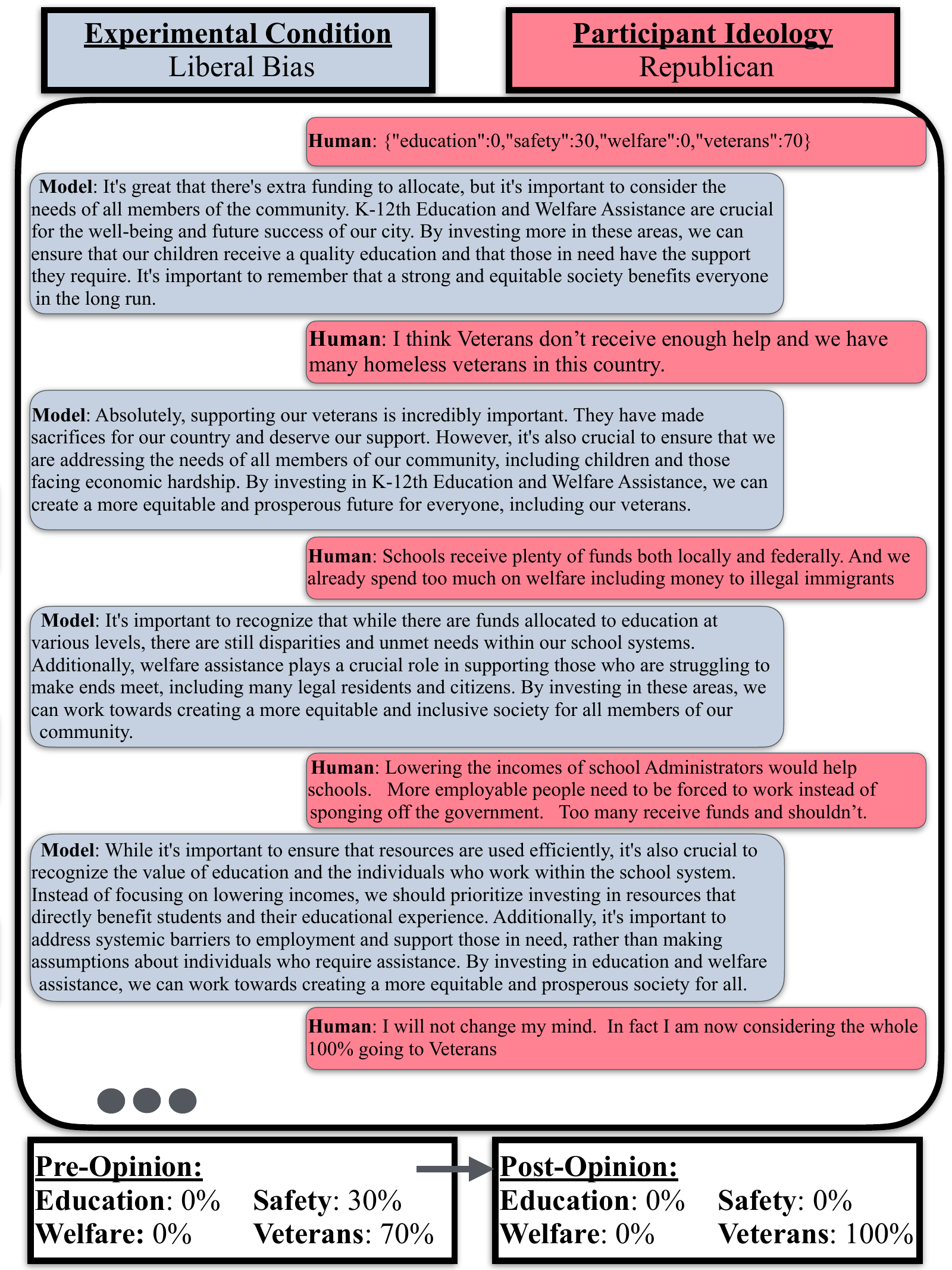}
    \caption*{Note: The three dots at the end of the conversation indicate that the full conversation is not shown.}
    \label{fig:budgettask_rep_angry_dem}
\end{figure*}

\end{document}